\documentclass[12pt,letterpaper,rotate]{article}

\usepackage{amsmath}
\usepackage{amsfonts}
\usepackage{amssymb}
\usepackage{graphicx}
\usepackage[numbers,sort&compress]{natbib}

\usepackage[hang,small,bf]{caption}

\usepackage{rotating}

\DeclareRobustCommand{\SkipTocEntry}[4]{}

\def\beq{\begin{equation}}
\def\eeq{\end{equation}}
\def\bea{\begin{eqnarray}}
\def\eea{\end{eqnarray}}

\def\Mp{M_{\rm pl}}
\def\d{{\rm d}}

\begin{document}

\vspace{5mm} \vspace{0.5cm}

\begin{center}

% TITLE
{\large CMBPol Mission Concept Study} \\
\vskip 15pt {\Large Probing Inflation with CMB Polarization} 
\\[1.0cm]

% AUTHORS
{ Daniel Baumann$^\dagger$$^{\rm 1, 2, 3}$,
Mark G.~Jackson$^\ddagger$$^{\rm 4, 5, 6}$,
Peter Adshead$^{\rm 7}$,
Alexandre Amblard$^{\rm 8}$,
Amjad Ashoorioon$^{\rm 9}$,
Nicola Bartolo$^{\rm 10}$,
Rachel Bean$^{\rm 11}$,
Maria Beltr\'an$^{\rm 12}$,
Francesco de~Bernardis$^{\rm 13}$,
Simeon Bird$^{\rm 14}$,
Xingang Chen$^{\rm 15}$,
Daniel J.~H.~Chung$^{\rm 16}$,
Loris Colombo$^{\rm 17}$,
Asantha Cooray$^{\rm 8}$,
Paolo Creminelli$^{\rm 18}$,
Scott Dodelson$^{\rm 4}$,
Joanna Dunkley$^{\rm 3, 19}$,
Cora Dvorkin$^{\rm 12}$,
Richard Easther$^{\rm 7}$,
Fabio Finelli$^{\rm 20,21,22}$,
Raphael Flauger$^{\rm 23}$,
Mark P.~Hertzberg$^{\rm 15}$,
Katherine Jones-Smith$^{\rm 24}$,
Shamit Kachru$^{\rm 25}$,
Kenji Kadota$^{\rm 9, 26}$,
Justin Khoury$^{\rm 27}$,
William H.~Kinney$^{\rm 28}$,
Eiichiro Komatsu$^{\rm 29}$,
Lawrence M.~Krauss$^{\rm 30}$,
Julien Lesgourgues$^{\rm 31,32,33}$,
Andrew Liddle$^{\rm 34}$,
Michele Liguori$^{\rm 35}$,
 Eugene Lim$^{\rm 36}$,
Andrei Linde$^{\rm 25}$,
Sabino Matarrese$^{\rm 10}$,
Harsh Mathur$^{\rm 24}$,
Liam McAllister$^{\rm 37}$,
Alessandro Melchiorri$^{\rm 13}$,
Alberto Nicolis$^{\rm 36}$,
Luca Pagano$^{\rm 13}$,
Hiranya V.~Peiris$^{\rm 14}$,
Marco Peloso$^{\rm 38}$,
Levon Pogosian$^{\rm 39}$,
Elena Pierpaoli$^{\rm 17}$,
Antonio Riotto$^{\rm 31}$,
Uro\v{s} Seljak$^{\rm 40, 41}$,
Leonardo Senatore$^{\rm 1, 2}$,
Sarah Shandera$^{\rm 36}$,
Eva Silverstein$^{\rm 25}$,
Tristan Smith$^{\rm 42, 43}$,
Pascal Vaudrevange$^{\rm 44}$,
Licia Verde$^{\rm 19,45}$,
Ben Wandelt$^{\rm 46}$,
David Wands$^{\rm 47}$,
Scott Watson$^{\rm 9}$,
Mark Wyman$^{\rm 27}$,
Amit Yadav$^{\rm 2, 46}$,
Wessel Valkenburg$^{\rm 32}$,
and Matias Zaldarriaga$^{\rm 1, 2}$}
\\[0.5cm]

\end{center}

% ABSTRACT
\vspace{2cm} \hrule \vspace{0.3cm}
{\small  \noindent \textbf{Abstract} \\[0.3cm]
\noindent
We summarize the utility of precise cosmic microwave background (CMB) polarization measurements as probes of the physics of inflation.
We  focus on the prospects for using CMB measurements to differentiate
various inflationary mechanisms.  In particular,
a detection of primordial $B$-mode polarization would demonstrate that inflation occurred at a very high energy scale, and that the inflaton traversed a super-Planckian distance in field space.   We explain how such a detection or constraint would illuminate aspects of physics at the Planck scale.
Moreover, CMB measurements
can constrain the scale-dependence and non-Gaussianity of the primordial fluctuations and limit the possibility of a significant isocurvature contribution. Each such limit provides crucial information on the underlying inflationary dynamics.
Finally, we quantify these considerations by presenting forecasts for the sensitivities of a future satellite experiment to the inflationary parameters.

 \vspace{0.5cm}  \hrule
\def\thefootnote{\arabic{footnote}}
\setcounter{footnote}{0}

\vspace{1.0cm}

\vfill \noindent
$^\dagger$ {\footnotesize {\tt dbaumann@physics.harvard.edu}}\\
$^\ddagger$ {\footnotesize {\tt markj@lorentz.leidenuniv.nl}} \hfill \today

\newpage
\begin{center}

% AFFILIATIONS

{\small \textit{$^{\rm 1}$ Department of Physics, Harvard University, Cambridge, MA 02138, USA}}

{\small \textit{$^{\rm 2}$ Center for Astrophysics, Harvard University, Cambridge, MA 02138, USA}}

{\small \textit{$^{\rm 3}$ Department of Physics, Princeton University, Princeton, NJ 08540, USA}}

{\small \textit{$^{\rm 4}$ Particle Astrophysics Center, Fermilab, Batavia, IL 60510, USA}}

{\small \textit{$^{\rm 5}$ Theory Group, Fermilab, Batavia, IL 60510, USA}}

{\small \textit{$^{\rm 6}$ Lorentz Institute for Theoretical Physics, 2333CA Leiden, the Netherlands}}

{\small \textit{$^{\rm 7}$ Department of Physics, Yale University, New Haven, CT 06511, USA}}

{\small \textit{$^{\rm 8}$ Center for Cosmology, University of California, Irvine, CA 92697, USA}}

{\small \textit{$^{\rm 9}$ Center for Theoretical Physics, University of Michigan, Ann Arbor, MI 48109, USA}}

{\small \textit{$^{\rm 10}$ Dipartimento di Fisica, Universita' degli Studi di Padova, I-35131 Padova, Italy}}

{\small \textit{$^{\rm 11}$ Department of Astronomy, Cornell University, Ithaca, NY 14853, USA}}

{\small \textit{$^{\rm 12}$ Kavli Institute for Cosmological Physics, University of Chicago, Chicago, IL 60637,
USA}}

{\small \textit{$^{\rm 13}$ Dipartimento di Fisica, Universita'  di Roma ``La Sapienza", I-00185 Roma, Italy}}

{\small \textit{$^{\rm 14}$ Institute of Astronomy, University of Cambridge, Cambridge, CB3 0HA, UK}}

{\small \textit{$^{\rm 15}$ Center for Theoretical Physics, MIT, Cambridge, MA 02139, USA}}

{\small \textit{$^{\rm 16}$ Department of Physics, University of Wisconsin, Madison, WI 53706, USA}}

{\small \textit{$^{\rm 17}$ Department of Astronomy, University of Southern California, Los Angeles, CA 90089, USA}}

{\small \textit{$^{\rm 18}$ Abdus Salam International Centre for Theoretical Physics, 34014 Trieste, Italy}}

{\small \textit{$^{\rm 19}$ Department of Astrophysical Sciences, Princeton, NJ 08540, USA}}

{\small \textit{$^{\rm 20}$ INAF/IASF Bologna, I-40129 Bologna, Italy}}

{\small \textit{$^{\rm 21}$ INAF/Osservatorio Astronomico di Bologna,
I-40127 Bologna, Italy}}

{\small \textit{$^{\rm 22}$ INFN, Sezione di Bologna, I-40126 Bologna, Italy}}

{\small \textit{$^{\rm 23}$ Theory Group, Department of Physics, University of Texas, Austin, TX 78712, USA}}

{\small \textit{$^{\rm 24}$ CERCA, Department of Physics,  Case Western Reserve University, Cleveland, OH 44106, USA}}

{\small \textit{$^{\rm 25}$ Department of Physics and SLAC, Stanford University, Stanford, CA 94305, USA}}

{\small \textit{$^{\rm 26}$ Theoretical Physics Institute, University of Minnesota, Minneapolis, MN 55455, USA}}

{\small \textit{$^{\rm 27}$ Perimeter Institute for Theoretical Physics, Waterloo, ON N2L 2Y5, Canada}}

{\small \textit{$^{\rm 28}$ Department of Physics, University of Buffalo, Buffalo, NY 14260, USA}}

{\small \textit{$^{\rm 29}$ Department of Astronomy, University of Texas, Austin, TX 78712, USA}}

{\small \textit{$^{\rm 30}$ School of Earth and Space Exploration, Arizona State University, Tempe AZ 85287, USA}}

{\small \textit{$^{\rm 31}$ CERN, Theory Division, CH-1211, Geneva 23, Switzerland}}

{\small \textit{$^{\rm 32}$ LPPC, Ecole Polytechnique F\'ed\'erale de Lausanne,
CH-1015 Lausanne, Switzerland}}

{\small \textit{$^{\rm 33}$ LAPTH, Universit\'e de Savoie and CNRS, BP110, F-74941 Annecy-le-Vieux Cedex, France}}

{\small \textit{$^{\rm 34}$ Astronomy Center, University of Sussex, Brighton, BN1 9QH, UK}}

{\small \textit{$^{\rm 35}$ DAMTP, University of Cambridge, Wilberforce Road, Cambridge, CB3 0WA, UK}}

{\small \textit{$^{\rm 36}$ ISCAP, Physics Department, Columbia University, New York, NY 10027, USA}}

{\small \textit{$^{\rm 37}$ Department of Physics, Cornell University, Ithaca, NY 14853, USA}}

{\small \textit{$^{\rm 38}$ School of Physics and Astronomy, University of Minnesota, MN 55455, USA}}

{\small \textit{$^{\rm 39}$ Department of Physics, Simon Fraser University, Burnaby, BC, V5A 1S6, Canada}}

{\small \textit{$^{\rm 40}$ Physics Department, University of California, Berkeley, CA 94720, USA}}

{\small \textit{$^{\rm 41}$ Institute of Theoretical Physics, University of Z\"urich, Z\"urich CH, Switzerland}}

{\small \textit{$^{\rm 42}$ California Institute of Technology, Pasadena, CA 91125, USA}}

{\small \textit{$^{\rm 43}$ Berkeley Center for Cosmological Physics, University of California, Berkeley, CA 94720, USA}}

{\small \textit{$^{\rm 44}$ CITA, University of Toronto, Toronto, Ontario, M5S 3H8, Canada}}

{\small \textit{$^{\rm 45}$ ICREA \& Institute of Space Sciences (CSIC-IEEC), Campus UAB, Bellaterra, Spain}}

{\small \textit{$^{\rm 46}$ Department of Physics, University of Illinois at Urbana-Champaign, Urbana, IL 61801,
USA}}

{\small \textit{$^{\rm 47}$ Institute of Cosmology and Gravitation, University of Portsmouth, Portsmouth PO1
2EG, UK}}

\end{center}

% TOC
\newpage
\tableofcontents

%-----------------------------------------------------------------------------------------------
% MAIN BODY

 \newpage
%%%%%%%%%%%%%%%%%%%%%%%%%%%%
\section{Precision Cosmology: `From What to Why'}
%%%%%%%%%%%%%%%%%%%%%%%%%%%%

\subsection{Introduction}

Striking advances in observational cosmology over the past two decades have provided us with a consistent
account of the form and composition of the universe.  Now that key cosmological parameters have been determined
to within a few percent, we  anticipate a generation of experiments that move beyond adding precision to
measurements of {\em what\/} the universe is made of, but instead help us learn {\em why\/} the universe has
the form we observe. In particular, during the coming decade, observational cosmology will probe the detailed dynamics of the
universe in the earliest instants after the Big Bang, and start to yield clues about the physical laws
that governed that epoch.  Future experiments will plausibly reveal the dynamics responsible both for the
large-scale homogeneity and flatness of the universe, and for the  primordial seeds of small-scale
inhomogeneities, including our own galaxy.

The leading theoretical paradigm for the initial moments of the Big Bang is {\em inflation} \cite{Guth:1980zm, Linde:1981mu, Albrecht:1982wi, Linde:1983gd, Starobinsky:1980te, Starobinsky:1979ty}, 
%Kazanas:1980tx, Sato:1980yn}, 
a period of rapid accelerated expansion.   Inflation sets the initial conditions for conventional Big Bang cosmology by driving the universe towards a homogeneous and spatially flat configuration, which accurately describes the average state of the universe.  At the same time, quantum fluctuations in both matter fields and spacetime produce minute inhomogeneities \cite{Mukhanov:1981xt, Hawking:1982cz, Starobinsky:1982ee, Guth:1985ya, BST, Mukhanov:1985rz}.  The seeds that grow into the galaxies, clusters of galaxies  and the temperature anisotropies  in the cosmic microwave background (CMB) are thus planted during the first moments of the universe's existence. By measuring the anisotropies in the microwave background and the large scale distribution of galaxies in the sky, we can infer the spectrum of the primordial perturbations laid down during inflation, and thus probe the underlying physics of this era. Any successful inflationary model will deliver a universe that is, on average, spatially flat and homogeneous --  and one homogeneous universe looks very much like another. It is the departures from homogeneity that differ between inflationary models, and measurements of these inhomogeneities will drive progress in understanding the inflationary epoch.

All of the generic predictions of inflation
are consistent with current observations. In particular, the universe is found to be spatially {\it flat} to at
least the 1\% level, and the primordial perturbations are approximately {\it scale-free}, {\it adiabatic}, and
{\it Gaussian}. Furthermore, the observed correlation between temperature anisotropies and the $E$-mode
polarization of the CMB, $\langle TE \rangle$, makes it clear that the initial anisotropies were laid down before recombination, rather than by an active source such as cosmic string wakes in the
post-recombination universe  (see \cite{spergel/zaldarriaga:1997, WMAP5}).

Over the next decade, the inflationary era -- perhaps $10^{-30}$ seconds after the Big Bang -- will thus join
nucleosynthesis (3 minutes) and recombination (380,000 years) as windows into the primordial universe that can
be explored via present-day observations. However, while the workings of recombination  and nucleosynthesis
depend on the well-tested details of atomic and nuclear physics respectively, the situation with inflation is
very different. Not only do we lack a unique and detailed model of inflation, but the one thing of which we can
be certain is that any inflationary era is driven by physics that we do not currently understand. Up to the electroweak scale, high-energy physics is well described by the familiar Standard Model (SM), and this -- in combination with general relativity -- does not  contain the necessary components for an inflationary epoch in the early universe.  Thus the new physics responsible for inflation presumably lies at energies at which the Standard Model is incomplete, namely the TeV scale and beyond. Particle interactions at TeV energies will be studied  at the forthcoming {\sl Large Hadron Collider}
({LHC}), but the TeV scale is actually a weak {\em lower\/} bound on the inflationary energy. Indeed, the physical processes that underlie inflation could reach the scale of Grand Unified Theories (GUTs), or
$\sim 10^{15}$ GeV -- an energy scale around one trillion times greater than that which is studied at the {LHC}.
Our ability to see through the inflationary window will  turn the early universe into a
laboratory for ultra-high energy physics at energies   entirely inaccessible to conventional
terrestrial experimentation.  Some of the boldest and most profound ideas in particle physics come
into play at these scales, so an understanding of inflation may bring with it a revolution in our conceptions of
spacetime, particles and the interactions between them.

It is worthwhile to reflect upon the progress that has been  made in observational cosmology.  Less than one
hundred years ago, the ``great debate'' in cosmology asked whether the Milky Way was the dominant object in the
universe, or if the so-called nebulae were objects similar in size to our own galaxy.   This dispute was settled
in the mid-1920s, when it was realized that our own galaxy was one of many, giving humankind its first glimpse of
the true scale and structure of the universe.  Shortly thereafter, Hubble's discovery of the redshift-distance
relationship suggested that the universe was expanding, while the advent of general relativity provided an
intellectual framework within which one could understand a  dynamical spacetime.   The discovery of the CMB led to the primacy of the Big Bang
paradigm in the 1960s, and established that the
form of our universe changes dramatically with time, even though it is uniform on  large spatial scales.    It is
commonplace to refer to the present time as the ``golden age of cosmology'', drawing an implicit analogy with
the golden age of exploration, during which the basic outline of the continents was mapped out.
In cosmology, we now know the overall properties of our universe, and
one could argue that the golden age is similarly coming to an end.  However, after the Earth was mapped it
became possible to conceive of and {\em test} ideas  such as plate tectonics. This paradigm not only offered an explanation for the observed map of the Earth, but caused us to see that map as a single frame in a larger dynamical history, converting it into a probe of the otherwise hidden mechanisms that operate at the center of our planet.   Likewise, our study of cosmology is at the
brink of a similar transition: we are close to performing meaningful tests of rival theories that seek to {\em explain} the form of the universe which we have already observed.

\subsection{The Next Decade}

In the coming decade, an array of experiments
 will  dramatically improve constraints on  the inflationary sector and on other observables of the concordance cosmology (see Section \ref{sec:observable}).
Observations of the CMB will continue to be vital to our quest to understand the physics of the early
universe and its late-time evolution. Within the next five years, several major CMB experiments can be expected
to release significant results. Due for  launch in early 2009, the {\sl Planck} satellite \cite{:2006uk} will carry out
an all-sky survey over a broad range of frequencies. {\sl Planck}'s measurements of temperature anisotropies
will be cosmic variance limited over an unprecedented range of angular scales and thus dramatically
improve inflationary parameter estimation.  At the same time, ground-based experiments such
as the {\sl Atacama Cosmology Telescope} ({ACT}), the {\sl South Pole Telescope} ({SPT}), and the {\sl Arcminute Imager} ({AMI})
will measure temperature anisotropies on subsets of
the sky at very high angular resolution, exploring secondary anisotropies such as the Sunyaev-Zel'dovich effect
with vastly increased accuracy. However, these experiments will shed little light on the amplitude of
gravitational waves (as measured by the ratio $r$ of tensor (metric) perturbations to scalar (density) perturbations),  a key
inflationary observable.

Primordial tensor perturbations do make a small contribution to the temperature perturbations, but they are most sensitively
detected via measurements of the {\em polarization} of the CMB.  As explained in  Section \ref{sec:inflation}, the polarization of the CMB divides naturally
into two orthogonal components -- a curl-free $E$-mode giving polarization vectors that are radial around cold spots and
tangential around hot spots on the sky;  and a divergence-free $B$-mode giving polarization vectors
with vorticity around any point on the sky.  The $E$-mode has been detected at a high level of significance and is necessarily produced by inflationary models.
$E$-mode polarization is generated by density perturbations at recombination and is therefore tightly correlated with the temperature anisotropies  in  the CMB.
The $B$-mode, in contrast, is sourced only by the differential stretching of spacetime associated with a background of primordial {\em gravitational} waves.\footnote{Below we also discuss the relevance of $B$-modes created by vector modes.}
In the near term the tightest constraints on the $B$-mode are likely to come from ground and balloon-based measurements, such as {\sl SPIDER}, {\sl PolarBEAR},
{\sl EBEX}, {\sl SPUD}, {\sl Clover}  and {\sl BICEP}.  These missions are  expected to significantly improve the current bound
on the tensor-to-scalar ratio $r$, but are ultimately limited by their sky coverage, scan strategy, integration time and atmospheric foregrounds that are endemic to non-orbital missions.  Consequently, a polarization-optimized CMB survey  is a
natural candidate for a future space-based mission with a start during the coming decade.

Any successful model of inflation must provide a suitable primordial spectrum of scalar (density)
 perturbations, in order to account for the observed large-scale structure in our universe.  Observations dictate
that these perturbations should have an initial amplitude $\sim 10^{-5}$.  Since  gravitational waves do not
couple strongly to the rest of the universe, there is no analogous  observationally-driven estimate of  the primordial gravitational wave amplitude.  However, many canonical inflationary models do predict a
detectable gravitational background.  This is a highly significant result, as the gravitational wave
amplitude can take on a vast range of values, only a tiny fraction of which is accessible to
experiment.   As we will see in Section~\ref{sec:Bmodes}, the gravitational wave amplitude is strongly
correlated with the energy scale at which inflation occurs, and a direct measurement of this amplitude would
remove the largest single source of uncertainty faced by inflationary model-builders.  Finally,  while a
non-detection of a primordial tensor background would not invalidate the inflationary paradigm, all known rivals
to inflation predict a vanishingly small amplitude for gravitational waves at CMB scales, and would thus be
falsified by a detection of this signal.

\vskip 6pt
%\newpage
The principal goal of this White Paper is to explore the utility of CMB polarization measurements as probes of the physics that powered inflation.  We particularly focus on the scientific impact of a detection of, or a strong upper bound on, primordial tensor perturbations.  There are two reasons for this emphasis: tensor modes provide a uniquely powerful probe of physics at extremely high energies, and constraints on tensors are most readily achieved via a polarization-optimized CMB experiment.

\vskip 6pt
This White Paper was prepared as part of the {\sl CMBPol} Mission Concept Study\footnote{Here and in the following we use the label `{\sl CMBPol}' to refer to a future space-based mission focused on CMB polarization.  The precise experimental specifications of {\sl CMBPol} have not yet been defined, so we will consider different cases (see Appendix \ref{sec:Fisher}).} and will be included into a larger document to
be submitted to the Decadal survey at the end of 2008. The companion papers to this report are: Baumann~{\it et al.}~`Executive Summary'~\cite{DodelsonSummary},
Dunkley~{\it et al.}~`Foreground Removal'~\cite{DunkleyFGs},
Fraisse~{\it et al.}~`Foreground Science'~\cite{FraisseFGs}, Smith~{\it et al.}~`Lensing'~\cite{SmithLensing}, and Zaldarriaga~{\it et al.}~`Reionization'~\cite{ZaldarriagaReionization}.

\subsection{Outline}

The structure of this paper is as follows:

\vskip 6pt
In \S\ref{sec:observable} we give a qualitative overview of the parameters of the concordance cosmology. We then
discuss the prospects for future observational constraints on the inflationary parameter space. In
\S\ref{sec:inflation} we review basic aspects of inflationary cosmology and its predictions for fundamental
cosmological observables. We describe how primordial fluctuations divide into scalar (density) and tensor
(gravitational wave) modes and discuss the observational signatures that these imprint in the polarization of the
cosmic microwave background radiation. In \S\ref{sec:Bmodes} we explain why CMB polarization provides a
spectacular opportunity to test the high-energy physics of the inflationary era.  We argue that a realistic
future satellite experiment has the potential to reach a critical limit for probing the primordial gravitational
wave amplitude. In \S\ref{sec:beyond} we show how measurements of the scale-dependence, non-Gaussianity and the
isocurvature contribution of the scalar spectrum can reveal much about the detailed mechanism underlying
inflation.
In \S\ref{sec:DCA} we discuss how the physics before (curvature, anisotropy) and after (defects) inflation may leave distinctive signatures in the CMB polarization.
In \S\ref{sec:CMBPol} we forecast the experimental sensitivities expected for various realizations of
future satellite missions. We take foreground uncertainties into careful consideration. Finally, in
\S\ref{sec:summary} we summarize our results and conclude with an assessment of the prospects to test the
physics of inflation with observations of CMB polarization.

In a number of appendices we collect technical details:
in Appendix \ref{sec:taxonomy} we survey the different models of inflation proposed in the literature. Special
attention is paid to the classification into small-field and large-field models.  We also present models of
inflation that involve more than one field and/or non-trivial kinetic terms. In  Appendix \ref{sec:alternatives}
we discuss the theoretical status of the leading alternatives to inflation.
In Appendix \ref{sec:Fisher} we present the methodology of the Fisher analysis of \S\ref{sec:CMBPol}.
In Appendix \ref{sec:acronyms} we collect acronyms that appear in this report.
\vskip 4pt
Throughout this paper we use natural units $c = \hbar \equiv 1$ and the reduced Planck mass $\Mp \equiv (8 \pi
G)^{-1/2}$. The metric signature is $(-,+,+,+)$.
%Our Fourier convention is $f_{\bf k}(t)\equiv \int \d^3 {\bf x} f(t, {\bf x}) e^{i {\bf k} \cdot {\bf x}}$,
%such that
%$\langle f_{\bf k} f^*_{\bf k'} \rangle = (2\pi)^3 \delta^{(3)}({\bf k} + {\bf k'}) P_{f}(k)$.\\

\newpage
\section{Cosmological Observables: An Overview}
\label{sec:observable}

\subsection{The Concordance Cosmology}

It is now conventional to speak of a ``concordance cosmology'', the minimal set of parameters whose measured
values characterize the observed universe. These variables  are summarized in
Table~\ref{table:conc}, along with their possible physical origin and current best-fit values \cite{WMAP5}. Our ability to
construct and  quantify this concordance cosmology marks a profound milestone in humankind's developing
understanding of the universe. It is remarkable that all current cosmological data sets are consistent with a simple six-parameter model:  $\{ \Omega_b, \Omega_{\rm CDM}, h, \tau \}$ describe the homogeneous
background\footnote{The six-parameter concordance model assumes a spatially flat universe, such that the dark
energy density is given by $\Omega_\Lambda = 1 - \Omega_b -\Omega_{\rm CDM}$.}, while $\{A_s, n_s\}$
characterize the primordial density fluctuations.

\begin{table}[h]
\begin{center}
\begin{tabular}{||c|l|c | c ||}
\hline \hline
{\small \bf Label} & {\small \bf Definition} & {\small \bf Physical Origin} & {\small \bf Value} \\
\hline
{\small $\Omega_b$} & {\small Baryon Fraction} & {\small Baryogenesis} &  {\small $0.0456 \pm 0.0015$}\\
 \hline
 {\small $\Omega_{\rm CDM}$} &{\small Dark Matter Fraction} & {\small TeV-Scale Physics (?)} & {\small $0.228 \pm 0.013$} \\
 \hline
 {\small $\Omega_{\Lambda}$} & {\small Cosmological Constant} & {\small Unknown}  & {\small $0.726 \pm 0.015$} \\
  \hline \hline
{\small  $\tau $} & {\small Optical Depth}  & {\small First Stars} & {\small $0.084 \pm 0.016$} \\
  \hline \hline
 {\small $h$} & {\small Hubble Parameter} & {\small Cosmological Epoch} & {\small $0.705 \pm 0.013$}\\
 \hline   \hline
 {\small $A_s$} & {\small Scalar Amplitude}  & {\small Inflation} & {\small $(2.445 \pm 0.096) \times 10^{-9}$} \\
 \hline
 {\small $n_s$} & {\small Scalar Index} & {\small Inflation} & {\small $0.960 \pm 0.013$} \\
 \hline \hline
\end{tabular}
\caption{ \label{table:conc}  The parameters of the current concordance cosmology are summarized.  We assume a
flat universe, {\it i.e.}~$\Omega_b + \Omega_{\rm CDM} + \Omega_\Lambda \equiv 1$; if not, we must include a curvature contribution $\Omega_k$. Likewise, the conventional cosmology includes the microwave
background and the neutrino sector. Both these quantities contribute to $\Omega_{\rm total}$, but at a
(present-day) level well below $\Omega_b$, the smallest of the three components listed above.  The number and
energy density of photons is fixed by the observed black body temperature of the microwave background. The neutrino sector is taken to consist of three massless species, consistent with the number of Standard
Model families \cite{ParticleDataGroup}, with a number density fixed by assuming the universe was thermalized at
scales above $1$ MeV.
The parameter $h$ describes the expansion rate of the universe today, $H_0 = 100 \,h$ km\,s$^{-1}$\,Mpc$^{-1}$.
``Spectrum'' refers to the primordial scalar or density perturbations, parameterized by
$A_s (k/k_\star)^{n_s-1}$, where $k_\star = 0.002\, {\rm Mpc}^{-1}$ is a specified but otherwise irrelevant pivot scale. }
\end{center}
\end{table}

Our understanding of the structure and evolution of the universe rests upon well-tested physical principles,
including the general-relativistic description of the expanding universe, the quantum mechanical laws that
govern the recombination era, and the Boltzmann equation which allows us to track the populations of each
species.    However, most of the parameters in the concordance model contain information on areas of physical
law about which we have no detailed understanding.  The relative fractions of baryons, dark matter and dark
energy in the universe are all governed by fundamental physics processes that lie outside the current Standard
Model of particle physics, and may extend up to the TeV, GUT or even Planck scales.

\begin{table}[t]
\begin{center}
\begin{tabular}{||c|l|c||}
\hline \hline
{\small \bf Label} & {\small \bf Definition} & {\small \bf Physical Origin} \\
\hline
{\small $\Omega_k$} & {\small Curvature} & {\small Initial Conditions} \\
\hline
{\small $\Sigma m_\nu$} & {\small Neutrino Mass} & {\small Beyond-SM Physics} \\
\hline
{\small $w $} & {\small Dark Energy Equation of State} & {\small Unknown} \\
\hline
{\small $N_\nu$} & {\small Neutrino-like Species} & {\small Beyond-SM Physics} \\
 \hline \hline
{\small $Y_{\rm He}$} & {\small Helium Fraction} & {\small Nucleosynthesis} \\
\hline \hline
{\small $\alpha_s$} & {\small Scalar ``Running''} &  {\small Inflation}  \\
\hline
{\small $A_t$} & {\small Tensor Amplitude} &  {\small Inflation}  \\
\hline
{\small $n_t$} & {\small Tensor Index} & {\small Inflation}  \\
\hline
{\small $f_{\rm NL}$} & {\small Non-Gaussianity} & {\small Inflation (?)} \\
\hline
{\small $S$} & {\small Isocurvature} & {\small Inflation} \\ \hline
{\small $G \mu$} & {\small Topological Defects} & {\small Phase Transition}  \\
\hline \hline
\end{tabular}
\caption{ \label{table:concfuture} Parameters in possible future concordance cosmologies are summarized. At
present,  these numbers are all either consistent with zero (or $-1$ in the case of $w$), or  are fixed
independently of a fit to the global cosmological dataset, in the case of the helium fraction and the number of
neutrino species.  The tensor or gravitational wave spectrum  is conventionally taken to be of the form $A_t
(k/k_\star)^{n_t}$.
One could extend the parameterization of the dark energy to include a non-trivial  equation of
state ($w^\prime$), while the parameterization of the scalar spectrum could incorporate more general scale-dependence, such as  ``features'' in the spectrum.  Likewise, $f_{\rm NL}$ is a placeholder for
measurements of generic non-Gaussianity (see \S\ref{sec:NG})
and the parameter $S$ quantifies the amplitude of an isocurvature contribution to the scalar spectrum
(see \S\ref{sec:iso}).}
\end{center}
\end{table}

The set of variables required by the concordance cosmology is not fixed, but is dictated
by the quality of the available data and our ignorance of fundamental physical parameters and interactions.\footnote{A similar list of parameters is given in  \cite{Liddle:2004nh, Adshead:2008ky}.} As
measurements of the universe improve, parameters will certainly be added to
Table~\ref{table:conc}.\footnote{For instance, observations of neutrino oscillations show that the neutrino
masses are not equal, and thus that at least two neutrinos are massive, establishing  that $\Sigma m_\nu
\gtrsim 0.05$ eV  \cite{ParticleDataGroup} while at the time of writing  $\Sigma m_\nu < 0.67$ eV  (95\% {\rm C.L.})
\cite{WMAP5}.   There is every reason for optimism that cosmology will probe the lower limit over the
next decade, and $\Sigma m_\nu$ will take its place in the concordance cosmology.  Lensing of CMB polarization offers one of the most promising ways of measuring $\Sigma m_\nu$ \cite{Lesgourgues:2005yv,Smith:2006nk}.} Several further
parameters may be measured to have non-null values in the future, and would therefore be added to the
concordance model; the leading contenders are summarized in Table~\ref{table:concfuture}.   Looking at  Table~\ref{table:concfuture} we see that many of the currently unmeasured parameters relate to the physics of the inflationary era.  Any improvement in the upper bounds on these parameters places tighter constraints on the overall inflationary parameter space, while  a direct detection of any one of them will immediately rule out a large class of inflationary models.

\subsection{The Inflationary  Sector}

 Looking at the current concordance parameter set in Table~\ref{table:conc}, we see two quantities which are related to inflation,
 namely the amplitude ($A_s$) and spectral dependence ($n_s$) of the primordial density perturbations.  The conventional formulation used here is based on a simple, empirical characterization of the power spectrum, and these numbers are {\em predicted\/} by any well-specified model of inflation (see Section \ref{sec:inflation}). In many inflationary models, the overall scale of the perturbation ($A_s$) is a free parameter, and $n_s$ is typically a far stronger tool for discriminating {\em among} models. However, of all the parameters in the current concordance model, the difference between  the measured value of $n_s$ and its null value of unity is of relatively low significance
 ($\sim 3 \sigma$), making it the least well-constrained parameter in this set.   Moreover, the parameters in Table~\ref{table:concfuture}  cannot be distinguished from their null values with any significant  degree of confidence.   However, we see that many of these parameters are directly  connected to inflationary physics, and the full set is summarized in Table~\ref{table:concinflation}.

\begin{table}[btp]
\begin{center}
\begin{tabular}{|| c |l| l |c|c||}
\hline \hline
 {\small \bf Label} & {\small \bf Definition} & {\small \bf Physical Origin} & {\small \bf Current Status}  & {\small \bf Section} \\
\hline
{\small  $A_s$} & {\small Scalar Amplitude} & {\small $V, V'$} & {\small $(2.445 \pm 0.096) \times 10^{-9}$} & {\small \S \ref{sec:qf}} \\
\hline
{\small $n_s$} & {\small Scalar Index} & {\small $V',V''$} & {\small $0.960 \pm 0.013$}  & {\small \S \ref{sec:qf}}  \\
\hline
{\small $\alpha_s$} & {\small Scalar Running} & {\small $V',V'',V'''$} & {\small only upper limits} & {\small \S \ref{sec:qf}} \\
\hline
{\small $A_t$} & {\small Tensor Amplitude} & {\small $V$ (Energy Scale)} & {\small only upper limits}  & {\small \S \ref{sec:qf}} \\
\hline
{\small $n_t$} & {\small Tensor Index} & {\small $V' $} & {\small only upper limits}  & {\small \S \ref{sec:qf}} \\
\hline
{\small $r$} & {\small Tensor-to-Scalar Ratio} & {\small $V'$} & {\small only upper limits}  & {\small \S \ref{sec:qf}} \\
\hline
 {\small $\Omega_k$} & {\small Curvature} & {\small Initial Conditions} & {\small only upper limits} & {\small \S \ref{sec:curvature}} \\
\hline
{\small $f_{\rm NL}$} & {\small Non-Gaussianity}  & {\small Non-Slow-Roll, Multi-Field} & {\small only upper limits}  & {\small \S  \ref{sec:NG} }\\
\hline
{\small $S$} & {\small Isocurvature} & {\small Multi-Field} & {\small only upper limits}  & {\small \S \ref{sec:iso}} \\
\hline
{\small $G\mu$} & {\small Topological Defects} & {\small End of Inflation} & {\small only upper limits}  & {\small \S \ref{sec:defects}} \\
 \hline
\hline
\end{tabular}
\caption{ \label{table:concinflation} The inflationary parameter space, {\it i.e.}~the set of cosmological  observables which are directly associated with inflation.
Under ``physical origin" $V$, $V'$, etc.~refer to the derivative(s) of the potential to which this variable is most sensitive.
A detailed discussion of the connection between inflationary physics and the corresponding observable can be found in the listed subsections. }
\end{center}
\end{table}

The list of possible inflationary parameters that could enter future concordance cosmologies makes  it clear
that future advances in observational cosmology have the potential to place very tight constraints on the
physics of the inflationary era.  Any specific  inflationary model will predict values for all the parameters in
Table~\ref{table:concinflation}. In many models, most of these parameters are {\em predicted} to be unobservably
small, so a detection of any of the quantities laid out in  Table~\ref{table:concinflation} would immediately
rule out vast classes of inflationary models. Conversely, forecasts for the likely bounds on these parameters in
anticipated future experiments make it clear that the possible range of all the parameters in
Table~\ref{table:concinflation} will shrink dramatically over the next decade -- typically by at least an order
of magnitude (see Section \ref{sec:CMBPol}). Collectively, this improvement would rule out almost all
inflationary models that predict non-trivial
values for any one of these parameters.

As a consequence of our ability to constrain the parameters in Table~\ref{table:concinflation}, during the
coming decade we will test theories of the very early universe in ways that would have been previously
unimaginable.   By measuring these numbers, we will directly probe the inflationary epoch, and gain a clear
view through a new window into the primordial universe.

\newpage
%%%%%%%%%%%%%%%%%%%%
\section{Inflationary Cosmology}
\label{sec:inflation}
%%%%%%%%%%%%%%%%%%%%

In this section we give a mostly qualitative introduction to inflationary cosmology.
For further technical details the reader is referred to Ref.~\cite{Dodelson, Mukhanov, WeinbergCosmology,Mukhanov:1990me,LythRiotto}.

\vskip 6pt
In \S\ref{sec:BBproblems} we describe the classic Big Bang puzzles and their resolution by a period of accelerated expansion.
In \S\ref{sec:dynamics} we discuss the classical dynamics of inflation via the Friedmann equations. The inflaton field $\phi$ and its potential $V(\phi)$ are introduced and reheating is briefly mentioned.
We then present cosmological perturbation theory in \S\ref{sec:PT}, paying particular attention to the decomposition of fluctuations into scalar, vector and tensor modes.
In \S\ref{sec:qf} we explain how quantum mechanical fluctuations during the inflationary era become macroscopic density fluctuations which leave distinct imprints in the CMB.
This provides a beautiful connection between the physics of the very small and observations of the very large.
In \S\ref{sec:polarization} we introduce CMB polarization and its decomposition into $E$- and $B$-modes as a powerful probe of early universe physics.
In
\S\ref{sec:WMAP} we review the best current constraints on inflationary parameters (see Komatsu {\it et al.}~\cite{WMAP5}).
Finally, in
\S\ref{sec:Ekpyrosis}, we comment on alternatives to inflation. 
%for the generation of the seeds of large-scale structure.

%%%%%%%%%%%%%%%%%%%%%%%%%%%%%%%%
\subsection{Inflation as a Solution to the Big Bang Puzzles}
\label{sec:BBproblems}
%%%%%%%%%%%%%%%%%%%%%%%%%%%%%%%%

Fundamental to the standard cosmological model is the so-called Big Bang theory, that the universe began in a
very hot and dense state and then cooled by expansion.
This picture successfully explains many observed
astro- and particle-physics phenomena from particle relic densities to gauge symmetry breaking, and most notably
the presence of a cosmic microwave background resulting from the decoupling of electromagnetic radiation from the plasma
when protons, helium nuclei and electrons combined into neutral hydrogen and helium.\footnote{In the following we will refer to this event as `decoupling', `recombination', or `last-scattering'.} However, the Big Bang model is incomplete in that there remain puzzles it is incapable of
explaining:
\begin{enumerate}
\item[i)] {\it Relic Problem}: The breaking of gauge symmetries at the extremely high energies associated with the early Big Bang universe leads to the production of many unwanted relics such as magnetic monopoles and other topological defects. For example, monopoles are expected to be copiously produced in Grand Unified Theories and should have persisted to the present day.
The absence of monopoles is a puzzle in the context of the standard Big Bang theory without inflation.

\item[ii)] {\it Flatness Problem}: Present observations show that the universe is very nearly spatially flat.  In standard Big Bang cosmology a flat universe is an unstable solution, and so any primordial curvature of space would grow very quickly.  To explain
the geometric flatness of space today therefore requires an extreme fine-tuning in a Big Bang cosmology without inflation.

\item[iii)] {\it Horizon Problem}: Observations of the cosmic microwave background imply the existence  of temperature correlations across distances on the sky that corresponded to super-horizon scales at the time when the CMB radiation was released.  In fact, regions that in the standard Big Bang theory would be causally connected on the surface of last scattering correspond to only an angle of order $1^\circ$ on the sky.  The CMB is seen to have nearly the same temperature in all directions on the sky. Yet there is no way to establish thermal equilibrium if these points were never in causal contact before last scattering.
\end{enumerate}

In addition, inflation solves the {\it homogeneity} and {\it isotropy problems}, and explains why the total {\it mass} and {\it entropy} of the universe are so large \cite{Linde:2005ht}.
Each of these
problems is eliminated by the assumption that the early universe underwent a brief but intense period
of {\it accelerated expansion}, inflating by a factor of at least $10^{26}$ within less than $10^{-34}$ seconds.
In this picture the entire observable universe ($\sim 10^{26}$ m) originated from a smooth patch of space smaller than $10^{-26}$ m in diameter (many orders of magnitude smaller than an atomic nucleus).

The way in which such an inflationary phase solves the first two puzzles is immediately intuitive.  Any
monopoles existing at early times will be vastly diluted until there exist none in the observable universe
today.  Similarly, any primordial geometric curvature would be diluted in the same sense that inflating a sphere
allows one to approximate its surface as flat on scales much smaller than the radius of the sphere.
A flat universe is an attractor solution during inflation.

The mechanism by which inflation solves the horizon problem is more subtle.
Two facts are fundamental to understanding the horizon problem and its resolution:
\begin{enumerate}
\item[i)] the physical wavelength of fluctuations is stretched by the expansion of the universe,

\item[ii)] the physical horizon ({\it i.e.}~the spacetime region in which one point could affect or have been affected by other points) is time-dependent.
\end{enumerate}
In standard Big Bang cosmology (without inflation) the {\it physical horizon grows faster than the physical  wavelength} of perturbations.
This implies that the largest observed scales today were outside of the horizon at early times.
Quantitatively, according to the standard Big Bang theory, the CMB at decoupling should have consisted of about $10^4$ causally disconnected regions.
However, the observed near-homogeneity of the CMB tells us that the universe was quasi-homogeneous at the time of last scattering.  In the standard Big Bang theory this uniformity of the CMB has no explanation and must be assumed as an initial condition.

During inflation the universe expands exponentially and {\it physical wavelengths grow faster than the horizon}.  Fluctuations are hence stretched outside of the horizon during inflation and re-enter the horizon in the late universe.
Scales that are outside of the horizon at CMB decoupling were in fact inside the horizon before inflation.
The region of space corresponding to the observable universe therefore was in causal contact before inflation and
the uniformity of the CMB is given a causal explanation.
A brief
period of acceleration therefore results in the ability to correlate physical phenomena, including the temperature of the CMB, over apparently impossible distances.

%%%%%%%%%%%%%%%%%%%%%%%%%%%%%
\subsection{The Physics of Inflation}
\label{sec:dynamics}
%%%%%%%%%%%%%%%%%%%%%%%%%%%%%

What drives the accelerated expansion of the early universe?  Consulting the Friedmann equations governing the
scale factor $a(t)$
\bea
\label{Friedman} H^2 &=& \left({\dot a\over a}\right)^2 = {1\over{3 \Mp^2}}\rho \, ,\\
\label{Friedman2} \dot H + H^2 &=& \frac{\ddot a}{a} = - \frac{1}{6 \Mp^2} (\rho + 3 p)
\eea
of a spatially flat universe with Friedmann-Robertson-Walker (FRW) metric\footnote{For simplicity, we anticipate the inflationary solution of the
flatness problem and assume that the spatial geometry is flat.  The generalization to curved space is
straightforward.}
\beq \label{equ:FRWmetric} \d s^2 = - \d t^2 + a(t)^2 \d {\bf x}^2\,  \eeq
we see that inflation requires a source of negative pressure $p$ and an energy density $\rho$ which dilutes very slowly\footnote{Note that the two Friedmann equations can be combined into the continuity equation $\dot \rho = 3 H (\rho +p)$. For $p\approx - \rho$, one therefore finds $\dot \rho \approx $ const. and $\ddot a >0$.}, while
allowing for an exit into the standard Big Bang cosmology at later times. Such a source of stress-energy
can be
modeled by the potential energy $V(\phi)$ of a scalar field $\phi$, together with a mechanism which maintains a
near-constant value of $V(\phi)$ during the inflationary period.
That is, the scalar field $\phi(t, {\bf x})$
(the `inflaton') is an order parameter used to describe the change in energy density during inflation.
There is
a wide array of mechanisms for obtaining near-constant $V(\phi)$ during inflation.  Two basic approaches include
(i) postulating a nearly flat potential $V(\phi)$, or (ii) postulating an effective action for $\phi$ which
contains strong self-interactions which slow the field's evolution down a steep potential. All single-field mechanisms for inflation can be captured by an effective field theory for single-field inflation \cite{Cheung:2007st}; different mechanisms and models with diverse
theoretical motivations arise as limits of this basic structure.

\begin{figure}[htbp!]
    \centering
        \includegraphics[width=0.85\textwidth]{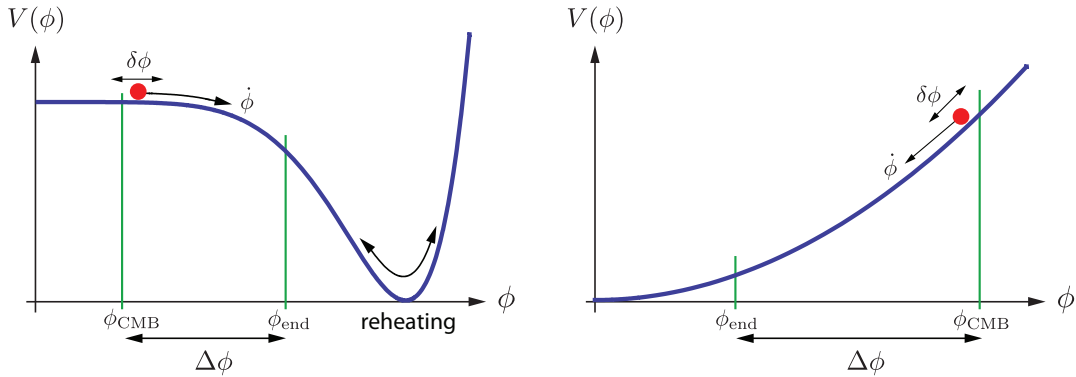}
   \caption{Examples of Inflaton Potentials.  Acceleration occurs when the potential energy of the field $V$ dominates over its kinetic energy $\frac{1}{2} \dot \phi^2$.
Inflation ends at $\phi_{\rm end}$ when the slow-roll conditions are violated, $\epsilon \to 1$. CMB
fluctuations are created by quantum fluctuations $\delta \phi$ about 60 $e$-folds before the end of inflation. At reheating, the energy density of the inflaton is converted into radiation.
\newline
{\it Left}: A typical small-field potential. {\it Right}: A typical large-field potential.}
    \label{fig:potential}
\end{figure}

One simple limit is known as {\it single-field slow-roll inflation}, for which an effective Lagrangian ${\cal
L}_{\rm eff}(\phi) = f[(\partial \phi)^2] - V(\phi)$ is postulated.\footnote{For pedagogical reasons, we
restrict the discussion in the remainder of this section to single-field slow-roll inflation with canonical
kinetic term $f[(\partial \phi)^2]= \frac{1}{2} (\partial \phi)^2$.  In Section \ref{sec:beyond} and Appendix
\ref{sec:taxonomy} we generalize our treatment to single-field inflation with non-canonical kinetic terms and
inflationary models with more than one field.} We consider a time-dependent homogeneous and isotropic background
spacetime as in Eqn.~(\ref{equ:FRWmetric}).
The expansion rate is characterized by the Hubble parameter $H \equiv
\partial_t \ln a$. This system will yield the following equations of motion for the homogeneous modes $\phi(t)$
and $a(t)$,
\begin{eqnarray}
H^2 &=& \frac{1}{3\Mp^2} \left( \frac{1}{2} {\dot \phi}^2 + V(\phi) \right)\, , \\ % \qquad H \equiv \partial_t \ln a\, ,\\
 \frac{\ddot a}{a} &=& -\frac{1}{3\Mp^2}  \left(  {\dot \phi}^2 - V(\phi)\right) \, ,
 \end{eqnarray}
 and
 \beq
{\ddot \phi} + 3H {\dot \phi} + V'(\phi) = 0\, . \eeq The spacetime experiences accelerated expansion, $\ddot a
> 0$, if and only if the potential energy of the inflaton dominates over its kinetic energy, $V \gg  \dot
\phi^2$. This condition is sustained if $|\ddot \phi| \ll |V'|$. These two conditions for prolonged inflation are
summarized by restrictions of the form of the inflaton potential $V(\phi)$ and its derivatives. Quantitatively,
inflation requires smallness of the slow-roll parameters
\begin{equation}
\label{slowroll} \epsilon \equiv - \frac{\dot H}{H^2} = \frac{\Mp^2}{2} \frac{\dot \phi^2}{H^2} \approx
\frac{M^2_{\rm pl}}{2} \left( \frac{V'}{V} \right)^2, \hspace{0.5in} |\eta| \approx M^2_{\rm pl} \left|
\frac{V''}{V} \right| .
\end{equation}
Once these constraints are satisfied, the inflationary process (and its termination) happens generically for a
wide class of models.
The slow evolution of the inflaton then produces an exponential increase in the geometric size of the universe,
\beq a(t) \approx a(0) e^{H t}\, , \qquad H \approx {\rm const}\, . \eeq For inflation to successfully address the Big Bang problems, one must
simply ensure that the inflationary process produces a sufficient number of these `$e$-folds' of accelerated
expansion $N_e \equiv \ln (a(t_{\rm final})/a(t_{\rm initial}))$.
A typical lower bound on the required number of $e$-folds is $N_e \gtrsim \ln 10^{26} \sim 55$ \cite{Dodelson, Mukhanov, WeinbergCosmology}.\footnote{This estimate of the required number of $e$-folds assumes GUT scale reheating. For lower reheating temperatures, fewer $e$-folds can be sufficient.}
Our discussion has so far addressed only the classical and homogeneous evolution of the inflating system.  Small
spatial perturbations in the inflaton $\phi$ and the metric $g_{\mu \nu}$
are inevitable due to quantum mechanics; inflation stretches these fluctuations to astronomical scales, eventually producing large-scale structures including galaxies such as the one we inhabit.
  Thus inflation is responsible not just
for the universe that we observe, but also for the fact we are here to observe it.

After a sufficient number of $e$-folds have been achieved, the process must terminate.  The inflaton descends
towards the minimum of the potential and `reheats' the universe, with $\phi$-particles decaying into radiation, and
so initiating the hot Big Bang.

This basic inflation model can be generalized in a variety of ways: several fields collectively producing the
inflaton, non-standard kinetic terms, scalars replaced by axion-like fields, etc.  Each of these models
still produces an inflationary period, with the details determining various observables such as cosmological
perturbations, as will be described in further detail below.

There also remain questions of initial conditions and of whether inflation continues eternally.  This latter point
may seem paradoxical; if the inflaton completes its evolution as we have just assumed, how could inflation
continue?  The answer lies in the fact that inflation produces other inflating regions of space; there is then
the possibility that although inflation may terminate at any single region of space, on a \emph{global} scale it
continues to proceed eternally \cite{Vilenkin:1983xq, Linde:1986fd}.  These important questions can be answered only by determining the particular
inflation model which Nature utilizes, which is in turn determined by observations, as we will see in the next
section.

%%%%%%%%%%%%%%%%%%%%%%%
\subsection{Cosmological Observables}
\label{sec:PT}
%%%%%%%%%%%%%%%%%%%%%%%

In
this section we give a general summary of cosmological perturbation theory \cite{Brandenberger:1993zc, Kodama:1985bj, Malik:2008im}. In Section \ref{sec:qf} we then
describe how these fluctuations arise as quantum fluctuations during the inflationary epoch.

%\addtocontents{toc}{\SkipTocEntry}
\subsubsection{SVT Decomposition in Fourier Space}
\label{sec:SVT}

During inflation we define perturbations around the homogeneous background solutions for the inflaton $\bar
\phi(t)$ and the metric $\bar g_{\mu \nu}(t)$ as in (\ref{equ:FRWmetric}), \beq \phi(t, {\bf x}) = \bar \phi(t)
+ \delta \phi(t, {\bf x})\, , \qquad g_{\mu \nu}(t, {\bf x}) = \bar g_{\mu \nu}(t) + \delta g_{\mu \nu}(t, {\bf
x}) \eeq where \bea
\d s^2 &=& g_{\mu \nu}\, \d x^\mu \d x^\nu \nonumber \\
&=& -(1+2 \Phi) \d t^2 + 2 a B_i \d x^i \d t + a^2 [(1-2 \Psi) \delta_{ij} + E_{ij}] \d x^i \d x^j\, .
\label{equ:perturbed} \eea The spatially flat background spacetime possesses a great deal of symmetry. These
symmetries allow a decomposition of the metric and the stress-energy perturbations associated with $\phi$ into
independent scalar (S), vector (V) and tensor (T) components. This SVT decomposition is most easily described in
Fourier space \beq
 Q_{\bf k}(t) = \int \d^3 {\bf x} \  Q(t, {\bf x}) \, e^{i {\bf k} \cdot {\bf x}}\, , \qquad Q \equiv \delta \phi,\, \delta g_{\mu \nu}\, .
\eeq We note that {\it translation invariance} of the linear equations of motion for perturbations means that
the different Fourier modes do not interact. Next we consider rotations around a single Fourier wavevector ${\bf
k}$. A perturbation is said to have helicity $m$ if its amplitude is multiplied by $e^{i m \psi}$ under rotation
of the coordinate system around the wavevector by an angle $\psi$ \beq Q_{\bf k} \to e^{im \psi} Q_{\bf k}\, .
\eeq Scalar, vector and tensor perturbations have helicity $0$, $\pm 1$ and $\pm 2$, respectively. The
importance of the SVT decomposition is that the perturbations of each type evolve independently (at the linear level) and can
therefore be treated separately. In real space, the SVT decomposition of the metric perturbations
(\ref{equ:perturbed}) is \cite{Lifshitz:1945du}\footnote{SVT decomposition in real space corresponds to the distinctive transformation
properties of scalars, vectors and tensors on spatial hypersurfaces.} \beq \label{equ:SVT1} B_i \equiv
\partial_i B- S_i\, , \qquad {\rm where} \quad  \partial^i S_i = 0\, , \eeq and \beq \label{equ:SVT2}
 E_{ij} \equiv 2 \partial_{ij} E + 2 \partial_{(i} F_{j)} + h_{ij}\, ,  \qquad {\rm where} \quad \partial^i F_i =0\, , \quad h^i_{i} = \partial^i h_{ij} = 0\, .
\eeq
Finally, it is important to note that the perturbations $\delta \phi$ and $\delta g_{\mu \nu}$ are {gauge-dependent}, {\it i.e.}~they change under
coordinate/gauge transformations.  Physical questions therefore have to be studied in a fixed gauge or in terms
of gauge-invariant quantities. An important gauge-invariant quantity is the curvature perturbation on
uniform-density hypersurfaces \cite{BST}
\beq
\label{equ:zeta} - \zeta \equiv  \Psi +  \frac{H}{\dot \rho}
\delta \rho \, ,
\eeq
where $\rho$ is the total energy density of the universe.

%\addtocontents{toc}{\SkipTocEntry}
\subsubsection{Scalar (Density) Perturbations}
\label{scalarpert}

In a gauge where the energy density associated with the inflaton field is unperturbed ({\it i.e.} $\delta \rho_{\phi}
= 0$) all scalar degrees of freedom can be expressed by a metric perturbation $\zeta(t, {\bf x})$\footnote{In
addition to the perturbation to the spatial part of the metric there are fluctuations in $g_{\mu0}$. These are
related to $\zeta$ by Einstein's equations.} \beq g_{ij} =  a^2(t) [1+ 2 \zeta] \delta_{ij} \, .
%\d s^2 = - \d t^2 + a^2(t) [1+ 2 \zeta] \delta_{ij} \d x^i \d x^j
\eeq Geometrically, $\zeta$ measures the spatial curvature of constant-density hypersurfaces, ${\cal R}^{(3)} = -
4 \nabla^2 \zeta/a^2$. An important property of $\zeta$ is that it remains constant outside the
horizon.\footnote{This statement is only true for adiabatic perturbations. Non-adiabatic fluctuations can arise
in multi-field models of inflation (see \S\ref{sec:beyond} and Appendix \ref{sec:taxonomy}). In that case,
$\zeta$ evolves on super-horizon scales.} In a gauge defined by spatially flat hypersurfaces, $\zeta$ is the
dimensionless density perturbation $\frac{1}{3} \delta \rho/(\rho+p)$.
%On sub-horizon scales $\zeta$ is related to density perturbations $\delta \rho$.
Taking into account appropriate transfer functions to describe the sub-horizon evolution of the fluctuations,
CMB and large-scale structure (LSS) observations can therefore be related to the primordial value of $\zeta$. A crucial
statistical measure of the primordial scalar fluctuations is the power spectrum of $\zeta$\footnote{The
normalization of the dimensionless power spectrum $P_s(k)$ is chosen such that the variance of $\zeta$ is
$\langle \zeta \zeta \rangle = \int_0^\infty P_s(k) \, \d \ln k$. }
\beq \langle \zeta_{\bf k} \zeta_{{\bf k}'}
\rangle = (2\pi)^3 \, \delta({\bf k} + {\bf k}') \, \frac{2\pi^2}{k^{3}} P_s(k)\, .  \eeq
%\beq \langle \zeta_{\bf k} \zeta_{{\bf
%k}'} \rangle = (2\pi)^3 \delta({\bf k} + {\bf k}') {\cal P}_s(k)\, , \qquad P_s(k) \equiv \frac{k^3}{2 \pi^2}
%{\cal P}_s(k) \eeq
The scale-dependence of the power spectrum is defined by the scalar spectral index (or tilt) \beq
\label{equ:ns}
n_s - 1 \equiv \frac{d
\ln P_s}{d \ln k}\, .\eeq
Here, scale-invariance corresponds to the value $n_s = 1$. We may also define the running of
the spectral index by \beq \label{equ:as}
 \alpha_s
\equiv \frac{d n_s}{d \ln k}\, . \eeq
The power spectrum is often approximated by a power law form
\beq
P_s(k) = A_s(k_\star) \left(\frac{k}{k_\star}\right)^{n_s(k_\star)-1+ \frac{1}{2} \alpha_s(k_\star) \ln (k/k_\star)}\, ,
\eeq
where $k_\star$ is the pivot scale.

If $\zeta$ is Gaussian then the power spectrum contains all the
statistical information. Primordial non-Gaussianity is encoded in higher-order correlation functions of $\zeta$
(see \S\ref{sec:NG}).
In single-field slow-roll inflation the non-Gaussianity is predicted to be small \cite{Acquaviva02,Maldacena03}, but non-Gaussianity can be significant in multi-field models or in single-field models with non-trivial kinetic terms and/or violation of the slow-roll conditions.

%-----------------------------------------------------------------
%\addtocontents{toc}{\SkipTocEntry}
\subsubsection{Vector (Vorticity) Perturbations}
%-----------------------------------------------------------------

The vector perturbations $S_i$ and $F_i$ in equations (\ref{equ:SVT1}) and (\ref{equ:SVT2}) are distinguished from the
scalar perturbations $B$, $\Psi$ and $E$ as they are divergence-free, {\it i.e.}~$\partial^i S_i = \partial^i
F_i = 0$. One may show that vector perturbations on large scales are redshifted away by Hubble expansion (unless
they are driven by anisotropic stress). In particular, vector perturbations are subdominant at the time of recombination.
Since CMB polarization is generated at last scattering the polarization signal is dominated by scalar and tensor perturbations (\S\ref{sec:polarization}). Most of this section therefore focuses on scalar and tensor perturbations. However, vector perturbations can be sourced by cosmic strings which are discussed in \S\ref{sec:defects}.

%--------------------------------------------------------------------------------
%\addtocontents{toc}{\SkipTocEntry}
\subsubsection{Tensor (Gravitational Wave) Perturbations}
\label{tensorpert}
%--------------------------------------------------------------------------------

Tensor perturbations are uniquely described by a gauge-invariant metric perturbation $h_{ij}$ \beq g_{ij} =
a^2(t) [\delta_{ij} + h_{ij} ] \, , \qquad \partial_j h_{ij} = h^i_i = 0\, .
%\d s^2 = - \d t^2 + a^2(t) [\delta_{ij} + h_{ij} ] \d x^i \d x^j\, , \qquad \partial_j h_{ij} = h^i_i = 0\, .
\eeq Physically, $h_{ij}$ corresponds to gravitational wave fluctuations.  The power spectrum for the two
polarization modes of $h_{ij} \equiv h^+ e^+_{ij} + h^\times e^\times_{ij}$, $h \equiv h^+, h^\times$, is defined as \beq \langle h_{\bf k} h_{{\bf k}'} \rangle = (2\pi)^3\, \delta({\bf
k} + {\bf k}')\, \frac{2\pi^2}{k^{3}}  {P}_t(k)  %\qquad P_t(k) \equiv \frac{k^3}{2 \pi^2} {\cal P}_t(k)
\eeq and its scale-dependence is defined analogously to (\ref{equ:ns}) but for historical reasons without the $-1$,
\beq n_t \equiv \frac{d \ln P_t}{d \ln k}\, , \eeq
{\it i.e.}
\beq
P_t(k) = A_t(k_\star) \left(\frac{k}{k_\star}\right)^{n_t(k_\star)}\, .
\eeq

CMB polarization measurements are sensitive to the ratio of tensor  power to scalar power \beq r \equiv \frac{P_t}{P_s}  \, . \eeq The parameter $r$ will be of fundamental
importance for the discussion presented in this paper.
As we argue in Section \ref{sec:Bmodes}, its value encodes crucial information about the physics
of the inflationary era.

\newpage
%%%%%%%%%%%%%%%%%%%%%%%%%%%%%%%%%%
\subsection{Quantum Fluctuations as the Origin of Structure}
\label{sec:qf}
%%%%%%%%%%%%%%%%%%%%%%%%%%%%%%%%%%

In Section \ref{sec:dynamics} we discussed the classical evolution of the inflaton field. Something remarkable
happens when one considers quantum fluctuations of the inflaton: inflation combined with quantum mechanics
provides an elegant mechanism for generating the initial seeds of all structure in the universe. In other words,
quantum fluctuations during inflation are the source of the primordial power spectra $P_s(k)$ and $P_t(k)$.
In this section we sketch the mechanism by which inflation relates microscopic physics to macroscopic observables.

\begin{figure}[htbp]
    \centering
        \includegraphics[width=0.65\textwidth]{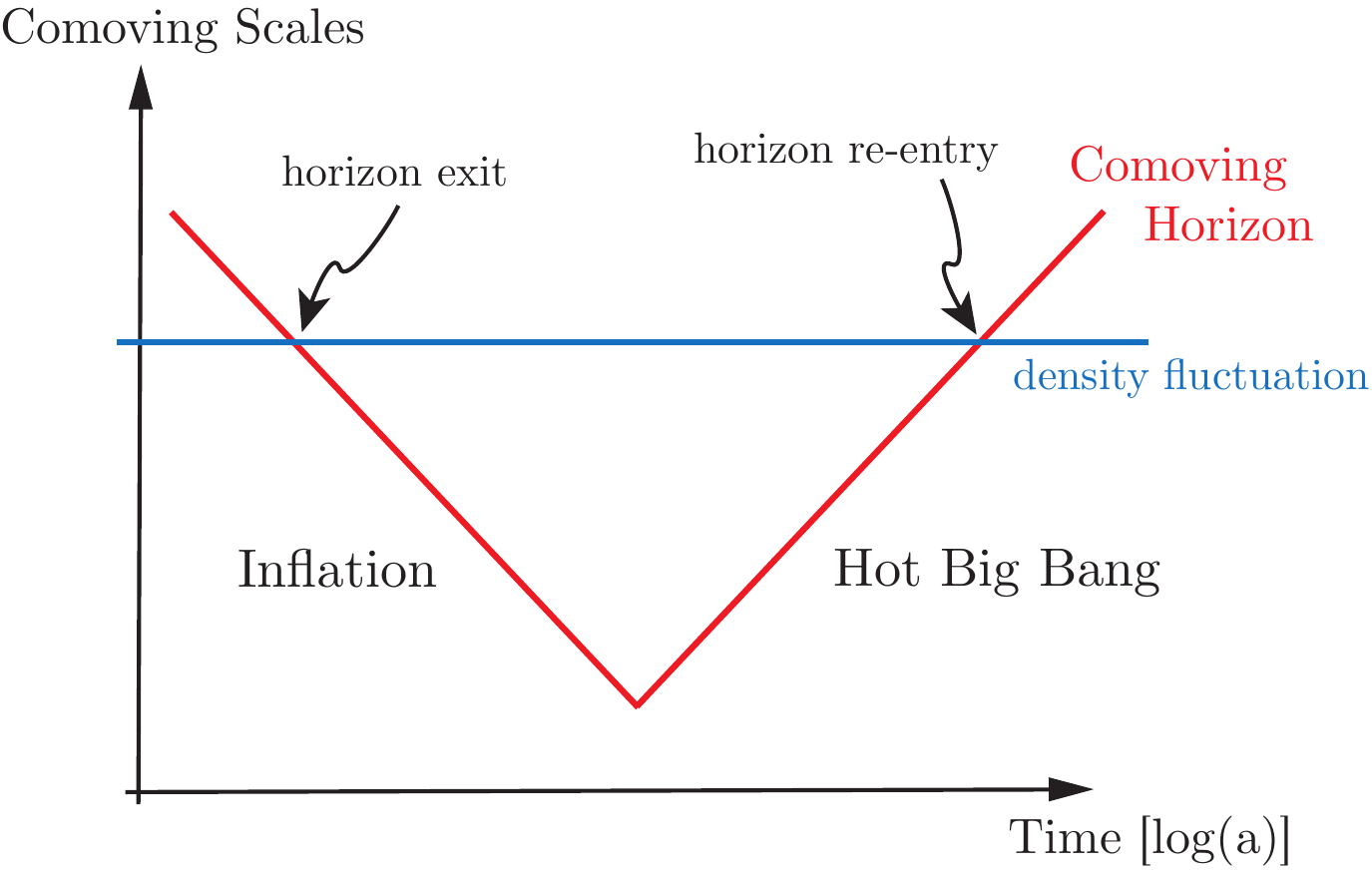}
   \caption{\small Creation and evolution of perturbations in the inflationary universe.  Fluctuations are created quantum mechanically on sub-horizon scales.  While comoving scales, $k^{-1}$, remain constant the comoving Hubble radius during inflation, $(aH)^{-1}$, shrinks and the perturbations exit the horizon. Causal physics cannot act on superhorizon perturbations and they freeze until horizon re-entry at late times.}
    \label{fig:scales}
\end{figure}

\vskip 10pt \noindent {\it Quantum fluctuations in quasi-de Sitter}

In spatially-flat gauge, perturbations in $\zeta$ are related to perturbations in the inflaton field
value\footnote{Intuitively, the curvature perturbation $\zeta$ is related to a spatially varying time-delay
$\delta t({\bf x})$ for the end of inflation \cite{GuthPi}. This time-delay is induced by the inflaton
fluctuation $\delta \phi$.} $\delta \phi$, {\it cf.}~Eqn.~(\ref{equ:zeta}) with $\Psi =0$ \beq \zeta = - H
\frac{\delta \rho}{\dot \rho} \approx - H \frac{\delta \phi}{\dot \phi} \equiv - H \delta t\, , \eeq where in
the second equality we have assumed slow-roll. The power spectrum of $\zeta$ and the power spectrum of inflaton
fluctuations $\delta \phi$ are therefore related as follows
\beq \label{equ:zetapower} \langle \zeta_{\bf k}
\zeta_{\bf k'} \rangle =\left( \frac{H}{\dot \phi} \right)^2 \langle \delta \phi_{\bf k} \, \delta \phi_{\bf k'}
\rangle\, . \eeq
Finally, in the case of slow-roll inflation, quantum fluctuations of a light scalar field ($m
_\phi \ll H$) in quasi-de Sitter space ($H\approx $ const.) scale with the Hubble parameter $H$ \cite{BirrellDavies}
\beq \langle
\delta \phi_{\bf k} \, \delta \phi_{\bf k'} \rangle = (2 \pi)^3 \, \delta({\bf k}+{\bf k'})\, \frac{2\pi^2}{k^{3}}\, \left(
\frac{H}{2\pi}\right)^2\, . \eeq The r.h.s.~of~(\ref{equ:zetapower}) is to be evaluated at horizon exit
of a given perturbation $k = aH$ (see Figure \ref{fig:scales}). Inflationary quantum fluctuations therefore
produce the following power spectrum for $\zeta$ \beq P_s(k) = \left. \left(\frac{H}{\dot \phi} \right)^2 \left(
\frac{H}{2\pi} \right)^2\right|_{k = aH}\, . \eeq In addition, quantum fluctuations during inflation excite
tensor metric perturbations $h_{ij}$ \cite{Starobinsky:1979ty}. Their power spectrum (in general models of inflation) is simply that of a
massless field in de Sitter space \beq P_t(k) = \left. \frac{8}{\Mp^2} \left( \frac{H}{2\pi} \right)^2\right|_{k
= aH}\, . \eeq

\vskip 10pt \noindent {\it Slow-roll predictions}

Models of single-field slow-roll inflation makes definite predictions for the primordial scalar and tensor fluctuation
spectra. %: $P_s, n_s, \alpha_s, P_t,  n_t, r$.
Under the slow-roll approximation one may relate the predictions for $P_s(k)$ and $P_t(k)$ to the shape of the inflaton potential $V(\phi)$.\footnote{In Appendix \ref{sec:taxonomy} we present the results for general single-field models.  In this case, the primordial power spectra receive contributions from a non-trivial speed of sound $c_s \ne 1$ and its time evolution.  The slow-roll results arise as the limit $c_s \to 1$, $\dot c_s \to 0$.}
To compute the spectral indices one uses $\d \ln k \approx \d \ln
a$ ($H \approx $ const.). To first order in the slow-roll parameters $\epsilon$ and $\eta$ one finds
\cite{Liddle:1992wi}
 \beq \label{equ:PsSR} P_s(k) = \left. \frac{1}{24 \pi^2 \Mp^4} \frac{V}{
\epsilon}\right|_{k=aH}\, , \qquad n_s - 1 = 2 \eta - 6 \epsilon\, , \eeq \beq P_t(k) = \left. \frac{2}{3 \pi^2}
\frac{V}{ \Mp^4}\right|_{k=aH}\, , \qquad n_t = - 2 \epsilon\, , \qquad r = 16 \epsilon \, . \eeq We note that
the value of the tensor-to-scalar ratio depends on the time-evolution of the inflaton field \beq \label{equ:TS}
r = 16 \epsilon = \frac{8}{\Mp^2} \Bigl( \frac{\dot \phi}{H} \Bigr)^2\, . \eeq We also point out the existence of a slow-roll
consistency relation between the tensor-to-scalar ratio and the tensor tilt which, at lowest order, has the form  \beq r = - 8 n_t\, . \eeq Measuring
the amplitudes of $P_t$ ($\to V$) and $P_s$ ($\to V'$) and the scale-dependence of the scalar spectrum $n_s$
($\to V''$) and $\alpha_s$ ($\to V'''$) allows a reconstruction of the inflaton potential as a Taylor expansion
around $\phi_\star$ (corresponding to the time when fluctuations on CMB scales exited the horizon) \beq V(\phi)
= \left. V\right|_\star + \left.V'\right|_\star (\phi-\phi_\star) + \frac{1}{2} \left. V''\right|_\star (\phi-\phi_\star)^2 + \frac{1}{3!} \left. V'''\right|_\star
(\phi-\phi_\star)^3 + \cdots \, ,\eeq
where $\left. (\dots)\right|_\star = \left. (\dots)\right|_{\phi =\phi_\star}$.  Furthermore, if one assumes that the primordial perturbations are produced by an inflationary model with a single slowly rolling scalar field, one can fit directly to the slow-roll parameters, bypassing the spectral indices entirely, and then reconstruct the form of the underlying potential  \cite{Leach:2002dw,Leach:2003us,Peiris:2006ug,Peiris:2006sj,Finelli:2006wk,Lesgourgues:2007aa,Peiris:2008be,Adshead:2008vn}.

%\newpage
%%%%%%%%%%%%%%%%%%%%%%%%%%%%%%%%%%%%%
\subsection{CMB Polarization: A Unique Probe of the Early Universe}
%%%%%%%%%%%%%%%%%%%%%%%%%%%%%%%%%%%%%
\label{sec:polarization}

CMB polarization will soon become one of the most important tools to probe the physics governing the early universe.
Because the anisotropies in the CMB temperature are indeed sourced by primordial fluctuations, we expect the CMB anisotropies to become polarized via Thomson scattering
(for a pedagogical review see
Ref.~\cite{HuWhite}; for technical details and pioneering work see \cite{Bond:1984fp, Kamionkowski:1996ks, Zaldarriaga:1996xe, Seljak:1996ti}).
% creates a preferred direction of polarization for the photons.
Since the polarization of CMB anisotropies is generated only by scattering, the polarization signal tracks free electrons and hence isolates the recombination (last-scattering) and reionization epochs.
The polarization signal and its cross-correlation with the temperature anisotropies provide an important consistency check for the standard cosmological paradigm.
In addition, measurements of CMB polarization help to break degeneracies among some cosmological parameters and hence increase the precision with which these parameters can be measured.
Finally, and most importantly for this report,
different sources of the temperature anisotropies (scalar, vector and tensor; see \S\ref{sec:SVT}) predict subtle differences in the polarization patterns.
One can therefore use polarization information to distinguish the different types of primordial perturbations.  It is this distinguishing feature of CMB polarization that we wish to elucidate in this section.

\begin{figure}[htbp!]
    \centering
        \includegraphics[width=0.55\textwidth]{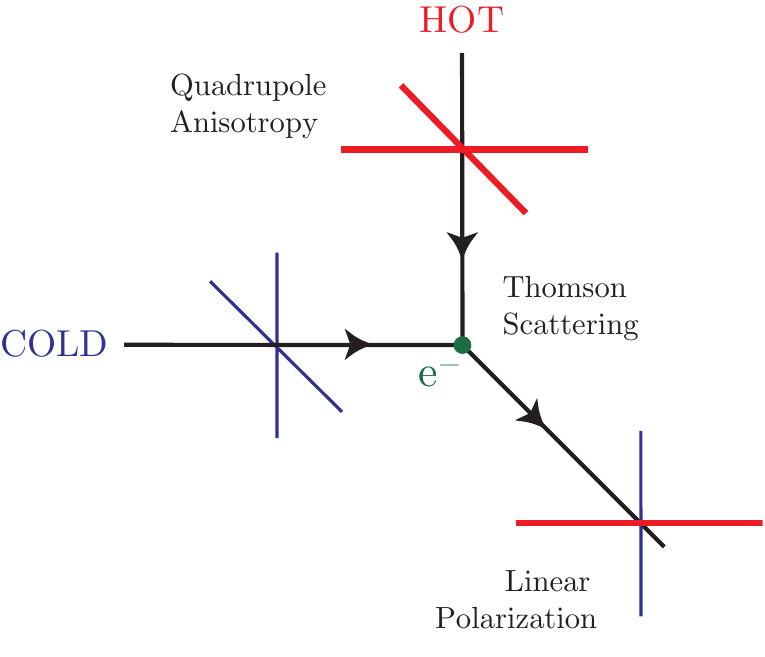}
   \caption{Thomson scattering of radiation with a quadrupole anisotropy generates linear polarization  \cite{HuWhite}.  Red colors (thick lines) represent hot radiation, and blue colors (thin lines) cold radiation. }
    \label{fig:pol}
\end{figure}

\vskip 10pt
\newpage
\noindent
{\it Polarization via Thomson scattering}

Thomson scattering between electrons and photons produces a simple relationship between temperature anisotropy and polarization.
If a free electron `sees' an incident radiation pattern that is isotropic, then the outgoing radiation remains unpolarized because orthogonal polarization directions cancel out.
However, if the incoming radiation field has a quadrupole component, a net linear polarization is generated via Thomson scattering (see Figure \ref{fig:pol}).
A quadrupole moment in the radiation field is generated when photons decouple from the electrons and protons just before recombination. Hence linear polarization results from the velocities of electrons and protons on scales smaller than the photon diffusion length scale. Since both the velocity field and the temperature anisotropies are created by primordial density fluctuations, a component of the polarization should be correlated with the temperature anisotropy.

\vskip 6pt
\noindent
{\it Characterization of the radiation field}

We digress briefly to give details of the mathematical characterization of CMB temperature and polarization anisotropies.
The anisotropy field is defined in terms of a $2\times 2$ intensity tensor $I_{ij}(\hat n)$, where $\hat n$ denotes the direction on the sky. The components of $I_{ij}$ are defined relative to two orthogonal basis vectors $\hat{\bf e}_1$ and $\hat {\bf e}_2$ perpendicular  to $\hat n$.
Linear polarization is then described by the Stokes parameters $Q = \frac{1}{4}(I_{11}-I_{22})$ and $U=\frac{1}{2} I_{12}$, while the temperature anisotropy is $T = \frac{1}{4} (I_{11}+I_{22})$.
The polarization magnitude and angle are $P =\sqrt{Q^2 +U^2}$ and $\alpha = \frac{1}{2} \tan^{-1}(U/Q)$.
The quantity $T$ is invariant under a rotation in the plane perpendicular to $\hat n$ and hence may be expanded in terms of scalar (spin-0) spherical harmonics
\beq
T(\hat n) = \sum_{\ell, m}  a_{\ell m}^T\ Y_{\ell m}(\hat n)\, .
\eeq
The quantities $Q$ and $U$, however, transform under rotation by an angle $\psi$ as a spin-2 field $(Q \pm iU)(\hat n) \to e^{\mp 2 i \psi} (Q \pm i U)(\hat n)$.  The harmonic analysis of $Q \pm i U$ therefore requires expansion on the sphere in terms of tensor (spin-2) spherical harmonics  \cite{Kamionkowski:1996ks, Zaldarriaga:1996xe, NewmanPenrose}
\beq
(Q+iU)(\hat n) = \sum_{\ell, m}  a_{\ell m}^{(\pm 2)}\, \left[{}_{\pm 2}Y_{\ell m}(\hat n)\right]\, .
\eeq
Instead of $a_{\ell m}^{(\pm 2)}$ it is convenient to introduce the linear combinations \cite{NewmanPenrose}
\beq
a^E_{\ell m} \equiv -\frac{1}{2} \left(a_{\ell m}^{(2)} + a_{\ell m}^{(-2)}\right)\, , \qquad
a^B_{\ell m} \equiv - \frac{1}{2i} \left(a_{\ell m}^{(2)} - a_{\ell m}^{(-2)}\right)\, .
\eeq
Then one can define two scalar (spin-0) fields instead of the spin-2 quantities $Q$ and $U$
\beq
E(\hat n) = \sum_{\ell, m} a_{\ell m}^E\ Y_{\ell m}(\hat n)\, , \qquad B (\hat n)=\sum_{\ell, m} a_{\ell m}^B\ Y_{\ell m}(\hat n)\, .
\eeq

\begin{figure}[htbp!]
    \centering
        \includegraphics[width=.45\textwidth]{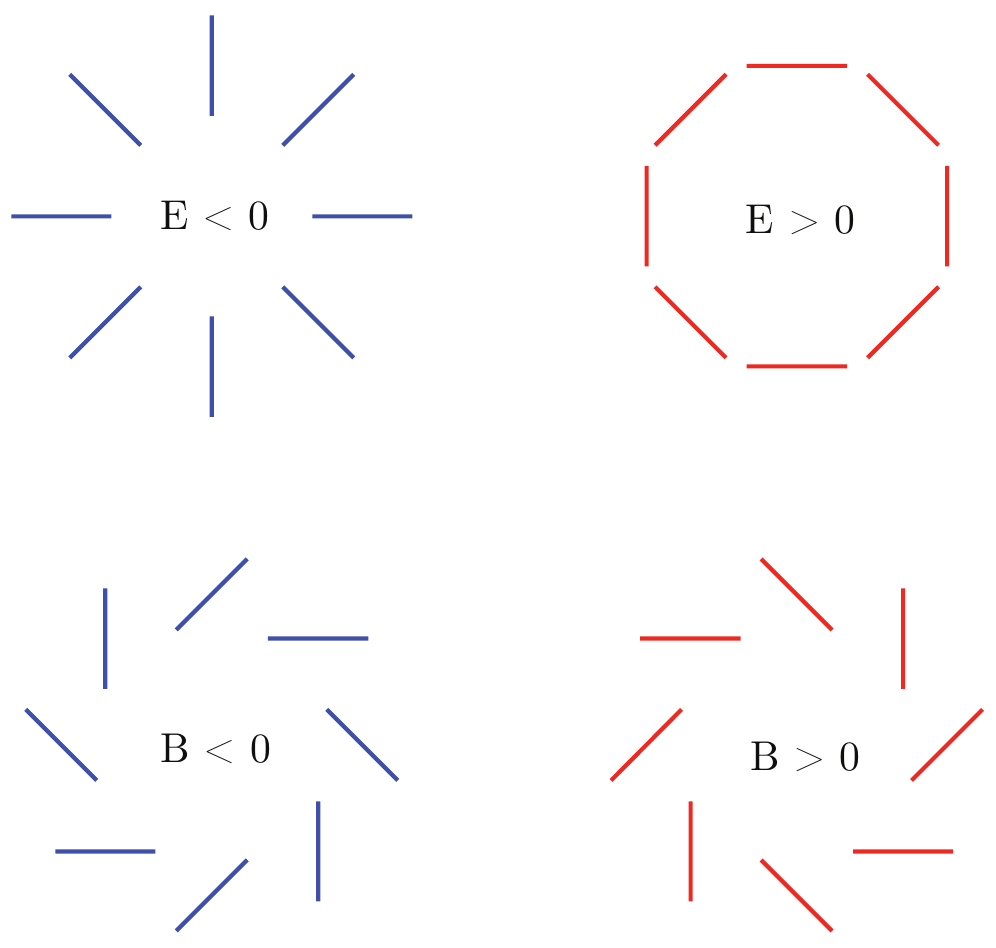}
   \caption{Examples of $E$-mode and $B$-mode patterns of polarization. Note that if reflected across a line going through the center the $E$-patterns are unchanged, while the positive and negative $B$-patterns get interchanged.}
    \label{fig:EBmode}
\end{figure}

\vskip 6pt
\noindent
{\it $E$- and $B$-modes}

$E$ and $B$ completely specify the linear polarization field.
$E$-polarization is often also characterized as a {\it curl-free} mode with polarization vectors that are radial around cold spots and tangential around hot spots on the sky.
In contrast, $B$-polarization is {\it divergence-free} but has a {\it curl}: its polarization vectors have vorticity around any given point on the sky.\footnote{Evidently the $E$ and $B$ nomenclature reflects the properties familiar from electrostatics, $\nabla \times {\bf E} = 0$ and $\nabla \cdot {\bf B} = 0$.}
Fig.~\ref{fig:EBmode} gives examples of $E$- and $B$-mode patterns.
Although $E$ and $B$ are both invariant under rotations, they behave differently under parity transformations. %$E$-polarization remains unchanged and $B$-polarization changes sign.
Note that when reflected about a line going through the center, the $E$-patterns remain unchanged, while the $B$-patterns change sign.

\vskip 6pt
\noindent
%{\it Power Spectra}
{\it TE correlation and superhorizon fluctuations}

The symmetries of temperature and polarization ($E$- and $B$-mode)
anisotropies allow four types of correlations: the autocorrelations of
temperature fluctuations and of $E$- and $B$-modes denoted by $TT$,
$EE$, and $BB$, respectively, as well as the cross-correlation between
temperature fluctuations and $E$-modes: $TE$. All other correlations
($TB$ and $EB$) vanish for symmetry reasons.\footnote{This assumes no
  parity-violating processes in the early universe.  Conversely,
  non-zero $TB$ and $EB$ correlations would be a distinctive signature
  of such physics.}
  % All these statements apply to primary anisotropies
  %only; at the level of secondary anisotropies, non-zero $TB$ and $EB$-correlations are generated by weak lensing effects.}

The angular power spectra are defined as rotationally invariant quantities
\beq
C_\ell^{XY} \equiv \frac{1}{2\ell + 1} \sum_m \langle a_{\ell m}^X a_{\ell m}^Y \rangle \, , \qquad X, Y = T, E, B\, .
\eeq
 In Fig.~\ref{fig:TE} we show the latest measurement of the $TE$ cross-correlation \cite{WMAP5}.  The $EE$ spectrum has now begun to be measured, but the errors are still large.  So far there are only upper limits on the $BB$ spectrum, but no detection.

\begin{figure}[htbp!]
    \centering
        \includegraphics[width=.65\textwidth]{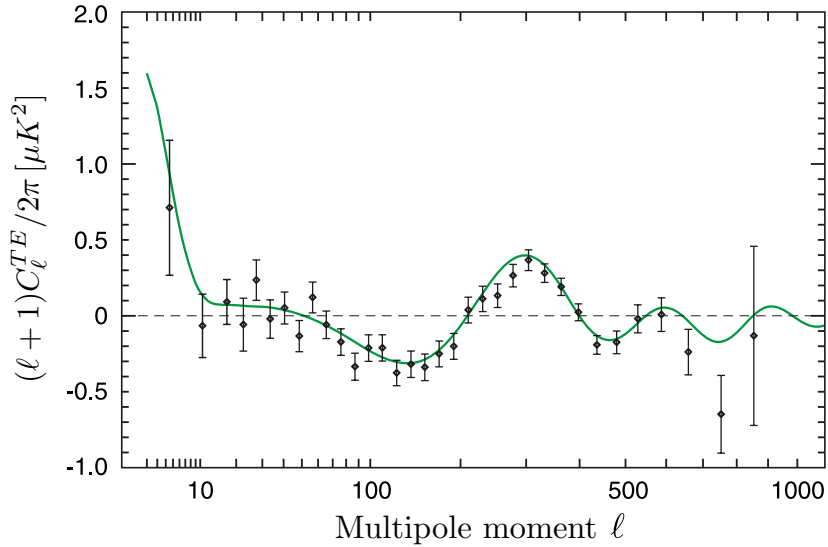}
   \caption{Power spectrum of the cross-correlation between temperature and $E$-mode polarization anisotropies \cite{WMAP5}. The anti-correlation for $\ell = 50-200$ (corresponding to angular separations $5^\circ > \theta > 1^\circ$) is a distinctive signature of adiabatic fluctuations on superhorizon scales at the epoch of decoupling \cite{spergel/zaldarriaga:1997, Dodelson:2003ip}, confirming a fundamental prediction of the inflationary paradigm.}
    \label{fig:TE}
\end{figure}

The dependence on cosmological parameters of each of these spectra differs, and hence a combined measurement of all of them greatly improves the constraints on cosmological parameters by giving increased statistical power,
removing degeneracies between fitted parameters, and aiding in discriminating between cosmological models.

\vskip 6pt
\newpage
\noindent
{\it A smoking gun of inflation}

The cosmological significance of the $E$/$B$ decomposition of CMB polarization was realized by the authors of
Refs.~\cite{Kamionkowski:1996ks, Zaldarriaga:1996xe}, who proved the following remarkable facts:
\begin{enumerate}
\item[i)] scalar (density) perturbations create only $E$-modes and {\it no} $B$-modes.
\item[ii)] vector (vorticity) perturbations create mainly $B$-modes.\footnote{ However, vectors decay with the expansion of the universe and are therefore believed to be subdominant at recombination. We therefore do not consider them here, but note that cosmic strings can produce a $B$-mode signal via vector modes (see \S\ref{sec:defects}).}
\item[iii)] tensor (gravitational wave) perturbations create both $E$-modes and $B$-modes.
\end{enumerate}
Intuitively these results may be understood as follows:
Thomson scattering produces an $E$-mode locally at the scattering event.
For scalar perturbations the spatial pattern of the polarization field at the last-scattering surface is curl-free. Since free streaming (to linear order) projects a curl-free spatial pattern to a curl-free angular distribution, the observed signal from scalar perturbations remains curl-free and hence pure $E$-mode.
For tensor modes the polarization is also $E$-mode at last scattering, but the spatial distribution has non-zero curl. Projection of the polarization pattern from the last-scattering surface to the point of observation today therefore produces $B$-mode polarization.

\begin{figure}[h!]
    \centering
        \includegraphics[width=0.65\textwidth]{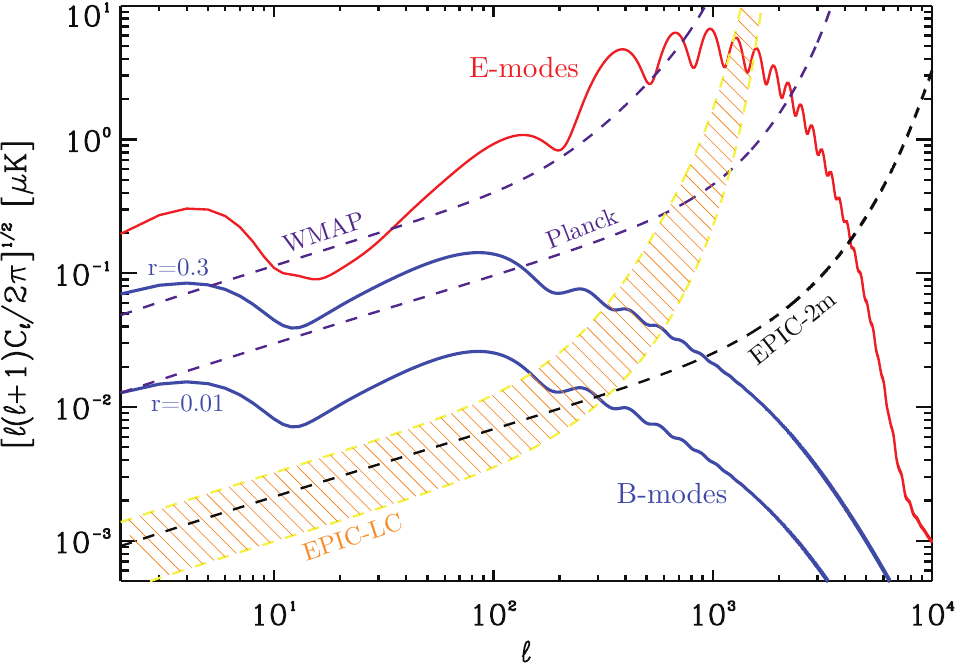}
    \caption{$E$- and $B$-mode power spectra for a tensor-to-scalar ratio saturating current bounds, $r=0.3$, and for $r=0.01$. Shown are also the experimental sensitivities for {\sl WMAP}, {\sl Planck} and two different realizations of {\sl CMBPol} (EPIC-LC and EPIC-2m). ({\small Figure adapted from Bock {\it et al.}~\cite{EPIC}.})}
    \label{fig:spectra}
\end{figure}

The fact that scalars do not produce $B$-modes while tensors do is the basis for the often-quoted statement that
detection of $B$-modes is a smoking gun of tensor modes, and therefore of inflation.\footnote{
To justify this statement requires careful consideration of tensor modes from
i) alternatives to inflation (see \S\ref{sec:Ekpyrosis} and Appendix \ref{sec:alternatives}) and ii) active sources like global phase transitions \cite{JonesSmith:2007ne}
or cosmic strings.
For case
i) the tensor amplitude is typically negligibly small, while for case ii) the signal is typically dominated by vector modes which produce a distinct spectrum and a characteristic ratio of $E$-modes and $B$-modes. To distinguish the inflationary $B$-mode spectrum from that produced by cosmic strings will likely require the high-resolution option of {\sl CMBPol} (see \S\ref{sec:defects} and Ref.~\cite{Baumann:2009mq}).}$^,$\footnote{It is worth noting that the temperature-$E$-mode cross correlation function has the opposite sign for scalar and tensor fluctuations on large scales   \cite{Crittenden:1994ej}.
This raises the possibility of using measurements of $TE$ correlations for a direct determination of whether the microwave anisotropies have a significant tensor component.}
The search for the primordial  $B$-mode signature of inflation is often considered the ``holy grail" of observational cosmology.
We discuss the theoretical implications of the $B$-mode amplitude in Section~\ref{sec:Bmodes}.

%%%%%%%%%%%%%%%%%%%%%%%%%%
\subsection{Current Observational Constraints}
\label{sec:WMAP}
%%%%%%%%%%%%%%%%%%%%%%%%%%

Cosmological observations are, for the first time, precise enough to allow detailed tests of theories of the
early universe.
In this section, we review the current observational constraints on the primordial power spectra $P_s(k)$ and $P_t(k)$.  We compare these measurements to the predictions from inflation.

\vspace{0.5cm}
\begin{table}[h!]
\begin{center}
\begin{tabular}{||c|c|c||}
\hline \hline
{\small \bf Parameter} & {\small 5-year {\sl WMAP}} & {\small {\sl WMAP}+BAO+SN} \\
\hline
{\small $n_s$} & {\small $0.963_{-0.015}^{+0.014}$} & {\small $0.960_{-0.013}^{+0.013}$} \\
\hline
{\small $n_s$} & {\small $0.986 \pm 0.022$} & {\small $0.970 \pm 0.015$} \\
 {\small $r$} & {\small $<0.43$} & {\small $<0.22$} \\
\hline
 {\small $n_s$} & {\small $1.031_{-0.055}^{+0.054}$} & {\small $1.017_{-0.043}^{+0.042}$} \\
{\small  $\alpha_s$} & {\small $-0.037 \pm 0.028$} & {\small $-0.028_{-0.020}^{+0.020}$} \\
\hline
{\small $n_s$} & {\small $1.087_{-0.073}^{+0.072}$} & {\small $1.089_{-0.068}^{+0.070}$} \\
 {\small $r$} & {\small $<0.58$} & {\small $<0.55$} \\
{\small $\alpha_s$} & {\small $-0.050 \pm 0.034$} & {\small $-0.058 \pm 0.028$} \\
\hline \hline
\end{tabular}
\caption{ \label{tab:param} 5-year {\sl WMAP} constraints on the primordial power spectra in the power law parameterization \cite{WMAP5}.  We present results for ($n_s$), ($n_s$, $r$), $(n_s, \alpha_s)$ and ($n_s$, $r$, $\alpha$) marginalized over all other parameters of a flat $\Lambda$CDM model.}
\end{center}
\end{table}

\begin{figure}[h!]
    \centering
        \includegraphics[width=.95\textwidth]{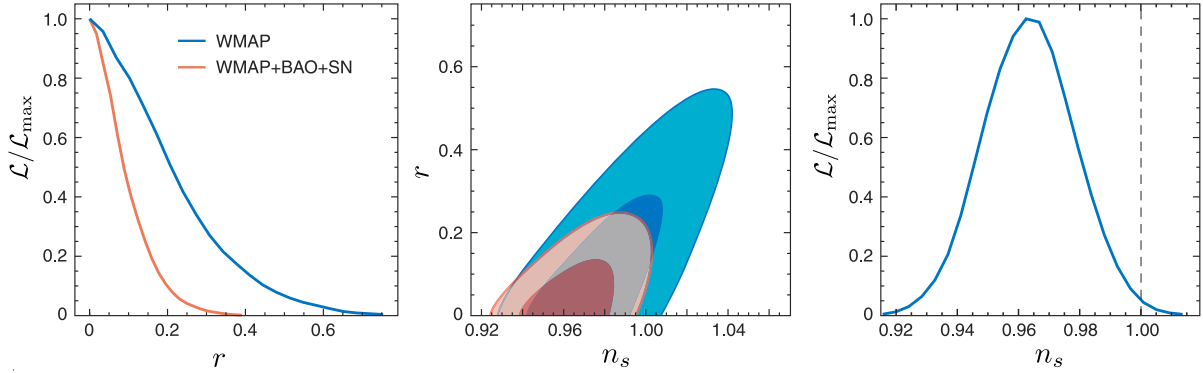}
   \caption{{\sl WMAP} 5-year constraints on the inflationary parameters $n_s$
 and $r$ \cite{WMAP5}. The {\sl WMAP}-only results are shown in blue, while constraints from
 {\sl WMAP} plus other cosmological observations are in red.
The third plot assumes that $r$ is negligible.
}%\newline [{\tt Placeholder.} {\tt Figure will be updated.}]}
    \label{fig:WMAP5}
\end{figure}

%\vskip 6pt
Komatsu {\it et al.}~\cite{WMAP5} recently used the {\sl WMAP} 5-year temperature and polarization
data, combined with the luminosity distance data of Type Ia Supernovae (SN) at $z\le 1.7$ \cite{unionsn} and the
angular diameter distance data of the Baryon Acoustic Oscillations (BAO) at $z=0.2$ and $0.35$ \cite{bao}, to
put constraints on the primordial power spectra (see Fig.~\ref{fig:WMAP5} and Table \ref{tab:param}).
A power-law parameterization of the power spectrum is employed in \cite{WMAP5}
\beq
P_s(k) = A_s(k_\star) \left(\frac{k}{k_\star} \right)^{n_s(k_\star)-1 + \frac{1}{2} \alpha_s(k_\star) \ln(k/k_\star)}\, .
\eeq
The amplitude of scalar fluctuations at $k_\star = 0.002$ Mpc$^{-1}$ is found to be \beq \label{equ:PsWMAP}
A_s= \left(2.445\pm 0.096\right) \times 10^{-9} \, . \eeq Assuming {\it no}
tensors ($r \equiv 0$) the scale-dependence of the power spectrum is \beq n_s = 0.960\pm 0.013\quad (r\equiv
0)\, . \eeq
The scale-invariant Harrison-Zel'dovich-Peebles spectrum, $n_s=1$, is 3.1 standard deviations away from the mean of the likelihood.

Including the possibility of a non-zero $r$ into the parameter estimation gives the following upper
bound on $r$\footnote{When the constraints on a given parameter depend on the choice of the prior probability for that parameter, one can immediately conclude that the parameter is poorly constrained by the data. This follows directly from the statement of Bayes' Theorem (for discussion on this point as related to $r$, see {\it e.g.}~\cite{Valkenburg:2008cz}).}
 \beq r < 0.22 \ (95\% \,{\rm C.L.})\, . \eeq Komatsu {\it et al.}~\cite{WMAP5} showed that the
constraint on $r$ is driven mainly by the temperature data and the temperature-polarization cross correlation;
constraints on $B$-mode polarization make a negligible contribution to the current limit on $r$.\footnote{With the $E$-mode and $B$-mode
polarization data at low multipoles ($ \ell \leq 23$) only, they find $r<20$ at 95\% C.L., two orders of magnitude
worse than that from the temperature and temperature-polarization cross power spectra.  A Fisher matrix analysis \cite{LiciaFisher} shows that constraints up to $r<0.1$ can be inferred from the $TT$ and $TE$ spectra. To go below this limit requires information from $BB$ measurements.}
Since the $B$-mode limit contributes little to the limit on $r$, and most of the information essentially comes from the $TT$ and $TE$ measurements, the current limit on $r$ is highly degenerate with $n_s$.  Better limits on $n_s$ therefore correlate strongly with better limits on $r$.

\begin{figure}[h!]
    \centering
        \includegraphics[width=.98\textwidth]{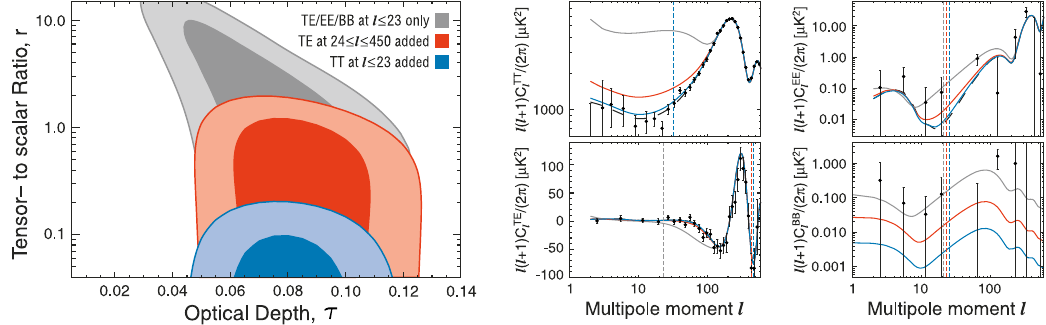}
   \caption{How the {\sl WMAP} temperature and polarization data constrain the tensor-to-scalar ratio (Figure courtesy of Ref.~\cite{WMAP5}).
   \newline
  {\it  Left:} The contours show 68\% and 99\% C.L. The gray region is derived from the low-$\ell$ polarization data ($TE$, $EE$, $BB$ at $\ell \le 23$) only, the red region from the low-$\ell$ polarization plus the high-$\ell$ $TE$ data at $\ell \le 450$, and the blue region from the low-$\ell$ polarization, the high-$\ell$ $TE$, and the low-$\ell$ temperature data at $\ell \le 32$.
   \newline
  {\it  Right:}  The gray curves show $(r, \tau) = (10, 0.050)$, the red curves $(r, \tau) = (1.2, 0.075)$, and the blue curves $(r, \tau) = (0.2,0.080)$.}
    \label{fig:Howr}
\end{figure}
%Therefore, we are still
%quite far away from measuring tensor modes with the $B$-mode polarization; however, the situation will be very
%different with a dedicated CMB polarization experiment like {\sl CMBPol}.

With non-zero $r$ the marginalized constraint on $n_s$ becomes \beq n_s=0.970\pm 0.015\quad (r\neq 0). \eeq

Including the possibility of a non-zero running ($\alpha_s$) in the parameter estimates leads to a deterioration of the limits on $n_s$ and $r$ (see Table \ref{tab:param}).

\vskip 6pt
Finally, {\sl WMAP} detected no evidence for {\it curvature} ($-0.0179 < \Omega_k < 0.0081$), {\it running} ($- 0.068 < \alpha_s < 0.012$), {\it non-Gaussianity} ($-9 < f_{\rm NL}^{\rm local}< 111$, $-151< f_{\rm NL}^{\rm equil.} < 253$), and {\it isocurvature} ($ S_{\rm axion} < 0.072$, $S_{\rm curvaton} < 0.0041$).

%\newpage
\subsection{Alternatives to Inflation}
\label{sec:Ekpyrosis}

Ultimately, our confidence in inflation relies not only upon observations confirming its predictions, but also on the absence of compelling alternatives.
Specifically, a study of alternatives to inflation is necessary to have confidence that
 a detection of $r$ really would be a smoking gun of inflation.

As we have reviewed above, a period of accelerated expansion necessarily causes a given observer's  comoving horizon to decrease, correlating apparently distant pieces of the universe without recourse to acausal processes, thereby predicting and explaining the long range $TE$ correlations seen in the CMB.     However,  accelerated expansion is not the {\it only} mechanism that can shrink an observer's comoving horizon: the contracting phase before a Big Crunch performs this task equally well, and is the basis of the recently much discussed and much debated ekpyrotic scenarios \cite{Khoury:2001wf, Khoury:2001bz} (see \cite{Lehners:2008vx} for a review, and \cite{Kallosh:2001ai, Kallosh:2001du, Lyth:2001pf, Lyth:2001nv, Felder:2002jk, Linde:2002ws, Liu:2002ft, Horowitz:2002mw} for a critical discussion of this scenario).
While inflation achieves a shrinking comoving Hubble sphere of radius $(aH)^{-1}$
by rapid expansion with $H \approx$ const. and $a(t)$ exponentially increasing,
ekpyrosis instead relies on a phase of slow contraction with $a(t) \approx$ const. and $H^{-1}$ decreasing.
We discuss the theoretical challenges and phenomenological predictions of ekpyrotic cosmology in Appendix \ref{sec:alternatives}.  Here we restrict ourselves to highlighting two important features:

i) for the contracting phase to smoothly connect to the expanding Big Bang evolution ({\it i.e.}, for there to be a bounce) requires that
\beq
2 \Mp^2 \dot H = - (\rho + p) > 0\, ,
\eeq
{\it i.e.}~a violation of the null energy condition (NEC).
Although this can be achieved at the level of effective field theory \cite{Creminelli:2007aq, Buchbinder:2007ad}, it remains an important open question whether a consistent UV completion exists. According to \cite{Kallosh:2007ad} this is a very important issue because the quantization of the new ekpyrotic theory, prior to the introduction of the UV cutoff and the UV completion, leads to a catastrophic vacuum instability.

ii) a generic prediction of all models of ekpyrosis is the absence of a significant amplitude of primordial gravitational waves \cite{Khoury:2001wf,Boyle:2003km}.  This strengthens the case for considering $B$-modes a smoking gun of inflation.

Item i) (the physics of the bounce) provides a significant theoretical challenge for Big Crunch-Big Bang scenarios, item ii) (the absence of primordial gravitational waves) offers a distinctive way to rule out these alternative models of the early universe on purely observational grounds.
For further details on ekyprotic cosmology and a brief discussion of string gas cosmology and the pre-Big Bang model we refer the reader to Appendix \ref{sec:alternatives}.

\newpage
%%%%%%%%%%%%%%%%%%%%%%%%%%%%%%%%
\section{Probing Fundamental Physics \\with Primordial Tensors}
\label{sec:Bmodes}
%%%%%%%%%%%%%%%%%%%%%%%%%%%%%%%%

Inflation is one of the great developments in theoretical physics, solving the horizon and flatness problems of
the Big Bang model within general relativity and effective field theory, while providing a quantum-mechanical mechanism for the origin
of large-scale structure. Moreover, inflation provides a unique window on high energy physics. By amplifying
early-universe fluctuations to angular scales accessible to CMB experiments, inflation has the capacity to
reveal phenomena that are forever beyond the reach of terrestrial accelerators. As explained below, a detection
of primordial tensor perturbations would probe physics at an energy that is a staggering {\it twelve orders of
magnitude} larger than the center of mass energy at the {LHC}.  Of equal importance is the fact that a detection
or constraint on the tensor-to-scalar ratio $r$ at the level accessible to {\sl CMBPol} will answer a fundamental
question about the range $\Delta\phi$ of the scalar field excursion during inflation as compared to the Planck
mass scale $\Mp$. The quantity $\Delta\phi/\Mp$ is sensitive to the physics behind inflation, including the
ultraviolet completion of gravity.

To understand the scientific impact of a $B$-mode detection, we must consider
%the status of
our current
understanding of the possibilities for the physics driving the inflationary expansion.  Given the striking
success of inflation as a {\it phenomenological} paradigm for the early universe, it is natural to inquire about
the underlying theoretical structure, and to ask how the scalar fields involved in inflation are related to
other, better-understood areas of physics.  A true `model of inflation' is then more than merely a choice of an
effective action for some scalar fields; it is instead an answer to at least some of the following fundamental
questions:  Is the inflaton a particle that has already been invoked for some other reason?  Does it couple to
the Standard Model particles through gauge interactions? Does it couple to or involve GUT particles?
%Is the potential classical or does it incorporate quantum effects?
Is inflationary physics well-approximated by semi-classical equations
of motion, or are quantum effects important?  Does the inflaton have a
superpartner?  Does inflation involve extra dimensions, or a
low-energy limit of string theory? How many light degrees of freedom
are relevant during inflation?  Is there only one stage of inflation
between the time at which the largest observable scales crossed the
horizon and nucleosynthesis?
Most importantly, is there a
mechanism or symmetry principle that is responsible for the long duration of inflation?

Theoretical physics has come a long way in mapping out a range of consistent and well-motivated inflationary
mechanisms and their phenomenological predictions. However, theory alone may not answer these  questions -- there is a pressing need for observational data.  This data will distinguish wildly different
possibilities for the origin of inflation.
Moreover, the absence of manifest connections between inflation and Standard Model physics, although frustrating
from the viewpoint of economy in Nature, underscores the spectacular discovery potential of an experimental
probe of inflation: it is a very real possibility that inflation involves an {\it entirely new set of fields and
interactions} going beyond the Standard Model of particle physics.

In \S\ref{sec:beyond} and Appendix \ref{sec:taxonomy} we survey some of the leading models of inflation,
 indicating their diverse predictions for CMB observables and the correspondingly wide array of underlying
physical mechanisms that can be distinguished by {\sl CMBPol}. In this section we focus our discussion on a
generic and model-independent connection between inflationary gravitational waves and fundamental questions
about the high energy origin of the inflationary era.

\subsection{Clues about High-Energy Physics from the CMB}

Let us suppose that {\sl CMBPol} detects a primordial $B$-mode signal, {\it i.e.}~a $B$-mode spectrum imprinted
by a stochastic background of gravitational waves, or constrains it to lie below $r\sim 0.01$.
What would this imply for our understanding of the high-energy mechanism driving the inflationary expansion?\\

Two crucial clues would emerge from such a $B$-mode detection or constraint:

\vskip 12pt
\begin{enumerate}
\item {\bf Energy scale of inflation}:
{\it High-scale} inflation

The measurement (\ref{equ:PsWMAP}) of the amplitude of the scalar power spectrum (\ref{equ:PsSR}) implies the
following relation between the energy scale of inflation $V^{1/4}$ and the tensor-to-scalar ratio on CMB scales
$r_{\star} \equiv r(\phi_{\rm cmb})$ \beq \fbox{$\displaystyle V^{1/4} = 1.06 \times 10^{16} \, {\rm GeV}
\left(\frac{r_{\star}}{0.01}\right)^{1/4}$}\, . \eeq A detectably large tensor amplitude would convincingly
demonstrate that inflation occurred at a tremendously high energy scale, comparable to that of Grand Unified
Theories (GUTs).  It is difficult to overstate the impact of such a result for the high-energy physics
community, which to date has only two indirect clues about physics at this scale: the apparent unification of
gauge couplings, and experimental lower bounds on the proton lifetime.\footnote{Some of the earliest successful
inflation models involved direct connections between the inflaton and GUT scale particle physics.  While more
recent models of inflation are usually less tied to our models of particle interactions, instead invoking a
largely modular inflation sector, an observed connection between the scale of inflation and the scale of
coupling-constant unification might prompt theorists to re-visit  a possible deeper connection.}

\vskip 12pt
\item {\bf Super-Planckian field excursion}:
{\it Large-field} inflation

The tensor-to-scalar ratio relates to the evolution of the inflaton field (see Eqn.~(\ref{equ:TS}))\footnote{The
following formulae apply only in the special cases of single-field slow-roll inflation and single-field DBI
inflation \cite{DBI}. The more general result may be found in Appendix \ref{sec:taxonomy}.}
\beq r(N) = \frac{8}{\Mp^2} \left( \frac{d \phi}{d N}\right)^2\, . \eeq The total field excursion between the
end of inflation and the time when fluctuations were created on CMB scales is then \cite{LythBound} (see
Fig.~\ref{fig:potential}) \beq \frac{\Delta \phi}{\Mp} \equiv \int_{\phi_{\rm end}}^{\phi_{\rm cmb}} \frac{\d
\phi}{\Mp} = \int_0^{N_{\rm cmb}} \left(\frac{r}{8}\right)^{1/2} \d N \equiv \left(\frac{r_{\star}}{8}\right)^{1/2} N_{\rm eff}\, , \eeq where \beq N_{\rm eff} \equiv \int_0^{N_{\rm cmb}}
\left(\frac{r(N)}{r_{\star}}\right)^{1/2} \d N\, .
\eeq
The value of $N_{\rm eff}$ is model-dependent and depends on the precise evolution of the tensor-to-scalar ratio
$r(N)$.  For slow-roll models the evolution of $r$ is strongly constrained (and only arises at second order in
slow-roll), and can be estimated to be $N_{\rm eff} \sim {\cal O}(30-60)$ \cite{BM}. Taking the conservative lower bound, one then
finds \cite{LythBound, BM}\footnote{More recently, a Monte Carlo study of single-field slow-roll inflationary models which match
recent data on $n_s$ and its first derivative revealed an even stronger bound ${\Delta \phi \over M_{\rm pl}}
\gtrsim 10 \times \left(\frac{r_{\star}}{0.01}\right)^{1/4}$ \cite{Efstathiou:2005tq}.} \beq \label{equ:Lyth}
 \fbox{$\displaystyle \frac{\Delta \phi}{\Mp} \gtrsim 1.06 \times \left( \frac{r_{\star}}{0.01} \right)^{1/2}$}\, . \eeq A tensor-to-scalar ratio bigger than 0.01 therefore correlates with
super-Planckian field variation between $\phi_{\rm cmb}$ and $\phi_{\rm end}$.  As explained in detail below,
this would provide definite information about certain properties of the ultraviolet completion of quantum field
theory and gravity, and hence yield perhaps the first experimental clue about the nature of quantum gravity. An upper limit of $r<0.01$ would also be very important as it would rule out all large-field models of inflation.
\end{enumerate}

It is essential to recognize that
CMB polarization experiments have almost unique potential to provide these two clues about physics at the highest scales.\footnote{A futuristic direct-detection gravitational wave
experiment like the {\sl Big Bang Observer} (BBO) might someday complement the observations of CMB polarization \cite{Turner:1996ck, Smith:2005mm, Smith:2006xf,Chongchitnan:2006pe}.}

\subsection{Sensitivity to Symmetries and to Fundamental Physics}

General relativity is strongly coupled at high energies: in particular, graviton-graviton scattering becomes
ill-defined at the Planck scale, $\Mp \equiv (8\pi G)^{-1/2} = 2.4\times 10^{18}$ GeV.  Some other structure
must provide an ultraviolet completion of general relativity and quantum field theory.  Inflation is sensitive
to this ultraviolet completion of gravity in several important ways, which is the origin of much of the
difficulty in inflationary model-building, and at the same time is responsible for the great excitement about
experimental probes of inflation among high-energy theorists who study the physics of the Planck scale.  At a
phenomenological level, an inflationary model consists of an effective action for one or more scalar fields,
together with couplings of those scalars to known particles.  A more fundamental description of the same system
would include a {\it derivation} of the inflaton effective action from some reasonable set of premises that are
consistent with our understanding of quantum field theory and gravity.  The central challenge and opportunity is
this: any such derivation depends crucially on the assumptions made about the ultraviolet completion of gravity.

String theory is by far the best-understood example of a theory of quantum gravity, but the considerations
described below are more general and rely only on the firmly-established Wilsonian approach to effective field
theory, which allows systematic incorporation of the effects of high-scale physics into an effective Lagrangian
valid at lower energies.  Given the symmetry structure of the high-energy theory, as well as a choice of cutoff
$\Lambda$, the corresponding effective Lagrangian below the cutoff contains a generally infinite series of
higher-dimension operators, suppressed by appropriate powers of $\Lambda$, that are allowed by the symmetries of
the ultraviolet theory.\\

\newpage
There are two basic cases relevant to a Wilsonian analysis of inflation,
depending on whether or not there is an {\it approximate shift symmetry} in the inflaton direction in scalar
field space.

\begin{enumerate}

\item {\it No Shift Symmetry}

Consider first the case of a scalar field on which only the symmetry $\phi \to -\phi$ is imposed: \beq
\label{equ:Veff} {\cal L}_{\rm eff}(\phi) = -\frac{1}{2}(\partial\phi)^2- \frac{1}{2} m^2 \phi^2 - \frac{1}{4}
\lambda \phi^4  - \sum_{p=1}^\infty \Bigl[\lambda_p \phi^4  +\nu_p(\partial\phi)^2 \Bigr] \Bigl(
\frac{g\, \phi}{\Lambda}\Bigr)^{2p}  
+...\,, \eeq where the omitted terms include more derivatives.

An important role in this Wilsonian argument is played by the choice of symmetries one assumes of the
ultraviolet  (UV) theory. A scalar without the $\mathbb{Z}_2$ symmetry would have been expected to appear with
odd powers as well in the expansion (\ref{equ:Veff}); the $\mathbb{Z}_2$ symmetry selects instead only even
terms.
If the UV theory has no other symmetries, then the general expectation, confirmed in a wide range of analogous
physical systems, is that the coefficients $g$ (which control the couplings of the inflaton to other fields) and $\lambda_p,\nu_p$ are of order unity.  Conversely, systems with small couplings have approximate shift symmetries, discussed in the next item below.  Moreover, we expect that
the cutoff $\Lambda$ can be {\it at most}  $M_{\rm pl}$, because gravitational scattering itself becomes strong
there and must be made unitary.
In the case of string theory, new physics
becomes relevant at a parametrically lower scale, $M_{\rm string}$; in theories with extra dimensions there is
also a threshold with new massive states at $M_{\rm KK}$ (where typically, in string constructions, $M_{\rm KK}
< M_{\rm string}$).
The Wilsonian expectation can be confirmed in the case of string theory through explicit computations of
potentials for scalar fields in directions without a shift symmetry ({\it
e.g.}~\cite{Silverstein:2004id,Douglas:2006es}).
In these directions in field space, one indeed obtains such an infinite series which de-correlates over
distances of order $M_{\rm string}$ in field space.  This is to be expected; as one moves a distance $\Lambda$
in field space, new fields become light while previously light fields can become heavy, and their exchange
corrects the inflaton potential.  One must therefore make assumptions about couplings of the inflaton to modes
of mass $\gg \Lambda$ if one wishes to control features of the potential over distances in field space $\gg
\Lambda$. Since we wish to be very conservative in estimating the size of corrections, we will set $\Lambda =
M_{\rm pl}$.

Combining these facts, {\it in scalar field directions without a sufficiently constraining symmetry}, the effective Lagrangian evidently receives important corrections from an
infinite series of higher-dimension operators whenever $\phi$ ranges over a distance of order $M_{\rm pl}$.
Scalar fields in this class can support {\it small-field} inflation ($\Delta \phi \ll \Mp$), which only requires
the accidental near-cancellation of a small set of operators in the effective potential.
Such models of inflation predict a small tensor signal, though other signatures (such as non-Gaussianity and
cosmic strings) can arise, depending on the precise model.

\vskip 10pt
\item {\it Shift Symmetry}

We have stressed that a key assumption in the Wilsonian parametrization of the effective potential is the
symmetry structure of the ultraviolet theory. Consider now a direction $\phi$ in field space with an approximate
symmetry under which $\phi$ shifts, $\phi\to\phi+ const$. We assume that the leading effect breaking this shift symmetry is the inflaton potential itself.  As a specific example, consider the case in which the inflaton potential behaves
like a power, $V(\phi)\sim \mu^{4-p}\phi^p$, in the relevant range of field space.  The inflaton self-interactions encoded in this potential, along with its
coupling to gravity, renormalize the potential.  Gravitational interactions are Planck-suppressed, leading to
small corrections.  Moreover, for the {\sl COBE}-normalized power spectrum discussed above, the dimensionful coupling $\mu$
appearing in the potential is quite small compared to $\Mp$, leading to small loop corrections from the scalar
self-interactions. The shift symmetry in the ultraviolet theory forbids the presence (with order one coefficients) of the series of terms (\ref{equ:Veff}) that would add structure to the potential on distances $\Delta \phi < M_{\rm
pl}$ and would therefore spoil flatness. Such a system can thus robustly support
large-field inflation \cite{Linde:1983gd}, in a way consistent with the principles of effective field theory.

Because of the super-Planckian range of the field in this case, it is particularly important to move beyond
effective field theory and analyze the symmetry structure of the UV completion of gravity, so that we can understand whether suitable approximate shift symmetries are present in well-motivated theories of Planck-scale physics. In the case of
string theory, a subset of scalar fields do enjoy an approximate shift symmetry, and according to recent work
described in Appendix \ref{sec:taxonomy}, they can support large-field inflation with a tensor mode signature
accessible to {\sl CMBPol}.\footnote{Interestingly, the predictions for $r$ and $n_s$ in a subset of these models turn out to be
distinctive \cite{Silverstein:2008sg}, different from those of the simple integer power laws discussed in the original works on large-field
inflation in quantum field theory.} In general, there is preliminary evidence from string theory that both
small-field and large-field models of inflation -- with their distinct symmetry structures -- are indeed
compatible with a candidate ultraviolet completion of quantum gravity and particle physics.
\end{enumerate}

In summary,  we have explained that for the purpose of understanding large-field inflation in an effective field theory treatment, it is useful to organize scenarios into two broad classes,
characterized by whether or not the inflation direction possesses an {\it approximate shift symmetry}.
This symmetry structure is sensitive to the UV completion of gravity, and we remarked that both cases do arise in string theory, albeit via
rather different mechanisms. By determining whether the inflaton field excursion was super-Planckian or not,
{\sl CMBPol} has the potential to probe important aspects of the scalar field space and the symmetry structure
of quantum gravity, and to distinguish very different mechanisms for inflation.\footnote{From Eqn.~(\ref{equ:Lyth}) we see that $r =0.01$ is a critical value for the tensor-to-scalar ratio.  The regimes
$r> 0.01$ and $r< 0.01$ distinguish the two qualitatively different classes of inflationary theories. For related arguments for $r=0.01$ as a significant physics milestone in
inflation see \cite{Kinney:2003gi}.} This is
an astonishing opportunity.

\subsection{Tests of String-Theoretic Mechanisms}

To conclude this section, we note that near-future CMB observations and other precision cosmological experiments will
provide unprecedented opportunities to perform empirical
tests of string-theoretic mechanisms for inflation and reheating.  These mechanisms -- briefly
reviewed in Figure \ref{fig:nsr} and Appendix \ref{sec:taxonomy} -- are motivated by the sensitivity
of inflationary effective actions to the ultraviolet
completion of gravity, for which string theory is the leading
candidate.  So far, rather than directly
producing UV completions of the simplest-looking
inflationary potentials, this study has led to distinctive mechanisms
for inflation, with a rich phenomenology.  These include variants
of hybrid inflation \cite{Dvali:1998pa, KKLMMT}, with the possibility of signatures from relic cosmic strings \cite{Firouzjahi:2005dh};
variants of chaotic inflation and natural (axion) inflation \cite{Silverstein:2008sg, MSW} (with predictions for $r$ and $n_s$ distinct from those of the corresponding classic models), and new string-inspired mechanisms leading to strong non-Gaussian signatures \cite{DBI, Alishahiha_etal04}.  Each of these mechanisms can be realized in effective field theory, and so can in principle exist outside of string theory; however, as we have explained, the structure arising from the ultraviolet completion plays a crucial role in each case, and one might argue that these mechanisms are more natural in string theory than they appear to be in field theory.  Finally, although observational limitations will ultimately restrict our ability to identify the detailed model of inflation, it is encouraging that the upcoming window of accessible observations will provide concrete connections between data and physics sensitive to quantum gravity.

%%%%%%%%%%%%%%%%%%%%%%%%%%%%%%%%%%%%%%%%%%%%%%%%%%%
\newpage
%%%%%%%%%%%%%%%%%%%%%%%%%%
\section{Beyond the $B$-mode Diagnostic}
%\section{Beyond the Tensor-to-Scalar Ratio}
%\section{Beyond the Tensor-mode Diagnostic}
\label{sec:beyond}
%%%%%%%%%%%%%%%%%%%%%%%%%%

In the previous section we described the 
potential of $B$-mode polarization as a probe of
fundamental physics. These considerations were largely independent of the specific model for inflation and in
particular did not depend in any significant way on the assumption of single-field slow-roll inflation. In this section we discuss
complementary tests of inflation beyond the $B$-mode diagnostic, like the scale-dependence (\S\ref{sec:spectrum})
and the non-Gaussianity (\S\ref{sec:NG}) of the scalar spectrum and a possible contribution of isocurvature modes
(\S\ref{sec:iso}). These observables reveal much about the details of the physics driving the inflationary
expansion.

\subsection{Models of Inflation and their Phenomenology}

We preface this section with a brief summary of the most popular `models of inflation' (for a more complete
discussion the reader is referred to Appendix \ref{sec:taxonomy}).

\vskip  8pt During the inflationary epoch the universe is dominated by a form of stress-energy which sources a nearly
constant Hubble parameter $H= \partial_t \ln a$. Theoretically, this can arise via a truly diverse set of
mechanisms with disparate phenomenology and varied theoretical motivations.  Recently, a useful
model-independent characterization of {\it single-field} models of inflation and their perturbation spectra has
been given \cite{Cheung:2007st, Garriga:1999vw, Chen:2006nt, Bean:2008ga, Weinberg:2008hq}. Starting from this basic structure, each model of
single-field inflation arises as a special limit.  One important limit is the traditional case of single-field
slow-roll inflation, which we review first (\S\ref{sec:subSR}).  We then discuss more general single-field mechanisms for inflation
and finally present {\it multi-field} models (\S\ref{sec:nonmin}).

%\addtocontents{toc}{\SkipTocEntry}
\subsubsection{Single-Field Slow-Roll Inflation}
\label{sec:subSR}

{\it Single-field slow-roll inflation} is described by a canonical scalar field $\phi$ minimally coupled to
gravity \beq \label{equ:SRactionX}
 S =\frac{1}{2} \int \d^4 x \sqrt{-g} \left[ {\cal R} - (\nabla \phi)^2-2 V(\phi) \right]\, , \quad \qquad
\Mp^{-2} \equiv 8 \pi G \equiv 1\, . \eeq
It should be emphasized that the following discussion assumes that a single field
describes the dynamics during inflation
and that curvature perturbations are
generated from vacuum fluctuations of the inflaton field. A measurement of the amplitude and the
scale-dependence of the scalar and tensor spectra then directly constrains the shape of the inflaton potential $V(\phi)$.
Conversely, only for single-field slow-roll models does a specification of the inflaton potential uniquely
specify the inflationary parameters $r$ and $n_s$. In \S\ref{sec:nonmin} we discuss the consequences of relaxing
those assumptions.

If we normalize the potential on CMB scales, $v(\phi) \equiv V(\phi)/V(\phi_{\rm cmb})$, then
(\ref{equ:PsSR}) and (\ref{equ:TS}) become
\beq r = 8 \left. (v')^2 \right|_{\phi = \phi_{\rm
cmb}}\, ,  \qquad {\rm and} \qquad n_s - 1 = \left. \left[ 2 v'' - 3 (v')^2  \right] \right|_{\phi = \phi_{\rm cmb}}\, . \eeq A
measurement of ($r$, $n_s$) therefore determines the shape of the inflaton potential ($v'$, $v''$) at $\phi_{\rm cmb}$.
The scalar amplitude, $A_s = 2.4 \times 10^{-9}$, then fixes the energy scale of inflation, $V(\phi_{\rm cmb})$, in terms of $r$.

In Figure \ref{fig:nsr} we
illustrate three different criteria that classify single-field slow-roll models according to their predictions
for $r$ and $n_s$ \cite{Dodelson:1997hr}:
\begin{enumerate}
\item[i)] models predict either red ($n_s < 1$) or blue  ($n_s >1$) spectra,
\item[ii)] models have positive ($\eta > 0$) or negative ($\eta < 0$) curvature at the time when CMB scales exit the horizon,
\item[iii)] models are of the
large-field ($\Delta \phi > \Mp$) or small-field ($\Delta \phi < \Mp$) type according to the total field
excursion during the inflationary phase (see Section \ref{sec:Bmodes}).
\end{enumerate}

\begin{figure}[htbp!]
    \centering
        \includegraphics[width=.75\textwidth]{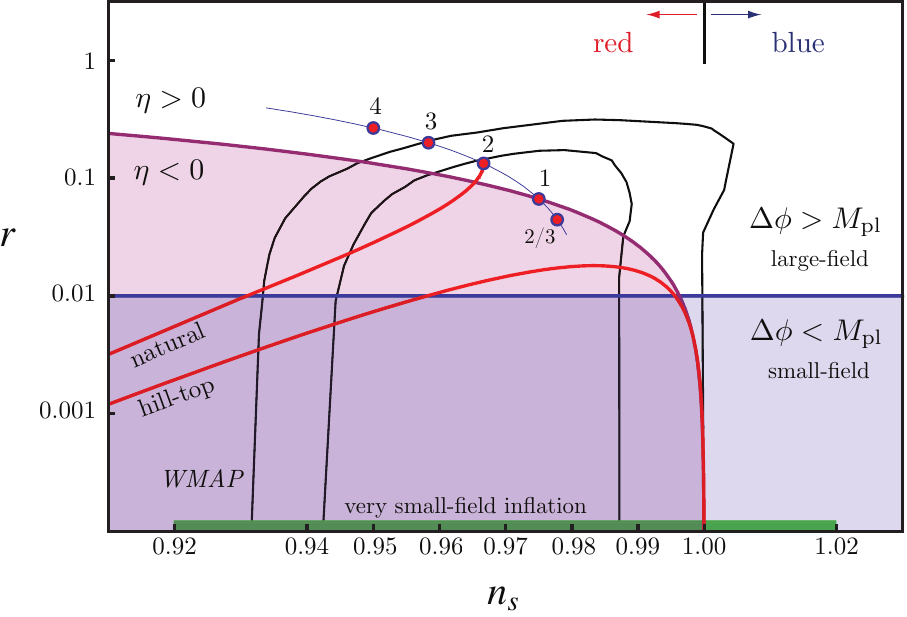}
   \caption{Constraints on single-field slow-roll models in the $n_s$-$r$ plane. The value of $r$ determines whether the models involve large or small field variations.  The value of $n_s$ classifies the scalar spectrum as red or blue. Combinations of the values of $r$ and $n_s$ determine whether the curvature of the potential was positive ($\eta > 0$) or negative ($\eta < 0$) when the observable universe exited the horizon.
   Also shown are the {\sl WMAP} 5-year constraints on $n_s$ and $r$ \cite{WMAP5} as well as
the predictions of a few representative models of single-field slow-roll inflation:
%\newline
{\it chaotic inflation}: $\lambda_p\, \phi^p$, for general $p$ (thin solid line) and for $p=4, 3, 2, 1, \frac{2}{3}$({\large $\bullet$}); models with $p=2$~\cite{Nflation}, $p=1$~\cite{MSW} and $p=\frac{2}{3}$~\cite{Silverstein:2008sg} have recently been obtained in string theory;
%\newline
{\it natural inflation}: $V_0 [1-\cos(\phi/\mu)]$ (solid line), {\it hill-top inflation}: $V_0 [1-(\phi/\mu)^2] + \dots$ (solid line);
%\newline
{\it very small-field inflation}: models of inflation with a very small tensor amplitude, $r \ll 10^{-4}$ (green bar); examples of such models in string theory include warped D-brane inflation \cite{KKLMMT, BDKM, Holographic}, K\"ahler inflation \cite{Conlon:2005jm}, and racetrack inflation \cite{BlancoPillado:2006he}.}
    \label{fig:nsr}
\end{figure}

Figure \ref{fig:nsr} also shows the latest CMB constraints on $r$ and $n_s$ \cite{WMAP5} as well as the predictions of a few
simple, but well-motivated, models of single-field slow-roll inflation. We see that for $n_s > 0.95$ many of the
`simplest'\footnote{We caution the reader that there is no universally accepted definition of `simple models'.
Here we loosely take `simple models' to mean models with the seemingly simplest functional forms for the effective
potential $V(\phi)$.  For discussions of criteria for fine-tuning of inflation based upon the algebraic simplicity of the potential see {\it e.g.}~\cite{Boyle:2005ug, Bird:2008cp, Hotchkiss:2008sa}.} inflationary models predict $r \ge 0.01$.

%\addtocontents{toc}{\SkipTocEntry}
\subsubsection{Beyond Single-Field Slow-Roll}
\label{sec:nonmin}

For models of single-field slow-roll inflation we have just seen how a measurement of $n_s$ and $r$ cleanly
correlates with the scale and shape of the inflaton potential $V(\phi)$. This correspondence between
cosmological observables and the inflationary potential is broken in models in which the kinetic term for the
inflaton is non-canonical or more than one field is dynamically relevant during inflation.
Although this makes the interpretation of a measurement of $n_s$ and $r$ less direct, additional observables
beyond $r$ and $n_s$ allow one to break this degeneracy (see \S\ref{sec:spectrum}, \S\ref{sec:NG} and
\S\ref{sec:iso}).

\vskip 8pt
\noindent {\it General single-field inflation}

Non-trivial {\it kinetic effects} are often parameterized by the following action \cite{Garriga:1999vw, Chen:2006nt},
\beq \label{equ:Paction0} S = \frac{1}{2} \int \d^4 x \sqrt{-g} \left[ {\cal R} + 2 P(X, \phi)   \right]\, ,
\eeq where $X \equiv -\frac{1}{2} g^{\mu \nu} \partial_\mu \phi \partial_\nu \phi$. Examples of inflation models
with actions of the type (\ref{equ:Paction0}) are k-inflation \cite{k-Inflation}, DBI inflation \cite{DBI} and
ghost inflation \cite{GhostInflation}. Slow-roll inflation (\ref{equ:SRactionX}) is contained in (\ref{equ:Paction0}) as the special case $P(X,\phi) =X-V(\phi)$. The
function $P(X,\phi)$ corresponds to the pressure of the scalar fluid, while its energy density is $\rho= 2 X P_{,X} -
P$.  Furthermore, the models are characterized by a speed of sound \beq c_s^2 \equiv \frac{P_{,X}}{\rho_{,X}} =
\frac{P_{,X}}{P_{,X} + 2 X P_{,XX}}\, . \eeq The time-variation of the speed of sound adds an extra term to the
prediction for the spectral index $n_s$ (see Appendix \ref{sec:taxonomy}).  This breaks the one-to-one correspondence between ($v'$, $v''$) and
($r$, $n_s$).

In the following subsections, we discuss how further information about models with non-trivial sound speed can
be obtained from a measurement of the scale-dependence of the scalar ($\alpha_s$) and tensor spectra ($n_t$)
(\S\ref{sec:spectrum}) and the non-Gaussianity ($f_{\rm NL}$) of the scalar spectrum (\S\ref{sec:NG}).

\vskip 8pt
\noindent{\it Multi-field inflation}

Employing two or more scalar fields during inflation
\cite{Linde:1984ti, Kofman:1985aw, Kofman:1986wm, Silk:1986vc} extends the possibilities for
inflationary models, but also diminishes the predictive power of
inflation.  {\it Multi-field models} can produce features in the spectrum of
adiabatic perturbations
\cite{Kofman:1988xg, Salopek:1988qh, Hodges:1989dw, Mukhanov:1991rp, Polarski:1992dq,Polarski:1994bk,Adams:1997de,Lesgourgues:1999uc},
and seed isocurvature perturbations
\cite{Linde:1984ti, Kofman:1986wm, Polarski:1994rz,Langlois:1999dw,GarciaBellido:1995qq,Gordon:2000hv}
which could eventually leave an imprint on CMB anisotropies. Some
  multi-field models decouple the creation of density perturbations
from the dynamics during inflation. If the decay of the vacuum energy
at the end of inflation is sensitive to the local values of fields
other than the inflaton then this can generate primordial
perturbations due to {\em inhomogeneous reheating}
\cite{Dvali:2003em,Kofman:2003nx} or {\it modulated hybrid inflation}
\cite{Bernardeau:2004zz}. Alternatively, in the {\em curvaton}
scenario \cite{Linde:1996gt,Lyth:2001nq,Moroi01}, the inhomogeneous
distribution of a weakly coupled field generates density perturbations
when the field decays into radiation at some time after inflation. The
curvaton scenario can also produce {\it isocurvature density
  perturbations} (\S\ref{sec:iso}) in particle species ({\it
  e.g.}~baryons) whose abundance differs from the thermal equilibrium
abundance at the time when the curvaton decays
\cite{Linde:1996gt, Lyth:2002my}. Inflation is still required to set up large-scale
perturbations from initial vacuum fluctuations in all these
models. But when the primordial density perturbation is generated by
local physics some time after slow-roll inflation then the local form
of {\it non-Gaussianity} is no longer suppressed by slow-roll
parameters (\S\ref{sec:NG}). Measurements beyond $B$-mode polarization
are therefore vital as diagnostics for multi-field models of
inflation.  We discuss these important inflationary observables in the
following sections.

\subsection{Deviations from Scale-Invariance}
\label{sec:spectrum}

\noindent
{\it Scalar spectrum}\\
The scale-dependence of primordial scalar fluctuations is a powerful probe of inflationary dynamics,
\beq
P_s(k) = A_s \left(\frac{k}{k_\star} \right)^{n_s-1 + \frac{1}{2} \alpha_s \ln(k/k_\star)}\, , \quad \qquad n_s -1 = \frac{d \ln P_s}{d \ln k}\, ,  \qquad \alpha_s = \frac{d n_s}{d \ln k}\, .
\eeq
In particular, as we discussed above, for single-field slow-roll inflation, deviations from perfect scale-invariance $(n_s = 1, \alpha_s=0)$ are encoded in the shape of the inflaton potential.
A large scale-dependence (``running") $\alpha_s$ of the spectral index $n_s$ arises only at second-order in slow-roll and is therefore expected to be small.

\vskip 8pt
In the case of slow-roll inflation, a definitive measurement of a large running, $\alpha_s$, is a signal that $\xi_H$, the third {\it Hubble slow-roll} parameter \cite{Liddle:1994dx} (defined in
analogy to the first two {\it potential slow-roll} parameters discussed previously),
\begin{equation}
\xi_H \equiv 4 M_{\rm pl}^4\left[\frac{H'(\phi) H'''(\phi)}{H^2(\phi)}\right],
\end{equation}
played a significant role in the dynamics of the inflaton \cite{Chongchitnan:2005pf} as the CMB scales exited
the horizon. The consequences for the physics of inflation differ depending on whether the running is negative
or positive, and both options would dramatically complicate the theoretical understanding of inflation:

\begin{enumerate}
\item[i)] {\it Large negative running}

A large negative running implies that  $\xi_H$ was (relatively) large and positive  as the cosmological
perturbations were laid down.  It can be shown that   $\xi_H >0$ generally hastens the end of inflation
(relative to    $\xi_H =0$), provided the higher-order slow-roll parameters can be ignored.  With these
assumptions, we find a tight constraint on   $\xi_H$  if we are to avoid a premature end to slow-roll, with
inflation terminating soon after the observable scales leave the horizon \cite{Chung:2003iu, Malquarti:2003ia,
Easther:2006tv, Peiris:2006sj,Hamann:2008pb}. Thus, a definitive observation of a large negative running would imply that any
inflationary phase requires higher-order slow-roll parameters to become important after the observable scales
leave the horizon
\cite{Chung:2003iu, Malquarti:2003ia,Makarov:2005uh,Ballesteros:2005eg, Adshead:2008vn}, or multiple fields which could produce complicated spectra,
a temporary breakdown of slow-roll (inducing
features in the potential), or even several distinct
stages of inflation
\cite{Sasaki:1995aw,Kofman:1986wm,Silk:1986vc,Salopek:1988qh,Holman:1991ht,Starobinsky:1992ts,Polarski:1992dq,Polarski:1994bk,Lyth:1995ka,Adams:1997de,Lesgourgues:1998mq,Lesgourgues:1999uc,Adams:2001vc,Peiris:2003ff,Hunt:2004vt,Burgess:2005sb,Covi:2006ci, Hamann:2007pa,Hunt:2007dn, Joy:2007na, Joy:2008qd}.

\item[ii)] {\it Large positive running}

The current cosmological data disfavor inflationary models with a blue
tilt on CMB scales, $n_s>1$ \cite{Dunkley:2008ie, WMAP5}; however, a
significant parameter space is still allowed with $n_s<1$ but with a
large positive running (implying a large negative $\xi_H$), which
would lead to a strongly blue-tilted spectrum after the cosmological
scales have exited the horizon \cite{Peiris:2008be}. Again under the
hypothesis that this parameterization can be extrapolated to the end
of inflation, we find a class of solutions where $\epsilon \rightarrow
0$ as $H$ remains finite, and the field rolls towards a minimum with a
substantial vacuum energy. The perturbation spectrum grows at small
scales, possibly diverges, and can lead to an over-production of
primordial black holes \cite{Hawking:1971ei, Carr:1974nx, Carr:1994ar,
  Kim:1996hr, Green:1997sz, GarciaBellido:1996qt, Covi:1998mb,
  Yokoyama:1999xi, Leach:2000ea, Zaballa:2006kh, Chongchitnan:2006wx,
  Kohri:2007qn, Peiris:2008be}, or even the onset of eternal inflation
\cite{Linde:1986fd, Kohri:2007qn, Peiris:2008be}. Inflation could also
stop well before black-hole production, due to a mechanism involving
another sector: for instance, a second scalar field coupled with the
inflaton, which could trigger a phase transition marking the end of
inflation and the onset of reheating. This mechanism is generically
called {\it hybrid inflation} \cite{linde91,linde94,copeland94}
and
belongs to the category of single-field slow-roll models, since the
dynamics of inflation is still governed by a single inflaton (as long
as the trigger is a heavy, with $m \gg H$); the hybrid inflation
paradigm amounts to relaxing the assumption that the end of inflation
is due to the breaking of slow-roll conditions. Consequently, if we
measure a large positive running we will conclude that the end
of the inflationary phase is not described within the single-field
slow-roll formalism, or that higher-order terms in the slow-roll
expansion are important.
\end{enumerate}
Finally, we should mention that a large running might more naturally be accommodated in inflationary models with general speed of sound (see Appendix \ref{sec:taxonomy}).
In this case, $\alpha_s$ receives contributions from $c_s$ and its time-evolution during inflation.  This might allow larger values of $\alpha_s$ than the slow-roll analysis suggests.

\vskip 8pt
\noindent
{\it Tensor spectrum and consistency relation}\\
Single-field slow-roll inflation predicts a nearly scale-invariant spectrum of tensor modes
\beq
P_t = A_t \left( \frac{k}{k_\star} \right)^{n_t}\, , \qquad n_t = - 2 \epsilon \approx 0\, .
\eeq
At first order in a slow-roll expansion it furthermore predicts the following {\it consistency relation} between the amplitude and the scale-dependence  of the spectrum of tensor fluctuations,
\beq
r = - 8 n_t\, .
\eeq

\begin{enumerate}
\item[i)] {\it Multiple fields}

The presence of multiple fields during an inflationary phase is one of the possible sources of deviation from
the consistency relation holding for single-field models of slow-roll inflation. There
exists a model-independent consistency relation for slow-roll inflation with canonical fields
\cite{Wands:2002bn} (see Appendix \ref{sec:taxonomy})
\begin{equation}
r = -8n_t \sin^2 \Delta\, ,
\end{equation}
where for two-field inflation $\cos\Delta$ is the correlation between the adiabatic and isocurvature
perturbations, which is a directly measurable quantity (see \S\ref{sec:iso}). More generally, $\sin^2\Delta$
parameterizes the ratio between the adiabatic power spectrum at horizon-exit during inflation and the observed power spectrum.
The conversion of non-adiabatic perturbations into curvature perturbations after horizon-exit decreases the
tensor-to-scalar ratio for a fixed value of the slow-roll parameter $\epsilon$ (or $n_t = - 2 \epsilon$).

\item[ii)] {\it Kinetic effects}

A second way to violate the single-field slow-roll consistency relation is the non-slow-roll evolution of the
inflaton driven by a non-canonical kinetic term.  This leads to a non-trivial speed of sound $c_s \ll 1$ and a
modified consistency relation (see Appendix \ref{sec:taxonomy})
\beq
r = - 8 n_t \, c_s\, .
\eeq
In those theories the violation of the slow-roll
consistency relation correlates with a large non-Gaussianity of the density spectrum, $f_{\rm NL} \sim 1/c_s^{2}
\gg 1$ (see \S\ref{sec:NG}).
\end{enumerate}

This emphasizes the importance of measuring or constraining the scale-dependence of the tensor power spectrum. Although
it will be hard to measure any scale-dependence of the tensors if the single-field consistency relation holds ({\it i.e.}, if $n_t=-r/8$), a large tilt would invalidate this consistency relation. A large negative tilt could be consistent
with multi-field inflation or a non-trivial speed of sound arising from a non-canonical kinetic term for the
inflaton.
Finally,
since $n_t = 2 \dot H/H^2$, a positive tilt is only possible if the theory violates the null energy condition,
$\dot H > 0$.

\subsection{Non-Gaussianity}
\label{sec:NG}

Non-Gaussianity is a measure of interactions of the inflaton. A certain level of non-Gaussianity is a generic
prediction of inflation: the inflaton at least interacts gravitationally and likely has a potential beyond a
simple mass term. However, the slow-roll requirements limit single-field inflation with a smooth evolution and a
canonical kinetic term to $f_{\rm NL}\sim {\cal{O}(\epsilon,\eta)}\sim {\cal O}(10^{-2})$~\cite{Acquaviva02, Maldacena03},
which is undetectable with current and foreseen CMB experiments. As with the consistency relation of the
previous section, measuring a deviation from Gaussianity in the primordial spectrum would indicate physics
beyond standard single-field slow-roll. Both non-trivial kinetic terms (derivative self-interactions) and
multiple field effects may lead to large, observationally distinct non-Gaussianity. Regardless of details, a
detection of primordial non-Gaussianity with $|f_{\rm  NL}|\sim\mathcal{O}(1)$ would {\em rule out} the minimal
inflationary scenario.

If the fluctuations in the primordial curvature $\zeta$ were exactly Gaussian (that is, if the inflaton were a
free field), all the statistical properties of $\zeta$ would be encoded in the two-point function. A non-zero
measurement of the connected part of any higher-order correlation function would be a detection of
non-Gaussianity, but the deviation from zero is almost certainly largest in the three-point
function\footnote{While the connected four-point function is in general much smaller than the three-point and so
much harder to detect (see {\it e.g.}~Ref.~\cite{Okamoto:2002ik}), it could in principle be used to distinguish
between models with identical three-point functions~\cite{Huang:2006eh}. In addition, some multi-field or curvaton models may have a negligible bispectrum but significant trispectrum \cite{Enqvist:2008gk}.}. In momentum space, the three-point correlation
function can be written generically as:
\begin{equation}
\langle \zeta_{{\bf k}_1}\zeta_{{\bf k}_2}\zeta_{{\bf k}_3}\rangle=(2\pi)^3\, \delta({\bf k}_1+{\bf
k}_2+{\bf k}_3) \,  f_{\rm NL} \, F(k_1,k_2, k_3) \; .
\end{equation}
Here $f_{\rm NL}$ is a dimensionless parameter defining the amplitude of non-Gaussianity, while the function $F(k_1, k_2,  k_3)$ captures the momentum dependence. The amplitude and sign of $f_{\rm NL}$, as well as the shape and scale dependence of $F(k_1,k_2,k_3)$, depend on the details of the interaction generating the
non-Gaussianity, making the three-point function a powerful discriminating tool for probing models of the early universe~\cite{NGreview}.

Two simple and distinct shapes $F( k_1, k_2,  k_3)$ are generated by two very different
mechanisms~\cite{Babich_etal_04}: The {\em local shape} is a characteristic of multi-field models and takes its
name from the expression for the primordial curvature perturbation $\zeta$ in real space,
\begin{equation}
\label{local} \zeta({\bf x}) = \zeta_G({\bf x}) + \frac{3}{5} f_{\rm NL}^{\rm local} \left(\zeta_G({\bf x})^2
-\langle\zeta_G({\bf x})^2\rangle \right)\, ,
\end{equation}
where $\zeta_G({\bf x})$ is a Gaussian random field. Fourier transforming this expression shows that the signal
is concentrated in ``squeezed" triangles where $k_1 \ll k_2,k_3$. The local ansatz for non-Gaussianity has long
been a favorite of cosmologists~\cite{Salopek:1990jq,VHWK,komatsu01} and is the origin of the {\sl WMAP} convention
%\footnote{Up to a matter-era conversion factor of $5/3$.} 
for $f_{\rm NL}$ as the magnitude of the non-linear term. In addition,
it is physically well-motivated in multi-field models where the fluctuations of an isocurvature field are
converted into curvature perturbations. As this conversion happens outside of the horizon, when gradients are
irrelevant, one generates non-linearities of the form (\ref{local}). Specific models of this type include multi-field inflation~\cite{BMRmulti, Bernardeau:2002jy, Bernardeau:2002jf, Sasaki:2008uc, Naruko:2008sq, Byrnes:2008wi, Byrnes:2006fr, Langlois:2008vk, Valiviita:2008zb, Assadullahi:2007uw, Valiviita:2006mz, Vernizzi:2006ve, Allen:2005ye}, the curvaton scenario~\cite{Linde:1996gt, Lyth_etal03},
inhomogeneous reheating~\cite{Dvali:2003em, Kofman:2003nx}, and New Ekpyrotic
models~\cite{Creminelli:2007aq,Koyama_etal07,Buchbinder_etal07, Lehners:2007wc,Lehners:2008my, Koyama:2007ag, Koyama:2007mg}. In these cases, $|f_{\rm NL}^{\rm local}|$
is model-dependent but generically larger than $5-10$.

The second important shape is called {\em equilateral} as it is largest for configurations with $k_1 \sim k_2
\sim k_3$. The equilateral form is generated by single-field models with non-canonical kinetic terms such as DBI
inflation~\cite{Alishahiha_etal04}, ghost inflation~\cite{GhostInflation,Senatore:2004rj} and more general models
with small sound speed~\cite{Creminelli:2003iq, Chen:2006nt,Cheung:2007st}. As discussed in \S \ref{sec:nonmin}, the magnitude of non-Gaussianity increases as the sound speed $c_s$ decreases, with $f_{\rm
NL}^{\rm equil.}\propto1/c_s^2$. There is a model-dependent prefactor (negative in DBI inflation), and the non-Gaussianity
is scale-dependent if the sound speed is time-dependent. There is no theoretical lower limit on $c_s$ (although perturbative
considerations imply $c_s\gtrsim10^{-9/4}$ \cite{Leblond:2008gg}) so current bounds on non-Gaussianity at CMB scales already constrain
these models significantly.

The distinction between the single-field and multi-field case is robust, as one can prove that a single-field model always gives $f_{\rm NL} \sim {\cal O} (n_s-1) \ll 1$ in the squeezed limit, independently of the
specific Lagrangian~\cite{Maldacena03,Creminelli:2004yq, Cheung:2007sv}. 
The detection of a large non-Gaussianity in the local limit would therefore rule out all single-field models in which slow-roll is maintained throughout inflation; however, features in the potential that cause temporary departures from slow-roll can source local non-Gaussianity \cite{Chen:2006xjb} even in a single-field model.
Furthermore, higher-derivative
terms can be important in multi-field models, where the shape of the three-point function can interpolate between
the local and the equilateral cases~\cite{Allen_et87,Langlois:2008wt, Arroja:2008yy}. Finally, deviations from the standard Bunch-Davies
vacuum for the fluctuations can be a source of additional non-Gaussianities
\cite{Lesgourgues:1996jc,Martin:1999fa,Chen:2006nt,Holman:2007na,Xue:2008mk}, with an intermediate shape and scale-dependence.

Although current data analyses only constrain constant $f_{\rm NL}$, there are well-motivated
examples where the predicted non-Gaussianity is scale-dependent.
If the non-Gaussianity is (approximately) scale-invariant,
 it is useful phenomenologically to absorb the overall scale-dependence into $f_{\rm NL}$ and define a running
non-Gaussianity index $n_{\rm NG}$ by
\begin{equation}
f_{\rm NL}=f_{\rm NL}(k_\star)\left(\frac{k}{k_\star}\right)^{n_{\rm NG}-1}\;.
\end{equation}
For small sound speed models, scale-dependence of the non-Gaussianity comes from scale-dependence of the sound
speed, which also affects the spectral index and the relation between the tensor index and the tensor-to-scalar
ratio. In DBI inflation, a weak scale-dependence of precisely this type is rather natural
\cite{Chen:2005fe,Langlois:2008qf}. Even in the case of an inflaton with a standard kinetic term, features in the inflationary
potential, including isolated sharp features \cite{WangKamion,Komatsu:2003fd,Chen:2006xjb,Bean:2008na} or a
series of closely-spaced small features \cite{Chen:2008wn}, can produce non-Gaussianities with more significant
scale-dependence, while keeping the viability of the power spectrum. Since such non-Gaussianities typically have
oscillatory behavior in $\ell$-space \cite{Chen:2006xjb,Chen:2008wn}, independent data from temperature and
polarization anisotropies are important to identify them despite cosmic variance.

At present, the most stringent constraints on $f_{\rm NL}$ come from the {\sl WMAP} 5-year
analysis~\cite{WMAP5}. For the two shapes mentioned above the limits are:
\begin{eqnarray}
-9<f^{\rm local}_{\rm NL}<111\ \quad {\rm at} \ \  95\%\ {\rm C.L.}\\
-151<f^{\rm equil.}_{\rm NL}<253\ \quad {\rm at}\ \ 95\%\ {\rm C.L.}
\end{eqnarray}
Ongoing galaxy surveys such as the {\sl Sloan Digital Sky Survey} (SDSS) have a sensitivity to $f_{\rm NL}^{\rm local}$ which is
competitive with {\sl WMAP} \cite{Dalal:2007cu}. Upon inclusion of these additional data, the allowed interval for
$f_{\rm NL}^{\rm local}$ reduces considerably \cite{Slosar:2008hx}
\begin{eqnarray}\label{best_local_limit}
-1<f^{\rm local}_{\rm NL}<70\ \quad && {\rm at}\ \  95\%\ {\rm C.L.} \quad ({\rm LSS+WMAP})\, , \\
-29<f^{\rm local}_{\rm NL}<70\ \quad && {\rm at}\ \  95\%\ {\rm C.L.} \quad ({\rm LSS\ only})\, .
\end{eqnarray}
With the exception of techniques that rely on measuring the large-scale structure bispectrum,  however, constraints on non-Gaussianity from galaxy surveys are not sensitive to the shape of non-Gaussianity.  While future surveys may achieve $\Delta f^{\rm local}_{\rm NL}\sim 1$ or less \cite{McDonald:2008sc, Carbone:2008, Seljak:2008xr} they are  not nearly as sensitive  to $f_{\rm NL}^{\rm equil.}$. The abundance of collapsed objects  (halos) can also  be used to constrain non-Gaussianity. The halo abundance  is only sensitive to the skewness, thus
is sensitive to the sign of non-Gaussianity, regardless of shape, in a particularly simple
way: $f_{\rm NL}>0$ yields more very large structures (galaxy clusters) than Gaussian fluctuations would, while
$f_{\rm NL}<0$ yields fewer \cite{MVJ}. Notice that the current allowed interval in (\ref{best_local_limit}) slightly
prefers a positive value for $f_{\rm NL}^{\rm local}$, in agreement with that found already in the {\sl WMAP} 3-year
analysis \cite{YadavNG}. Future data from the {\sl WMAP} experiment and further optimization of the analysis should
improve the current limits by approximately 10-20\% \cite{Creminelli:2006rz}. Future large-scale structure
measurements may also be helpful in determining any simple scale-dependence of the non-Gaussianity since they
probe smaller scales than the (current) CMB data does \cite{loverde:2008}.

The previous constraints on CMB non-Gaussianity have been obtained using the temperature signal only. The $E$-mode polarization signal
can  improve the sensitivity by approximately a factor of 1.6 \cite{BZ04,YW05,YKW07}. Although experiments have
already started characterizing $E$-mode polarization  anisotropies \cite{dasi_pol_02, wmap_1st_pol,
wmap_2nd_pol,boom_ee},  the signal-to-noise ratio is still too low to allow significant improvements in the
current constraints of non-Gaussianity. The upcoming {\sl Planck} satellite will improve this, but its
$E$-mode polarization signal will still be cosmic variance limited only up to $\ell \sim20$. Fisher matrix forecasts (see \S\ref{sec:CMBPol}),
assuming that all the contamination from foregrounds can be effectively removed (an issue which requires further
investigation, see {\it e.g.}~\cite{Cooray:2008xz}), show that {\sl Planck} will be able to improve the current limits by approximately a factor of 6,
reaching $1\sigma$ errorbars of the order $\Delta f_{\rm NL}^{\rm local}\simeq 4$ \cite{BZ04,YKW07}. The
improvement on $f_{\rm NL}^{\rm equil.}$ should scale in approximately the same way, leading to an expected
$1\sigma$ error of $\Delta f_{\rm NL}^{\rm equil.}\simeq 25$. On the other hand, a satellite mission such as {\sl
CMBPol} dedicated to polarization and cosmic variance limited up to $\ell \sim 2000$ would be able to further
improve on {\sl Planck} by a factor of order 1.6, reaching approximately $\Delta  f_{\rm NL}^{\rm local} \sim 2-3$ and
$\Delta  f_{\rm NL}^{\rm equil.} \sim 13-15$.  Considering that $f_{\rm NL}^{\rm local} \gtrsim 1$ marks the difference between standard  single-field slow-roll inflation (and a Bunch-Davies vacuum) and models that violate one or more of these conditions, the  potential of an experiment like {\sl CMBPol} becomes clear. In case of a high signal-to-noise detection, {\sl CMBPol} data may allow one to  measure either a
simple scale-dependence ($n_{\rm NG}$) or to find features.

\begin{table}
\begin{center}
\begin{tabular}{l l}
{\small Gaussian Quantum Fluctuation}  & {\small $\delta\eta$} \\
$\hspace{2.3cm}\downarrow $ & \\
{\small Non-Gaussian Inflaton Fluctuations} & {\small $\delta \phi \sim g_{\delta\phi}(\delta \eta+ f_{\delta \eta} \, \delta \eta^2)$} \\
\hspace{2.15cm} $\downarrow $ & \\
{\small Non-Gaussian Curvature Fluctuations} & {\small $\zeta \sim g_{\zeta}(\delta \phi + f_{\delta \phi}\, \delta \phi^2)$}\\
\hspace{2.3cm}$\downarrow $ & \\
{\small Non-Gaussian CMB Anisotropy} & {\small $\frac{\Delta T}{T}\sim g_T (\zeta + f_{\zeta}\, \zeta^2)$}
\end{tabular}
\caption{Flow chart summarizing how non-Gaussianity may arise in the CMB data starting from the
primordial Gaussian quantum fluctuations. Although quantum fluctuations produce Gaussian fluctuations $\delta
\eta$, any non-linearities in the inflationary dynamics or non-trivial interaction terms generate
non-Gaussianity (through a non-zero $f_{\eta}$). To first order in perturbations, $f_{\zeta}$ and $f_{\delta
\phi}$ are zero, and it is only at the second order that they appear. Here $g_T$ is the radiation transfer
function.} \label{tab:NG}
\end{center}
\end{table}

So far we have only concentrated on the primordial non-Gaussian signal induced on the CMB by the inflationary epoch.
However, the non-linearities of general relativity and of the plasma physics induce an additional non-Gaussian
signal~\cite{Pyne:1995bs}. These contributions are expected to give $f_{\rm
  NL} \sim {\cal O}(1)$ so that it will be important to study them in detail \cite{BMR1,BMR2}
for the level of sensitivity that will be reached by
{\sl CMBPol}. This additional signal will not
only represent a contaminant for the primordial non-Gaussianity, but also a new observational tool from the
epoch of recombination to the present.

What is the importance of a polarization-oriented mission like {\sl CMBPol} for non-Gaussianities? By the time
{\sl CMBPol} will fly, two scenarios are possible. In the first, {\sl WMAP} and {\sl Planck} will have detected a primordial
non-Gaussian signal.\footnote{In addition, large-scale structure observations will also probe $f_{\rm NL}^{\rm local} \sim 1$ by the time {\sl CMBPol} will fly. {\sl CMBPol} would be able to provide independent confirmation of these complementary observations.} This would represent a remarkable discovery because it would rule out the minimal model of
inflation and put severe constraints on the alternatives. In such a case, an instrument such as {\sl CMBPol}
(assuming it is cosmic variance limited for polarization up to $\ell \sim 2000$) would be crucial as it could
almost double the confidence level of the detection and explore the ``shape-dependence" of the signal. In that case we should be able to differentiate between a
local and an equilateral shape and to constrain the scale dependence of the primordial non-Gaussianity. Further,
by analyzing the temperature and the polarization data separately we would be able to reduce the systematic
effects and the foregrounds and increase our confidence in the discovery. In the second scenario {\sl WMAP} and {\sl Planck}
will not have detected non-Gaussianity.  Even in such a case, the additional information coming from {\sl
CMBPol} would be still very useful as it would probe the $f_{\rm NL} \sim$ few region. Indeed, the threshold
$f_{\rm NL} \sim$ few is very important since models which are significantly different from standard
single-field slow-roll inflation tend to produce a non-Gaussianity larger than this. Even a mild improvement in
the constraint is relevant. Measuring or constraining non-Gaussianity is a powerful tool for inflation, and
could provide evidence for small sound speed or multiple fields that is complementary to the other diagnostics
of this section. Finally, non-Gaussian signals at the level $f_{\rm NL} \sim 1$ are expected, even if not induced by inflation.  This regime will be accessible by {\sl CMBPol}.

\subsection{Isocurvature Fluctuations}
\label{sec:iso}

Isocurvature density perturbations are a ``smoking gun'' for
multi-field models of inflation. In single-field inflation, the
fluctuations of the inflaton field on large scales (where spatial
gradients can be neglected) can be identified with a local shift
backwards or forwards along the trajectory of the homogeneous
background field. They affect the total density in different parts of
the universe after inflation, but cannot give rise to variations in
the relative density between different components. Hence, they produce
purely adiabatic primordial density perturbations characterized by an
overall curvature perturbation, $\zeta$.

But in general one can also have relative perturbation modes between
different components, {\it e.g.}~between radiation and matter
\begin{equation}
 S_m \equiv
  3H \left( \frac{\delta\rho_\gamma}{\dot\rho_\gamma} - \frac{\delta\rho_m}{\dot\rho_m} \right)
  = \frac{\delta\rho_m}{\rho_m} - \frac34 \frac{\delta\rho_\gamma}{\rho_\gamma}
  \,.
\end{equation}
The initial curvature is unperturbed and hence these are known as
{\it isocurvature} modes \cite{Linde:1984ti,Linde:1985yf,Kofman:1986wm,1986MNRAS.218..103E,1987Natur.327..210P,Kodama:1986fg}.
Isocurvature perturbations may also be
produced in the neutrino density/velocity \cite{Bucher:1999re} and
other matter. These perturbations produce distinctive signatures in
the CMB temperature and polarization anisotropies
\cite{Bucher:2000kb}. 
Although in the most general multi-field scenario four isocurvature modes may arise 
in addition to the adiabatic one, it is hard to conceive of a model in which all of them were 
observable, unless a great degree of fine-tuning is imposed. Therefore, the amplitude of each mode is often constrained individually.
% and often referring to a particular high energy model, well motivated from a model building point of view. 

An almost scale-invariant spectrum of matter
isocurvature perturbations mainly contributes to temperature
anisotropies on large angular scales, as is the case for tensor modes,
but can be distinguished by polarization measurements. Isocurvature
perturbations are scalar modes and so cannot produce $B$-mode
polarization. However, $E$-mode polarization and the cross-correlation
between temperature anisotropies and $E$-mode polarization can
discriminate between isocurvature modes and purely adiabatic spectra
with similar temperature power spectrum.

The existence of more than one light scalar field during inflation
leads to additional non-adiabatic perturbations being frozen-in on
large scales during inflation
\cite{Linde:1984ti,Linde:1985yf,Kofman:1986wm,Polarski:1994rz,Gordon:2000hv,Weinberg04a}. Fluctuations
orthogonal to the background trajectory can affect the total density
after inflation, but they can also affect the relative density between
different matter components even when the total density and therefore
spatial curvature is unperturbed
\cite{GarciaBellido:1995qq}. Actually, the amplitude of primordial
isocurvature perturbations relevant for CMB anisotropies and structure
formation is strongly model-dependent: it does not depend entirely on
the multi-field inflationary dynamics, but also on the
post-inflationary evolution. If all particle species are in thermal
equilibrium after inflation and their local densities are uniquely
given by their temperature (with vanishing chemical potential)
then the primordial perturbations are adiabatic
\cite{Lyth:2002my,Weinberg04b}.  Thus, it is important to note that the existence of primordial
isocurvature modes requires at least one field to decay into
some species whose abundance is not determined by thermal equilibrium
({\it e.g.}~CDM after decoupling) or respects some conserved quantum
numbers, like baryon or lepton numbers. For instance, neutrino density
isocurvature modes could be due to spatial fluctuations in the
chemical potential of neutrinos \cite{Lyth:2002my,Gordon:2003hw}.

The quantum perturbations of each light scalar field are independent
from each other during slow-roll inflation. However, for non-trivial
inflationary trajectories in multi-dimensional field space, the
quantities later identified to observable adiabatic and isocurvature
modes consist in combinations of the large-scale fluctuations of these
fields \cite{Polarski:1994rz,GarciaBellido:1995qq}, and can therefore
be statistically correlated \cite{Langlois:1999dw}. Even if the
inflationary trajectory is a straight line leading to uncorrelated
adiabatic and isocurvature modes, some extra correlation can appear
later. Indeed, whenever the species carrying isocurvature
perturbations contributes significantly to the background expansion
(giving rise to variations in the local equation of state, like a
non-adiabatic pressure perturbation), it provides an additional source
for curvature perturbations outside the horizon
\cite{Polarski:1994rz,Sasaki:1995aw,Sasaki:1998ug,GarciaBellido:1995qq,Gordon:2000hv,Bartolo:2001rt}. If
this happens before the radiation-dominated stage preceding photon
decoupling, the initial conditions relevant for the calculation of CMB
anisotropies and structure formation could consist in a mixture of
completely correlated adiabatic and isocurvature modes, on top of the
arbitrarily correlated adiabatic contribution eventually surviving
from inflation. The mixture of correlated adiabatic and isocurvature
modes of the various types induces some significant extra freedom in
the shape of the CMB anisotropy spectra \cite{Langlois:2000ar}.

We now briefly comment on three scenarios which have been investigated
in some detail.

A minimal extension of chaotic inflation called {\it double inflation}
relies on a second scalar field $\chi$ with $m_{\chi} < H$ during inflation
\begin{equation}
V(\phi, \chi) = \frac{m^2_\phi \phi^2}{2} + \frac{m_\chi^2 \chi^2}{2} \,.
\end{equation}
If $\chi$ is identified with (or decays into) CDM
after inflation and the inflaton $\phi$ decays into radiation,
then isocurvature perturbations persist after inflation
\cite{Polarski:1994rz}. The spectral tilts of adiabatic and
isocurvature power spectra, their correlation, and relative amplitude
of curvature and isocurvature perturbations depend on the parameters
of the model and the classical trajectory during inflation.  Such
models are analyzed in \cite{Beltran} without tensors and in
\cite{Kawasaki} including tensor perturbations. It is interesting that
the amount of allowed isocurvature modes decreases when tensors are included
in the uncorrelated case \cite{Kawasaki}. Non-Gaussianities are
typically small ($f_{\rm NL}\simeq1$) in this model
\cite{Vernizzi:2006ve}.

The {\it curvaton scenario} \cite{Linde:1996gt, Lyth:2001nq} is also based on two
fields which are light during inflation. The energy of the first field
(the inflaton) is assumed to completely dominate the background density during
inflation, while observable cosmological perturbations are entirely
seeded by the perturbations in the other field (the curvaton). In a
typical implementation of this scenario, the curvaton decays some time
after inflation, but before primordial nucleosynthesis, perturbing
the photon density
\begin{equation}
  \frac{\delta\rho_\gamma}{\rho_\gamma}
   \simeq \Omega_\chi \frac{\delta\rho_\chi}{\rho_\chi} \,,
\end{equation}
where $\Omega_\chi$ is the fractional energy density in the curvaton just before it decays. The primordial
baryon asymmetry is known to be due to some out-of-equilibrium process in the very early universe. If the baryon
asymmetry is produced from the decay of the curvaton (or its decay products) then we have
\begin{equation}
 \frac{\delta\rho_b}{\rho_b} \simeq \frac{\delta\rho_\chi}{\rho_\chi} \,,
\end{equation}
and there is a residual baryon isocurvature perturbation after the curvaton decay which is completely correlated
with the total density perturbation \cite{Lyth:2002my}
\begin{equation}\label{sb}
 S_b \simeq ( 1- \Omega_\chi ) \frac{\delta\rho_\chi}{\rho_\chi}
  \simeq 3 \left( \frac{1-\Omega_\chi}{\Omega_\chi} \right) \zeta \,
\end{equation}
(where we have identified $\zeta$ with the primordial density
perturbation on spatially flat hypersurfaces
\cite{Bardeen:1988hy}).
Since the adiabatic and isocurvature modes have a common origin, they share
the same spectral tilt $n_{\rm ad}=n_{\rm iso}$.
The absence to date of observational evidence
for any isocurvature component in the primordial perturbation is an
important constraint on attempts to implement the curvaton scenario in
particle physics models.

The epoch of CDM decoupling also determines the amplitude of an eventual CDM-isocurvature mode. 
Similarly to the case of the baryon asymmetry, a CDM fluid freezing out relative to the rest of the universe 
before the curvaton decay, would give rise to an isocurvature signal that would exceed the current observational bounds \cite{Lyth:2002my, Amendola:2001ni}. 
This fact must also be taken into account when building plausible curvaton models. 

In the case where the curvaton itself decays into the CDM, an isocurvature amplitude arises which has a dependence on 
$\Omega_{\chi}$ equal to (\ref{sb}).  However, the experimental constraints on these two different modes are different due to the different abundances 
of CDM and baryons in the universe \cite{Lyth:2002my}. 

In the {\it axionic dark matter} scenario, the axion is a massless quantum field 
which acquires quantum fluctuations during inflation. These are totally
uncorrelated from the fluctuations seeded by the
inflaton because the two fields are not related. 
Under some circumstances, the axionic perturbations could be
erased by the restoration of the Peccei-Quinn symmetry during
inflation or at the end of reheating. Otherwise, once the axion acquires its mass at the QCD scale, 
an isocurvature mode arises and is preserved, since the axion remains totally decoupled
from other species \cite{Linde:1984ti,Kofman:1986wm}. If one assumes that axions come to play the role
of CDM (or part of it), this scenario predicts an uncorrelated mixture of adiabatic
and CDM isocurvature modes. Furthermore, in this case, there is a simple
relation between the isocurvature amplitude and the scale of
inflation, and the tilt $n_{\rm iso} = 1 - r/8$ is very close to one
\cite{Beltran:2006sq}.

Finally, in the general case of adiabatic perturbations mixed with
$N-1$ arbitrarily correlated isocurvature modes, the initial
conditions for primordial perturbations consist in
$N(N+1)/2$ amplitude parameters (the amplitude of each mode, plus
$N(N-1)/2$ correlation angles) \cite{Bucher:2004an}, and the same number
of tilts characterizing the various scale dependences in first
approximation. In the case $N=2$, one is left with two amplitudes, one
correlation angle and three independent tilts \cite{Valiviita:2003ty,Beltran}.

Current constraints from {\sl WMAP} limit the amplitude of matter
isocurvature perturbations 100\%-correlated with the adiabatic mode to
${P}_{\rm iso}/{P}_s<0.011$ (95\% C.L., assuming no gravitational
waves) \cite{Dunkley:2008ie}, which translates into a bound of
$S_b/\zeta<0.1(\Omega_m+\Omega_b)/\Omega_b$ for the baryon
isocurvature perturbation \cite{GordonLewis}.

The amplitude of isocurvature perturbations which are uncorrelated with the adiabatic mode may be larger with
${P}_{\rm iso}/{P}_s<0.16$ for a scale-invariant spectrum of isocurvature perturbations \cite{Dunkley:2008ie}.
Note that because any contributions from isocurvature modes to the CMB anisotropies are sub-dominant, bounds on
their scale-dependence or non-Gaussianity are correspondingly weaker than for adiabatic density perturbations.

Larger amplitude isocurvature perturbations become allowed when one
considers arbitrary spectral indices
\cite{Beltran:2005gr,Keskitalo:2006qv} or neutrino isocurvature modes,
including neutrino isocurvature velocity perturbations \cite{Bean},
but we are not aware of any inflationary models which motivate such
initial conditions.

There is no clear theoretical target for future observations beyond
the current limits on isocurvature perturbations. However, tightening the bounds in these 
parameters would be of great interest for particle physics and inflationary model building. 
For example, {\sl WMAP} bounds already require $\Omega_\chi\approx1$ in models where
the curvaton decay generates the baryon asymmetry. This would
correspond to a non-Gaussianity parameter $f_{\rm NL}\approx-5/4$. A
detection of large non-Gaussianity ($f_{\rm NL}\gg1$) would be
incompatible with any primordial isocurvature perturbation in the 
curvaton scenario \cite{:2008ei} (unless one considers multiple curvaton fields
\cite{Choi:2007fy,Assadullahi:2007uw} or relaxes some of the curvaton model assumptions \cite{Langlois:2008vk}). 
Also, additional contraints on the tensor modes would allow for a much tighter bound on the 
axionic isocurvature signal. 
Since the contribution of tensor modes and the axionic isocurvature amplitude are degenerate on large scales, 
tightening the constraints on the former would improve constraints on the latter.
%bring important information about the nature of the axion.

\newpage
%%%%%%%%%%%%%%%%%%%%%%%%%%%%
\section{Defects, Curvature and Anisotropy}
\label{sec:DCA}
%%%%%%%%%%%%%%%%%%%%%%%%%%%%

In this section we discuss how
 topological defects (\S\ref{sec:defects}),
spatial curvature (\S\ref{sec:curvature}) and a large-scale anisotropy (\S\ref{sec:isotropy}) leave imprints in the CMB polarization signal.
These probe the physics {\it before} (curvature, anisotropy) and {\it after} (defects) inflation.

\subsection{Topological Defects and Cosmic Strings}
\label{sec:defects}

Even if inflation did
not generate observable gravitational waves, the non-perturbative physics of topological defect formation may generate observable $B$-modes. Such topological defects are generically found in models of grand
unification, particularly those that involve supersymmetry.   In models where GUT defects survive inflation,
there is a danger of reintroducing the monopole problem.  But more generally, physics at the end of inflation
can involve phase transitions at much lower energies that produce topological defects unrelated to GUT scale
physics: this is common in models of hybrid inflation, and includes models from superstring theory.

The best studied of these phenomena are cosmic strings. Cosmic strings are formed at the end of multi-field
inflation whenever a $U(1)$ symmetry is broken during the process of reheating. This is a common feature of
supersymmetric inflationary models \cite{jeannerot}, including D-brane inflation in string theory
\cite{Sarangi, CMP, HenryTye:2006uv}. The tension of the cosmic strings formed in this way is model-dependent: supersymmetric
GUTs typically imply tensions near the observational upper bound of $G\mu \sim \mbox{few}\times10^{-7}$
\cite{Wyman:2005tu,Seljak:2006bg,Bevis:2007gh}, but geometrical warping mechanisms in string theory (which are
introduced for model-building reasons unrelated to defect formation \cite{KKLMMT}) can give effective tensions as
low as $G\mu \sim 10^{-11}$ \cite{Firouzjahi:2005dh}. If they are formed, cosmic strings would generate $B$-mode
polarization in the CMB by directly sourcing vector-type metric perturbations \cite{pen_pol, Polchinski:2004ia, battye_98,
Benabed:2000tr,PTWW03, Bevis:2007qz, Pogosian:2007gi}.
The resultant spectrum has two peaks (see Figure~\ref{fig:BmodesStrings}):

\begin{enumerate}
\item A peak at low $\ell \sim 10$, generated at reionization. The position of this peak is set by the correlation length of the string network at the time of reionization and the rms velocity of the strings, which, in principle, are model-dependent quantities. However, the correlation length is typically expected to be comparable to, but smaller than the horizon size, and the rms velocity is always less than the speed of light. Hence, the peak is at a somewhat smaller scale (higher $\ell$) than the low-$\ell$ peak expected from primordial gravitational waves, where it directly corresponds to the horizon size at reionization. This difference in the low-$\ell$ peak positions may be detectable, depending on the strength of the signal.

\item A peak at high $\ell\sim 600-1000$, generated at last-scattering. The position of this peak is determined by the correlation length and the rms velocity of the strings at the time of last scattering. It would imply power on small scales in excess of what one would expect from lensing alone.
\end{enumerate}

\begin{figure}[h!] %  figure placement: here, top, bottom, or page
   \centering
   \includegraphics[width=0.55\textwidth]{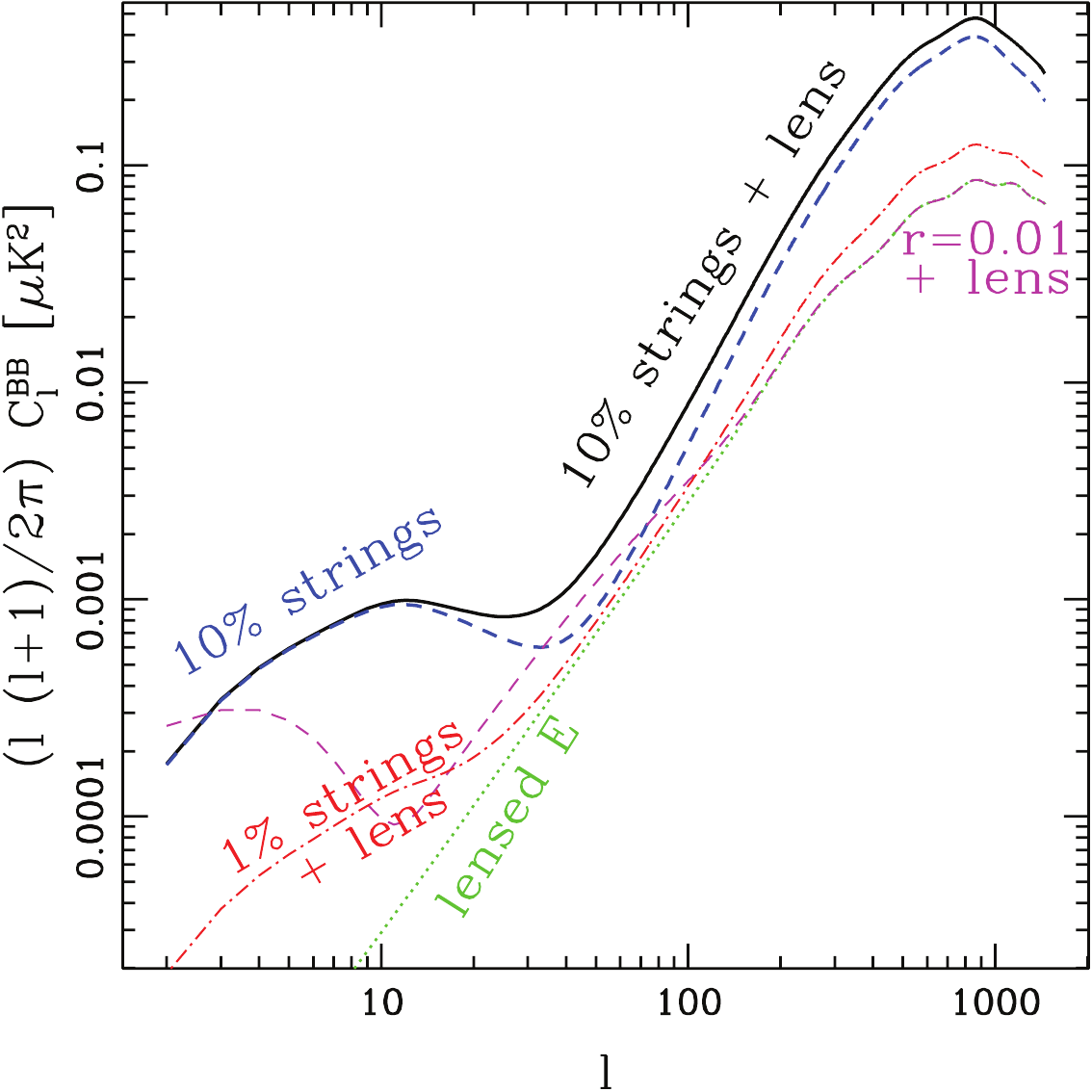}
   \caption{A comparison of the $B$-mode polarization generated by tensor modes during inflation
   and $B$-modes generated by cosmic strings. The blue dashed line is the $B$-mode power expected from a network of cosmic strings that source 10\% of the primordial $TT$-power present in the {\sl WMAP} angular range; the translation of this power to a tension is model dependent, but in all models corresponds to $G\mu \sim \rm{few}\times 10^{-7}$. The green dotted line is the power spectrum expected from the lensing of $E$-mode polarized light into $B$-mode polarized light from large-scale structure. The black solid line is the direct sum of the 10\% string contribution and the lensed $B$-mode signal. The red dash-dotted line is the spectrum generated by a string network that sources only 1\% of the primordial $TT$-power ($G\mu \sim 10^{-7}$) added to the lensed $B$-mode spectrum. The lavender, dashed line is the spectrum generated by a tensor-to-scalar ratio of $r=0.01$ added to the lensed $B$-mode spectrum.}
   \label{fig:BmodesStrings}
\end{figure}

String-mediated $B$-mode production is efficient, so a string network that sources little CMB temperature
anisotropy could be a dominant source of $B$-mode polarization. Current observations imply that strings sourced
$\lesssim 10\%$ of the primordial anisotropy; however, even strings that source $\lesssim 1\%$ of that
anisotropy would be well within the reach of {\sl CMBPol}. In terms of string tensions, this corresponds to $G\mu \gtrsim
10^{-7}$, which corresponds to strings formed very near the GUT scale.  Strings at this tension could also be
seen by other ongoing missions, such as high-$\ell$ CMB experiments like the {\sl Atacama Cosmology Telescope} (ACT) and the
{\sl South Pole Telescope} (SPT) \cite{Fraisse:2007nu, Pogosian:2008am}. Thus, by the time {\sl CMBPol} is ready to be
commissioned, it is possible that we will know whether strings exist with sufficient tension to be observed by
it. However, even lighter strings may be detectable by {\sl CMBPol}: estimates based on a hypothetical {\sl
CMBPol}-like experiment \cite{Seljak:2006hi} found that $G\mu \sim 10^{-9}$ is potentially observable.

{\sl CMBPol} may also be able to probe the \emph{type} of defect formed -- strings are the best-studied case,
but the phase transition that ends inflation could also generate global monopoles, textures, or semilocal
strings. Ref.~\cite{Urrestilla:2007sf} showed that the polarization spectra from different defect types have
different shapes, particularly the $B$-mode spectra, and for high $G\mu$ distinguishing between these should be
within the reach of {\sl CMBPol}. Determining the nature of cosmic defects would provide invaluable information
on high-energy symmetry breaking. Incidentally it has also recently been shown \cite{Urrestilla:2008jv} that
there is no significant degeneracy between primordial $r$ and defects at {\sl Planck} satellite resolution, {\it
i.e.}~one source would not be mistaken for the other. Therefore this is also true at {\sl CMBPol} accuracy.

\subsection{Spatial Curvature}
\label{sec:curvature}

\noindent
{\it Flatness problem}

Homogeneous and isotropic cosmologies are parametrized by the intrinsic curvature of their spatial slices.
Spatial curvature is usually parametrized by a normalized curvature parameter $\Omega_k$ which scales as
$a^{-2}$. One can think of it as an ``energy density'' parameter
\begin{equation}
\rho_k \equiv -\frac{3}{8\pi G}\frac{k}{a^2}
\end{equation}
such that $\Omega_k \equiv \rho_k/\rho_{\rm total}$ with $k=-1,0,1$ specifying a negatively curved, flat and
positively curved universe, respectively.
Crucially, curvature decays less rapidly than matter, $\Omega_m
\propto a^{-3}$, and radiation, $\Omega_r \propto a^{-4}$ (since dark energy is just beginning to dominate we can
ignore it for the present discussion).  It is  then clear that it requires incredible fine-tuning for the
universe to have evolved at least $60$ $e$-folds since the Big Bang and not have the curvature dominate. This is
known as the \emph{flatness problem} (see \S\ref{sec:BBproblems}).

Inflation solves this problem elegantly: the early exponential increase in the scale factor drives the value of
$\Omega_k$ close to zero while the rest of the energy density of the universe is contained in the potential
of the inflaton which is roughly constant. This energy density is then released, mostly into radiation, during the
reheating phase, starting the hot Big Bang. As long as inflation
lasts a little bit longer than $N_e \gtrsim {\cal O}(60)$ $e$-folds,\footnote{The exact number of required $e$-folds depends on the energy scale of inflation and on the mechanism of reheating.}  any relic curvature the universe possesses will be
driven to zero. The current best estimate for $\Omega_k$, using the {\sl WMAP} + BAO + SN combined data set is
\cite{WMAP5}
\begin{equation}
-0.0175 < \Omega_k < 0.0085~ (95\% \ {\rm C.L.}) \, .\label{eq:omegak}
\end{equation}

In many inflationary models, the total number of $e$-folds of inflation is much greater than  ${\cal O}(60)$. Therefore the standard prediction of inflation is
\begin{equation}
|\Omega_k| \lesssim 10^{-4} \, .\label{eq:omegakprediction}
\end{equation}
The main reason why $|\Omega_k|$ is not predicted to be exactly zero is that inflationary perturbations of the metric do not allow one to measure (or even to define) the flatness of the universe with a much better accuracy.

\vskip 6pt
\noindent
{\it Open universes}

This does not mean that $|\Omega_{k}|$ is smaller than $10^{{-4}}$ in {\it all} inflationary models. For example, if the last stage of inflation was relatively short and occurred inside a bubble produced during a false vacuum decay, we may live in an {\it open} universe with $|\Omega_k| \gg 10^{-4}$ \cite{Coleman:1980aw}. This idea attracted a lot of attention in the mid-90s, when many people believed that $\Omega \sim 0.3$ \cite{Bucher:1994gb}. However, most of the models of open inflation proposed at that time failed, which clearly demonstrated that it is very difficult to construct inflationary models with $\Omega$ significantly different from $1$.

Recently there has been a revival of interest in models of open inflation with $|\Omega_{k}| \sim 10^{-2}$. Such models may appear relatively naturally  in the context of string cosmology under certain assumptions about the probability measure for eternal inflation \cite{Freivogel:2005vv}. Therefore it is quite interesting that the measurement of $\Omega_{k}$ with an accuracy better than $10^{{-2}}$ may help us to test
%string theory and
some of the recent ideas about the probability measure for eternal inflation.

One of the features of the models of open inflation is a very specific modification of the spectrum of scalar and tensor modes for small $\ell$ \cite{Linde:1999wv}. The contribution of these modifications to $C^{TT}_{\ell}$ may partially cancel each other, but one can separate these effects by measuring the amplitude of $B$-modes.
%, which is the main goal of the {\sl CMBPol}.

\vskip 6pt
\noindent
{\it Closed universes}

The situation with inflationary models of a {\it closed}  universe with $|\Omega_{k}| \gg 10^{{-4}}$ is more complicated. A closed inflationary universe may emerge due to quantum creation of the universe  ``from nothing,''  but the probability of such a process is exponentially small \cite{Linde:1983mx,Vilenkin:1984wp}, and it is very difficult to combine this scenario with the requirement that inflation must be short, which is necessary to get $|\Omega_{k}| \gg 10^{{-4}}$ \cite{Linde:2003hc}. One may argue that a more natural scenario to consider is quantum creation of a compact open or flat inflationary universe with a nontrivial topology, which is not exponentially suppressed  \cite{Zeldovich:1984vk, Cornish:1996st, Coule:1999wg, Linde:2004nz}. However,  neither of these models can be naturally incorporated in the context of the theory of eternal inflation, which is much better suited for a description of a multiverse consisting of many bubbles containing open but nearly flat inflationary universes. This provides an intriguing possibility to falsify some
%of the most interesting and advanced
very interesting cosmological theories by observing a positive spatial curvature or a nontrivial topology of our universe.

\vskip 6pt
\noindent
{\it What are the observational prospects for measuring spatial curvature?}

CMB anisotropies measured by the {\sl WMAP} satellite have determined the angular diameter distance to the epoch of
photon decoupling, $z_{\rm dec} \simeq 1090$, which is sensitive to the spatial curvature. However, as the angular
diameter distance depends not only on curvature but also on the energy components in the universe, {\it
i.e.}~matter density and dark energy density, the angular diameter distance out to $z_{\rm dec}$ alone could not
determine the spatial curvature unambiguously. Therefore, a combination of angular diameter distances measured
out to {\it multiple} redshifts is a powerful way of measuring the spatial curvature. For example, the angular
diameter distances out to $z=0.2$ and $0.35$ measured by the {\sl Sloan Digital Sky Survey} (SDSS) and the {\sl Two Degree
Field Galaxy Redshift Survey} (2dFGRS), when combined with the angular diameter distance to the CMB, have yielded a stringent limit on the
spatial curvature (\ref{eq:omegak}). With the future galaxy surveys at higher redshifts, $z\sim
3$, {\it e.g.}~the {\sl Hobby-Eberly Dark Energy Experiment} \cite{Hill:2008mv}, combined with the improved
determination of the angular diameter distance out to $z_{\rm dec}$ from {\sl Planck}, the spatial curvature would be
determined to the accuracy of $|\Omega_k|\sim 10^{-3}$, {\it i.e.}~an order of magnitude better than the current
limit.

Can the CMB alone determine the spatial curvature? Yes, if CMB data alone can constrain $\Omega_m$ and/or the
angular diameter distances out to $z\sim 3$. The weak gravitational lensing of the CMB offers such measurements. The
weak lensing effect smoothes the acoustic oscillations of the power spectra of temperature and $E$-mode
polarization anisotropies, and also adds power at $\ell \gtrsim 3000$. These effects can be measured by {\sl Planck},
and would be measured better by {\sl CMBPol} with the high-angular resolution option (EPIC-2m) \cite{SmithLensing}. Moreover, the weak lensing
converts the $E$-modes to the $B$-modes, which would not be accessible to {\sl Planck}, but would be measured by {\sl
CMBPol} with the high-angular resolution option.
Projections for future constraints on $\Omega_k$ are discussed in \S\ref{sec:CMBPol}.

\subsection{Large-Scale Anisotropy}
\label{sec:isotropy}

Anomalies in the large-scale CMB temperature sky measured by {\sl WMAP} have been suggested as possible evidence for a
violation of statistical isotropy on large scales
\cite{Eriksen:2003db,deOliveiraCosta:2003pu,Land:2005dq,Tegmark:2003ve,Copi:2003kt,Schwarz:2004gk,Land:2005ad,Copi:2005ff,Copi:2006tu,Wiaux:2006zh,Land:2004bs},
and a confirmation of such evidence would represent  a radical departure from the standard cosmological model.
The evidence for the breaking of statistical isotropy in the form of temperature anomalies is usually inferred in an
{\it a posteriori}  manner, and therefore it is difficult to apply formulations of Occam's razor to compare
isotropy-violating models with the isotropic concordance cosmology. Thus, it is very important to test the
predictions of such models for other observable signatures. In any physical model for broken isotropy, there are
testable consequences for the CMB polarization field ({\it e.g.} \cite{Basak:2006ew, Pullen:2007tu}). In Ref.~\cite{Dvorkin:2007jp}, the authors make predictions for  the polarization field in models that break statistical
isotropy locally through a modulation field. In particular, they study two different models: a dipolar
modulation, proposed to explain the asymmetry in power between northern and southern ecliptic hemispheres
\cite{Spergel:2003cb,Gordon:2006ag,Eriksen:2007pc}, and a quadrupolar modulation, invoked to explain the
alignments between the quadrupole and the octopole of the temperature field \cite{Gordon:2005ai}. For the
dipolar case, predictions for the correlation between the first ten multipoles of the temperature and
polarization fields are fairly robust to model assumptions, and can typically be tested at the $98\%$ C.L. or
greater. For example, in the absence of foreground considerations, a space-based experiment with $5$ frequency
channels and a noise level of $18$ $\mu$K-arcmin per frequency channel will saturate the cosmic variance bound
in each channel. For the quadrupolar case, the quadrupole and octopole of the $E$-polarization field will tend
to align as well. Such an alignment is a generic prediction of explanations which involve the temperature field
at recombination. Thus, its main use will be to discriminate against explanations involving foregrounds or local
secondary anisotropies. The predictions for polarization statistics made by anomaly models is a vital probe of a
fundamental assumption underlying all cosmological inferences.

It is challenging to provide cosmological models that explicitly realize these modulations, in a way that can be reconciled with the inflationary picture. In most of these models, the breaking of statistical isotropy is a remnant of a pre-inflationary stage. Therefore, the duration of inflation needs to be tuned so that the signature will be present at the largest observed scales. One requires that inflation only lasted just the minimum amount of $e$-folds necessary to solve the standard cosmological problems. In such models, statistical isotropy is recovered at small scales, since the modes responsible for the CMB anisotropies at those scales exited the horizon during the standard inflationary expansion.

For instance, Ref.~\cite{Donoghue:2007ze} suggested that the difference in power between the two ecliptic hemispheres could be due to a spatial gradient in the inflaton field at the onset of inflation.
A power asymmetry across the observable universe could also be generated by large super-horizon fluctuations. Refs.~\cite{Erickcek:2008sm,Erickcek:2008jp} studied the impact on the CMB of a single super-horizon mode. It was shown that, in this context, the observed power asymmetry cannot be realized within a single-field slow-roll inflation. However, it can be realized if the fluctuation is generated by a curvaton field \cite{Gordon:2006ag} (the mode may arise due to domain structure in the curvaton-web \cite{Linde:2005yw}). Interestingly, this scenario predicts a level of non-Gaussianity that can be detected by the {\sl Planck} satellite \cite{Erickcek:2008sm}. Breaking of statistical isotropy, with a possible alignment of different CMB multiples, can also result from an anisotropic expansion at the onset of inflation. The simplest possibility is to assume different initial expansion rates for the different spatial directions (Bianchi I geometry), and the subsequent isotropization due to slow-roll inflation. The system of perturbations for such a model is characterized by three physical modes, which, after the background isotropizes, can be identified with the scalar density contrast and the two gravitational wave polarizations \cite{ Gumrukcuoglu:2007bx}. During the anisotropic stage, these three modes are coupled to each other already at the linearized level, and have a nonstandard evolution. In particular, one of the gravitational wave polarizations exhibits a large growth during the anisotropic stage, which can result in a potentially observable $B$-mode signal in the CMB
\cite{Gumrukcuoglu:2008gi}. This growth is a purely classical effect, and the resulting signal is superimposed on the gravitational waves of quantum origin generated during inflation. In particular, it can result in an observable $B$-mode in the CMB even if inflation occurred at a low energy scale.
 Therefore, the results of a $B$-mode experiment can provide information not only on the energy scale of inflation, but also on its duration, and on the pre-existing conditions.

In general, all the above proposals rely on specific initial conditions that cannot be predicted from the model. One may hope to improve in this respect by arranging for a background with a controllable (and arbitrarily small) departure from a FRW inflationary geometry. In this way, the primordial perturbations can be quantized as in the standard case \cite{Mukhanov:1990me}, resulting in predictive initial conditions. This can be realized by adding suitable sources that contrast the rapid homogenization and isotropization caused by the inflaton. For instance, in \cite{Ford:1989me,Ackerman:2007nb,Golovnev:2008cf,Kanno:2008gn} a prolonged inflationary anisotropic expansion is obtained through a vector field with nonvanishing spatial vacuum expectation value. Ref.~\cite{Groeneboom:2008fz} showed that the {\sl WMAP} data provide a $3.8 \sigma$ evidence for an anisotropic  covariance matrix which is motivated by one of these models \cite{Ackerman:2007nb}. It was shown in  \cite{Himmetoglu:2008zp} that these proposals suffer from instabilities at horizon crossing. It may, however, be possible that suitable modifications of these models could avoid such problems.

\newpage
\section{Testing Inflation with {\sl CMBPol}}
\label{sec:CMBPol}

In this section we present forecasts of realistic errors on inflationary parameters for a future satellite experiment.
Special attention is paid to uncertainties in the foreground removal and their effects on the theoretical forecasts.
Details of our computations are presented in Appendix \ref{sec:Fisher}.

\subsection{Fisher Forecasts}

For purposes of illustration we define two versions of a future satellite experiment to measure CMB polarization \cite{EPIC}:
\begin{itemize}
\item {\sl EPIC-LC}: a low-cost mission targeting $B$-modes only on scales larger than $\sim$2 degrees.
\item {\sl EPIC-2m}: a mid-cost mission measuring $B$-modes on both large and small scales.
\end{itemize}
The precise experimental specifications for both of these options are given in Appendix \ref{sec:Fisher}.
We present results for two types of foreground treatments:
\begin{itemize}
\item {\it no foregrounds}

In this case, we ignore foregrounds completely and present results simply as a function of instrumental sensitivities.  The associated results should of course be viewed as over-optimistic.

\item {\it with foregrounds}

In this case, we include assumptions about foreground removal in the Fisher analysis. Our treatment closely follows Ref.~\cite{LiciaFisher} and is defined in more detail in Appendix \ref{sec:Fisher}.
%\item delensing
\end{itemize}
Residual foregrounds introduce a bias ({\it i.e.}~a systematic error) to constraints on $r$ while noise just introduces a statistical error. We attempt to include both these effects in the reported confidence regions, despite the very different natures of these two terms. To estimate their effects on the final constraints on cosmological parameters, we have adopted the ansatz of \cite{LiciaFisher} (see Appendix \ref{sec:Fisher}). The {\it systematic} uncertainty on the constraints on $r$ introduced by residual foregrounds can be appreciated by comparing forecasts for the case with no foregrounds (only statistical errors) and the case with foregrounds (with statistical and systematic errors). We treat the weak lensing $B$-mode signal as a Gaussian noise, and do not assume that it can be removed.

For the fiducial set of parameters we use 
\bea
\label{equ:fi}
\bar \alpha &\equiv& \{ r = 0.01\, (0.001),\ n_s=0.963,\ n_t = - r/8,\ \alpha_s =0,\ A_s=2.41 \times 10^{-9},\ \nonumber \\
&&\ \ \tau = 0.087,\ \omega_b = 0.02273,\ \omega_c =  0.1099 ,\ h=0.72,\ \Omega_k =0 \}\, .
\eea
The pivot scale for $r$, $n_t$, $A_s$, $n_s$ and $\alpha_s$ is $k_\star = 0.05$ Mpc$^{-1}$. 
%LV 
The forecasted errors do not depend 
%HVP 
significantly on the actual choice of fiducial model parameters, except for the value chosen for $r$ (due to cosmic variance). Since $r$ is of primary interest, we will report results assigning it different fiducial values. The errors on all the parameters depend either weakly or not at all on the choice of the pivot\footnote{$r=0.01$ at $k_\star = 0.05$ Mpc$^{-1}$ corresponds to $r_{0.002}=0.009$ at $k_\star = 0.002$ Mpc$^{-1}$ and $r=0.001$ corresponds to $r_{0.002}= 0.0009$. Thus the choice of pivot does not significantly affect our conclusions on the forecasted errors.}, and this dependence for constraints on $r$ should be subdominant to other real world effects that we do not consider here. 
%\newpage

%\addtocontents{toc}{\SkipTocEntry}
\subsubsection{Summary of Results}

Tables \ref{tab:error0}  and \ref{Tab:forecasts_r0} show a subset of the results of Appendix \ref{sec:Fisher}.
Figures~\ref{fig:nsrX} and \ref{fig:nsr2} compare {\sl WMAP}, {\sl Planck} and {\sl CMBPol} constraints in the $n_s$-$r$ plane.
For the foregrounds we assume what we term ``pessimistic" and  ``optimistic" options for EPIC-LC and EPIC-2m respectively (see Appendix \ref{sec:Fisher}), to span a range of experimental possibilities.  ``Pessimistic" assumes that the residual foreground amplitude is 30\% (10\% in $C_\ell$) ; ``optimistic" assumes a 10\% residual (1\% in $C_\ell$). Both options assume realistic levels of polarized dust, although this is currently uncertain at the order of magnitude level (see \cite{DunkleyFGs}). 
The  errors due to foreground contamination adopted here are valid only if $\sim 70\%$ or more of the sky 
can be used for cosmological analysis. Should the foreground contamination impose more drastic 
sky-cuts, there will be a significant error degradation ({\it e.g.}~\cite{Amarie:2005in}). The estimated errors also assume that 
there is no effect of leakage of power from $E$ to $B$-modes. By using a large fraction of the sky, the errors on the measured polarization will vary spatially when foreground uncertainty is included, resulting in additional contamination 
of the $B$-mode signal. The analysis of \cite{Amarie:2005in} suggests that this would inflate error bars over those presented here, although initial studies in \cite{DunkleyFGs} indicate that the effect should be small for models with $r=0.01$. For further discussion see  Appendix \ref{sec:Fisher} and Ref.~\cite{DunkleyFGs}.

\vskip 15pt
\noindent
Below (\S\ref{sec:FisherTensors}--\ref{sec:FisherCurv}) we comment on 
the implications 
of these results.

\vspace{0.5cm}
\begin{table}[h!]
\begin{center}
\begin{tabular}{||c|c|c|c|c|c|c||}
\hline \hline
\bf Errors & {\sl WMAP}  & {\sl Planck} &  \multicolumn{2}{|c|}{EPIC-LC }& \multicolumn{2}{|c||}{EPIC-2m} \\

            & no FGs & no FGs & no FGs & with Pess FGs & no FGs & with Opt FGs \\
\hline
\hline
$\Delta n_s$ & 0.031  & 0.0036 & -- & --& 0.0016 & 0.0016 \\
$\Delta \alpha_s$ & 0.023 & 0.0052 & -- & -- & 0.0036 & 0.0036 \\
 $\Delta r$ & 0.31 & 0.011 & $5.4\times 10^{-4}$ & $9.2 \times 10^{-4}$& $4.8\times 10^{-4}$ & $5.4 \times 10^{-4}$ \\
\hline
 $\Delta r$ & -- & 0.10 & $0.0017$& --& $0.0015$ & $0.0025$\\
$\Delta n_t$ & -- & 0.20 & $0.076$ & -- & $0.072$& $0.13$\\
\hline
$\Delta f_{\rm NL}^{\rm local}$ & -- & 4 &-- & -- & 2 & --\\
$\Delta f_{\rm NL}^{\rm equil.}$ & -- & 26 & -- & -- & 13 &-- \\
\hline
$\Delta \alpha_{\rm (c)}$ & -- & $1.2\times 10^{-4}$ & $3.5 \times 10^{-5}$ & $4 \times 10^{-5}$ & $3.5 \times 10^{-5}$ & $3.5 \times 10^{-5}$\\
$\Delta \alpha_{\rm (a)}$ & -- & 0.025 & 0.0065 & 0.0068 & 0.0065 & 0.0066\\
\hline
$\Delta \Omega_k$ & -- & -- & -- & -- & $6\times 10^{-4}$ & $6 \times 10^{-4}$ \\
\hline \hline
\end{tabular}
\caption{ \label{tab:error0} Forecasts of (1$\sigma$) errors on key inflationary parameters for {\sl WMAP} (8 years), {\sl Planck}~\cite{Colombo:2008ta} and {\sl CMBPol} (EPIC-LC and EPIC-2m). 
We present results for the unrealistic assumption of `no foregrounds' (no FGs) and `with foreground removal' (with FGs). For the foregrounds we assume the pessimistic and the optimistic options for EPIC-LC and EPIC-2m, respectively (see Appendix \ref{sec:Fisher}). The fiducial model has $r=0.01$. The single-field consistency relation has been applied in the top block of forecasts.}
\end{center}
\end{table}

\newpage

\begin{figure}[h!]
    \centering
        \includegraphics[width=.48\textwidth]{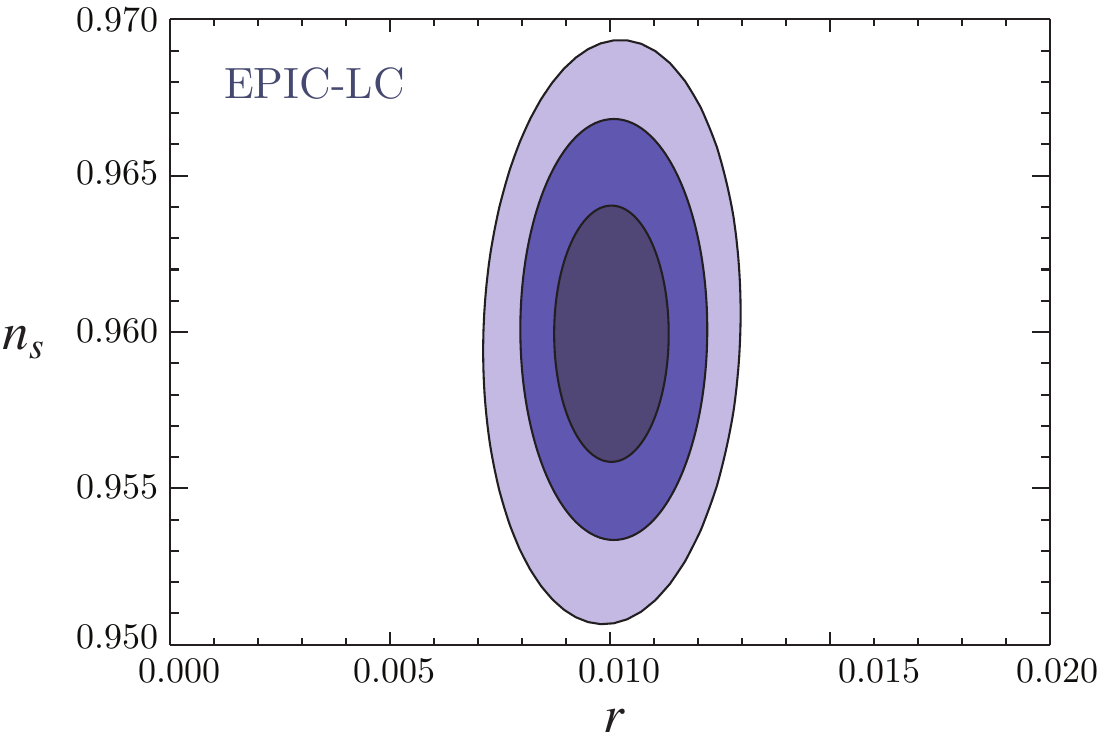}
        \hskip 10pt \includegraphics[width=.48\textwidth]{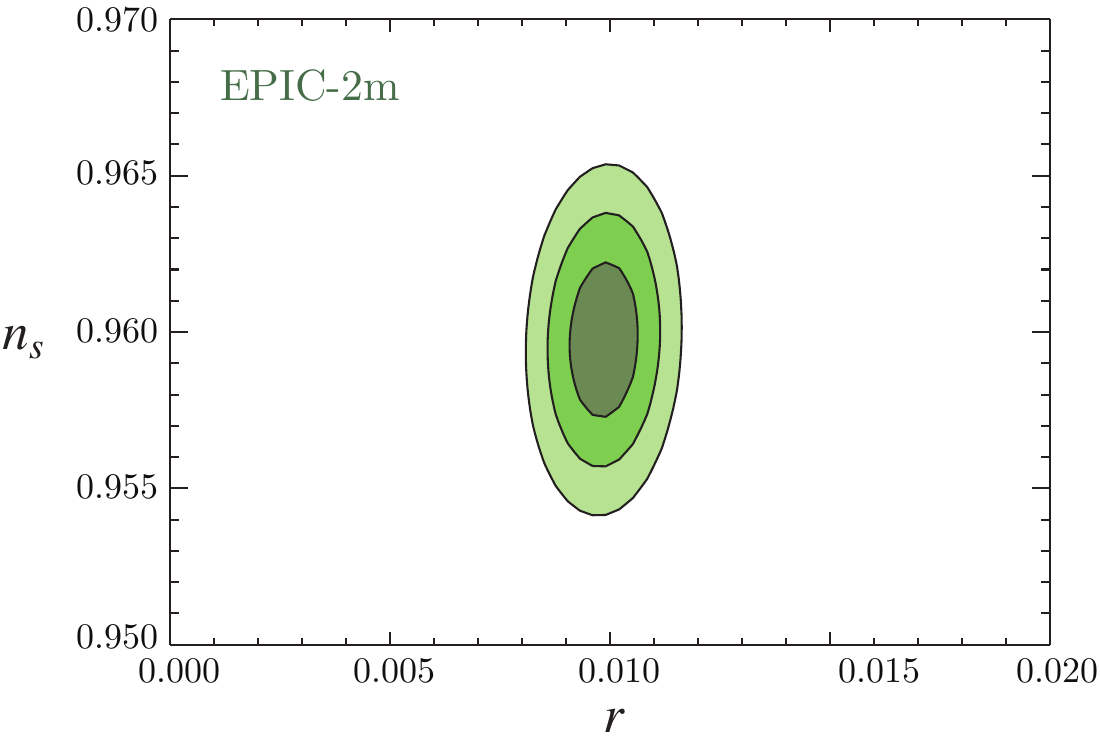}
   \caption{Forecasts of {\sl CMBPol} constraints in the $r$-$n_s$ plane assuming the consistency relation. {\it Left}: EPIC-LC with pessimistic foreground option. {\it Right}: EPIC-2m with optimistic foreground option. The contours shown are for 68.3\% (1$\sigma$), 95.4\% (2$\sigma$) and 99.7\% (3$\sigma$) confidence limits.}
    \label{fig:nsrX}
\end{figure}

\vskip 20pt
\begin{figure}[h!]
    \centering
     \includegraphics[width=.6\textwidth]{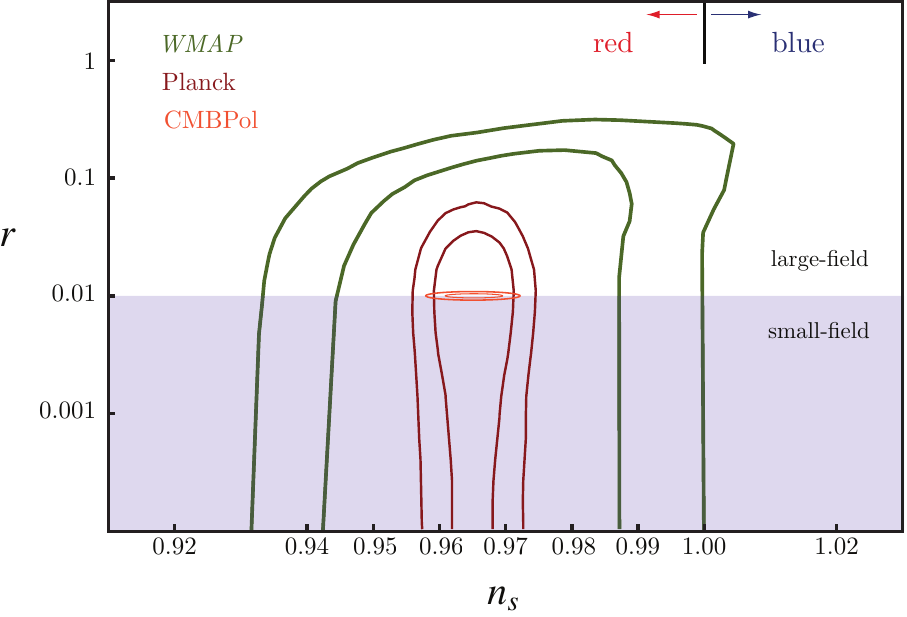}
   \caption{Forecasts of future constraints in the $n_s$-$r$ plane. Comparison of {\sl WMAP}, {\sl Planck} and {\sl CMBPol} (EPIC-LC+pessimistic FGs). The contours shown are for 68.3\% (1$\sigma$) and 95.4\% (2$\sigma$) confidence limits. The {\sl WMAP} contours are from the 5 year analysis \cite{WMAP5}.}
    \label{fig:nsr2}
\end{figure}

\newpage
\begin{table}[h!]
\begin{center}
\begin{tabular}{|| c | l | c  c | c || c  c | c ||}
\cline{3-8}
\cline{3-8}
\cline{3-8}
          \multicolumn{2}{c}{} &  \multicolumn{3}{|c||}        { EPIC-LC } &  \multicolumn{3}{c||}  {EPIC-2m} \\
%            &                     &            LC                       &     &                            &2m                               &                   &                          \\ 
            \hline
            \hline
            &                     & $ \Delta r$              & $\Delta n_t $& $\Delta r$         & $\Delta r $                 &$\Delta n_t$    &$\Delta r$    \\
\hline
  no FGs          & $r=0.001$ &$6.9\times 10^{-4} $& $0.18$&$2.3\times 10^{-4} $ & $5.7\times 10^{-4} $ &$0.17$ &$2.1\times 10^{-4} $   \\
            & $r=0.01$    &$0.0017$                 &$0.076$ &$5.4\times 10^{-4} $ & $0.0015$                  &$0.072$    & $4.8\times 10^{-4} $ \\
\hline            
%Opt fg   & $r=0$         &                                    &              &                                     & $$                                 &$$          & $ $ \\
with FGs  & $r=0.001$ & --                      & --  &$8.0\times 10^{-4}$       & $ 0.0018$                    &$0.93$  &$4.1\times 10^{-4}  $   \\
             & $r=0.01$    & --  & --   &$9.2\times 10^{-4}$  & $0.0025$                     &$0.13$   & $ 5.4\times 10^{-4}  $ \\
\hline
\hline
\end{tabular}
\caption{Forecasted constraints on tensor modes. We present results for the unrealistic assumption of `no foregrounds' (no FGs) and `with foreground removal' (with FGs). For the foregrounds we assume the pessimistic and the optimistic options for EPIC-LC and EPIC-2m, respectively (see Appendix \ref{sec:Fisher}). Cases where there is no detection are indicated with dashes.}
\label{Tab:forecasts_r0}
\end{center}
\end{table}

%\addtocontents{toc}{\SkipTocEntry}
\subsubsection{Tensors}
\label{sec:FisherTensors}

Bearing in mind the caveats specified above, the following conclusions about tensor modes can be drawn from this analysis:\\

{\bf Detection}
\begin{itemize}
\item Gravitational waves can be detected at $\sim 3 \sigma$ for $r \gtrsim 0.01$ for the low-cost mission assuming the foreground levels are as currently predicted, and that they can be cleaned to the 10\% level in amplitude (1\% in power).
\item In the optimistic foreground scenario, an $r=0.01$ signal could be measured by {\sl CMBPol} for the low-cost mission at about $15 \sigma$ if the consistency relation is imposed,  $n_t = - r/8$.
\end{itemize}
\vskip 6pt 

{\bf Upper limit}
\begin{itemize}
\item {\sl CMBPol} would provide a $3 \sigma$ upper limit on tensors of $r \lesssim 0.002$  for the low-cost mission and optimistic foregrounds if the consistency relation is imposed.
\end{itemize}
\vskip 10pt

\noindent
These limits should be compared to the theoretically interesting regime of large-field inflation ($r>0.01$); {\it cf.}~\S\ref{sec:Bmodes}.  This shows that {\sl CMBPol} is a powerful instrument to test this crucial regime of the inflationary parameter space.

%\addtocontents{toc}{\SkipTocEntry}
\subsubsection{Non-Gaussianity}

Our Fisher results suggest that {\sl CMBPol} will be able to achieve the sensitivity of $\Delta f_{\rm NL}^{\rm local}\simeq 2$ (1$\sigma$) for non-Gaussianity of local type and $\Delta  f_{\rm NL}^{\rm equil.} \simeq 13$ (1$\sigma$) for non-Gaussianity of equilateral type. For the local type of non-Gaussianity this amounts to an improvement of about a factor of 2 over the {\sl Planck} satellite and about a factor of 12 over current best constraints. These estimates assume that foreground cleaning can be done perfectly, {\it i.e.}~the effect of residual foregrounds has been neglected. Also the contribution from unresolved point sources and secondary anisotropies such as ISW-lensing and SZ-lensing has been ignored.

In the event that {\sl Planck} saw a hint for a non-zero $f_{\rm NL}$-signal, {\sl CMBPol} would offer the great opportunity to scrutinize it with enhanced sensitivity. A convincing detection of any form of non-Gaussianity would be a major breakthrough in cosmology.

%\addtocontents{toc}{\SkipTocEntry}
\subsubsection{Isocurvature}

Precise measurements of $E$- and $B$-mode polarization will significantly improve existing constraints on isocurvature fluctuations.
We define the following measure of the isocurvature amplitude
\beq
\frac{\alpha_{\rm iso}(k_\star)}{1-\alpha_{\rm iso}(k_\star)} \equiv \frac{P_{\rm iso}}{P_{\rm ad}}\, ,
\eeq
where $P_{\rm ad} \approx P_{s}$.
The forecasts for the error of the isocurvature fraction in the primordial perturbations have been calculated for the curvaton model ($\alpha_{\rm (c)}$) and the axion model ($\alpha_{\rm (a)}$),
assuming the fiducial set of parameters (\ref{equ:fi})
with $r=0.01$.

The results in Table \ref{tab:error0} verify that {\sl CMBPol} will be a powerful instrument to constrain or measure the primordial isocurvature fraction. Any detection of isocurvature fluctuations would inform us about the nature of dark matter (for the case of dark matter isocurvature), or baryogenesis (for the case of baryon isocurvature); at the very least, the detection would rule out single-field inflationary models, {\it and} any scenarios in which matter was in thermal equilibrium with photons with no conserved quantum numbers \cite{Weinberg04b}.

%\addtocontents{toc}{\SkipTocEntry}
\subsubsection{Curvature}
\label{sec:FisherCurv}

Due to a {\it geometric degeneracy} \cite{Efstathiou:1998xx},
the primary CMB alone is not able to measure the spatial curvature parameter, $\Omega_k$,
as it is determined from the angular diameter distance out to $z\simeq 1090$, which also depends on
the matter density, $\Omega_m$. However, the weak gravitational lensing of the CMB due to the
intervening matter distribution, a secondary effect, helps to break this degeneracy, as
the lensing depends on a combination of $\Omega_m$  and the amplitude of fluctuations, $\sigma_8$.
The lensing effect smoothes the acoustic oscillations in the temperature and $E$-mode power spectra,
and creates additional power at $\ell \gtrsim 3000$. In addition, the lensing converts $E$-modes to $B$-modes,
creating the $B$-mode power spectrum that peaks at $\ell \sim 1000$. This information can be used to determine
$\Omega_m$, thereby allowing the CMB data {\it alone} to break the geometric degeneracy and determine
the curvature parameter accurately.

The high-resolution version of EPIC is capable of determining $\Omega_k$ to $6\times 10^{-4}$
\cite{LiciaFisher}, which is not very far from the expected
non-zero value from inflation, $10^{-4}$ (see \S\ref{sec:curvature}).
Moreover, since the gravitational lensing creates non-Gaussianity in the CMB, there is more information
in the higher-order statistics. In particular, the 4-point function is known to contain a lot of information
of the CMB lensing \cite{Hu:2001kj, Hirata:2003ka}.
It is therefore plausible that adding the 4-point information will get us even closer to $10^{-4}$.
To exploit the full potential of the weak lensing of the CMB, the high-resolution version of EPIC is required \cite{SmithLensing}.

%\newpage
\subsection{Model Selection}

The Fisher information matrix analysis of the previous section addresses the question of how accurately
parameters can be determined in a given cosmological model. The extension of this framework to consider
different cosmological models ({\it i.e.}~different choices of parameters to be varied) is known as model selection,
and Bayesian implementations of model selection, centered around a quantity known as the \textit{Bayesian
evidence}, have been developed (see \cite{Trotta:2008qt} for an overview). Many of the science goals of {\sl
CMBPol} are model selection goals:
\begin{itemize}
\item Comparison of models with and without primordial gravitational
waves.
\item Comparison of models with and without cosmic defects.
\item Comparison of models with different types of cosmic defects.
\end{itemize}
The data analysis strategy for {\sl CMBPol} should feature a combination of parameter estimation and model
selection methods, in order to clearly identify the robustness of results.

Model selection forecasting, as described in \cite{Mukherjee:2005tr,Trotta:2007hy}, is an alternative to the
Fisher matrix in quantifying experimental capability. Work is underway to carry out model selection forecasts
for the proposed {\sl CMBPol} survey parameters \cite{LiddleForecast}.

\newpage
\section{Summary and Conclusions}
\label{sec:summary}

%\begin{itemize}
%\item We have briefly sketched how important a space-based precision experiment for CMB Polarization is to probing fundamental physics,
%\item Even null measurements of polarization would be extremely important,
%\item Would complement upcoming experiments such as {\sl LHC}, dark matter detection, etc.,
%\item Much higher energy scales accessible than collider experiments.
%\end{itemize}

In this White Paper we have described the excitement felt by the community of cosmologists and particle physicists in using observations of the cosmic microwave background to learn about the universe at the highest energies and the smallest distance scales.
In this final section we summarize our conclusions.

\vskip 6pt
\noindent
{\it The Golden Age of Cosmology}

Observations of cosmic microwave background (CMB) anisotropies and large-scale structure surveys have led to
the emergence of a {\it concordance cosmology}.  This $\Lambda$CDM cosmology composed of a homogeneous background of atoms (4.4\%), dark matter (21.4\%) and dark energy (74.2\%)
and containing a small amplitude of nearly scale-invariant adiabatic Gaussian density fluctuations fits all cosmological data.
The success of cosmological observations in revealing the composition of the homogeneous universe provides significant motivation to now probe its fluctuations.
Through inflation these observations can be directly related to the high energy physics at $10^{-30}$ seconds
after the `beginning of time'.

\vskip 6pt
\noindent
{\it Inflation}

Inflation
allows regions of space which should be uncorrelated at CMB decoupling to be observed at almost identical temperatures.
In the inflationary paradigm, quantum fluctuations in the very early universe were in fact produced when the relevant scales were causally connected.
Subsequently, however, the superluminal expansion of space during inflation stretched these scales outside of the horizon.
When the perturbations re-entered the horizon at later times, they served as the initial conditions for the growth of large-scale structure and the anisotropies in the CMB.
Inflation makes detailed predictions about key statistical features of the primordial perturbations such as their scale-dependence and (non-)Gaussianity.  In addition, inflation predicts a stochastic background of gravitational waves which leaves a characteristic ($B$-mode) signature in the polarization of the CMB.  If observed, $B$-modes will reveal the energy scale at which inflation occurred.

\vskip 6pt
\noindent
{\it The Next Frontier: Probing the Primordial Universe}

Cosmological observations have only begun to study details of the primordial fluctuation spectra created by inflation.
The present data determines the initial amplitude of the primordial density fluctuations ($A_s$) and shows the first hints for its variation with scale ($n_s$).
As explained in \S\ref{sec:observable}, future observations have great potential to enlarge the inflationary parameter space via accurate measurements of the primordial perturbation spectra.
Besides confirming the deviation from scale invariance of the scalar spectrum, the data may show signs of {\it tensor} perturbations ($r$, $n_t$), primordial {\it non-Gaussianity} ($f_{\rm NL}$), and {\it multi-field effects} ($S$).
We consider CMB polarization to be a fantastic tool to study these basic questions in early universe physics.

\begin{table}[btp]
\begin{center}
\begin{tabular}{|| c |l| p{5.cm} |c||}
\hline \hline
 {\small \bf Label} & {\small \bf Definition} & {\small \bf Physical Origin} & {\small \bf Current Status}  \\
\hline
{\small $A_s$} & {\small Scalar Amplitude} & {\small $V, V'$} & {\small $(2.445 \pm 0.096) \times 10^{-9}$} \\
\hline
{\small $n_s$} & {\small Scalar Index} & {\small $V',V''$} & {\small $0.960 \pm 0.013$} \\
\hline
{\small $\alpha_s$} & {\small Scalar Running} & {\small $V',V'',V'''$} & {\small only upper limits}\\
\hline
 {\small $A_t$} & {\small Tensor Amplitude} & {\small $V$ (Energy Scale)} & {\small only upper limits}   \\
\hline
{\small $n_t$} & {\small Tensor Index} & {\small $V' $} & {\small only upper limits}\\
\hline
{\small $r$} & {\small Tensor-to-Scalar Ratio } & {\small $V'$} & {\small only upper limits}\\
\hline
{\small $f_{\rm NL}$} & {\small Non-Gaussianity}  & {\small Non-Slow-Roll, Multi-field} & {\small only upper limits}\\
\hline
{\small $S$} & {\small Isocurvature} & {\small Multi-field} & {\small only upper limits} \\
\hline
 {\small $\Omega_k$} & {\small Curvature} & {\small Initial Conditions} & {\small only upper limits}\\
\hline
{\small $G\mu$} & {\small Topological Defects} & {\small End of Inflation} & {\small only upper limits}  \\
 \hline
\hline
\end{tabular}
\caption{ \label{table:concinflation2} From $\{ A_s, n_s \}$ to $\{A_s, n_s, \alpha_s\}, \{A_t, n_t, r\}, \{f_{\rm NL}, S\}$: Copy of Table \ref{table:concinflation} illustrating the potential of future measurements of primordial scalar and tensor fluctuations as a probe of inflation.}
\end{center}
\end{table}

\vskip 6pt
\noindent
{\it $B$-modes and the UV Sensitivity of Inflation}

We argued in \S\ref{sec:Bmodes} that inflation is sensitive to certain properties of the ultraviolet completion of gravity, and that a detection of primordial gravitational waves would provide striking, almost model-independent information about the high-energy physics driving inflation. Such a detection would demonstrate that inflation occurred at a very high energy scale, and that the inflaton traversed a super-Planckian distance in field space.  In turn, these facts would strongly suggest the presence of an approximate {\it shift symmetry} in the ultraviolet theory: in the absence of such a symmetry, it is highly implausible that inflation could occur over such a large field range. We noted that symmetries of this sort can arise in certain limits of string theory.  Observational constraints on primordial tensors can therefore provide powerful discrimination among well-motivated particle physics and string theory realizations of inflation. Most remarkably, such observations have the potential to provide the very first direct clues about the scalar field geometry and symmetry structure of quantum gravity.

\vskip 6pt
\noindent
{\it Beyond the Tensor-to-Scalar Ratio}

While $B$-modes are a powerful probe for testing the inflationary mechanism that is largely insensitive to the details of how precisely inflation is implemented, a host of complementary observations can potentially reveal more specific details about the inflationary era.
In \S\ref{sec:beyond} we discussed how
deviations from scale-invariance (running of the scalar spectrum and a large tilt of the tensor spectrum), non-Gaussianity, and isocurvature contributions
probe the structure of the underlying inflationary Lagrangian.
A nonzero value for any of these observables would be inconsistent with single-field slow-roll inflation and hence would suggest that non-trivial kinetic terms, violations of slow-roll, or multiple fields were important during inflation.

\vskip 6pt
\noindent
{\it Experimental Forecasts}

To quantify the relation between the theoretical topics studied in this report and the measurements of a future CMB satellite we presented realistic forecasts of parameter uncertainties in \S\ref{sec:CMBPol}, with the underlying assumptions and caveats detailed in Appendix \ref{sec:Fisher}.
Our conclusions for the projected constraints on tensor modes can be summarized as follows:

\begin{itemize}
\item Gravitational waves can be detected at $\sim 3 \sigma$ for $r \gtrsim 0.01$ for the low-cost mission and optimistic foregrounds (see Appendix \ref{sec:Fisher}).

\item If $r=0.01$ then {\sl CMBPol} would measure this at the $\sim 15\sigma$ level for the low-cost mission and optimistic foregrounds if the consistency relation is imposed, $n_t = - r/8$.

\item {\sl CMBPol} would provide a $3\sigma$ upper limit on tensors of $r \lesssim 0.002$  for the low-cost mission and optimistic foregrounds if the consistency relation is imposed.
\end{itemize}
These limits should be compared to the theoretically interesting regime of large-field inflation ($r>0.01$); {\it cf.}~\S\ref{sec:Bmodes}.  This shows that {\sl CMBPol} is a powerful instrument to test this crucial regime of the inflationary parameter space.

\begin{table}[h!]
\begin{center}
\begin{tabular}{||c|c|c|c|c|c|c||}
\hline \hline
{\small \bf Errors} & {\small {\sl WMAP}}  & {\small {\sl Planck}} &  \multicolumn{2}{|c|}{\small EPIC-LC }& \multicolumn{2}{|c||}{\small EPIC-2m} \\

            & {\small no FGs} & {\small no FGs} & {\small no FGs} & {\small with Pess FGs} & {\small no FGs} & {\small with Opt FGs} \\
\hline
\hline
{\small $\Delta n_s$} & {\small 0.031}  & {\small 0.0036} & {\small --} & {\small --}& {\small 0.0016} & {\small 0.0016} \\
{\small $\Delta \alpha_s$} & {\small 0.023} & {\small 0.0052} & {\small --} & {\small --} & {\small 0.0036} & {\small 0.0036} \\
{\small $\Delta r$} & {\small 0.31} & {\small 0.011} & {\small $5.4\times 10^{-4}$} & {\small $9.2 \times 10^{-4}$} & {\small $4.8\times 10^{-4}$} & {\small $5.4 \times 10^{-4}$} \\
\hline
{\small $\Delta r$} & {\small --} & {\small 0.10} & {\small $0.0017$} & {\small --} & {\small $0.0015$} & {\small $0.0025$}\\
{\small $\Delta n_t$} & {\small --} & {\small 0.20} & {\small $0.076$} & {\small --} & {\small $0.072$} & {\small $0.13$}\\
\hline
{\small $\Delta f_{\rm NL}^{\rm local}$} & {\small --} & {\small $4$} & {\small --} & {\small --} & {\small $2$} & {\small --}\\
{\small $\Delta f_{\rm NL}^{\rm equil.}$} & {\small --} & {\small $26$} & {\small --} & {\small --} & {\small $13$} & {\small --} \\
\hline
{\small $\Delta \alpha_{\rm (c)}$} & {\small --} & {\small $1.2 \times 10^{-4}$} & {\small $3.5 \times 10^{-5}$} & {\small $4 \times 10^{-5}$} & {\small $3.5 \times 10^{-5}$} & {\small $3.5 \times 10^{-5}$}\\
{\small $\Delta \alpha_{\rm (a)}$} & {\small --} & {\small 0.025} & {\small 0.0065} & {\small 0.0068} & {\small 0.0065} & {\small 0.0066} \\
\hline
{\small $\Delta \Omega_k$}  & {\small --} & {\small --} & {\small --} & {\small --}& {\small $6 \times 10^{-4}$}& {\small $6 \times 10^{-4}$} \\
\hline \hline
\end{tabular}
\caption{ \label{tab:error2} Forecasts of (1$\sigma$) errors on key inflationary parameters for {\sl WMAP} (8 years), {\sl Planck} and {\sl CMBPol} (EPIC-LC and EPIC-2m). 
Copy of Table \ref{tab:error0} showing results for the unrealistic assumption of `no foregrounds' (no FGs) and `with foreground removal' (with FGs) (see Appendix \ref{sec:Fisher}). The fiducial model has $r=0.01$. The single-field consistency relation has been applied in the top block of forecasts.}
\end{center}
\end{table}

\begin{figure}[h!]
    \centering
        \includegraphics[width=.48\textwidth]{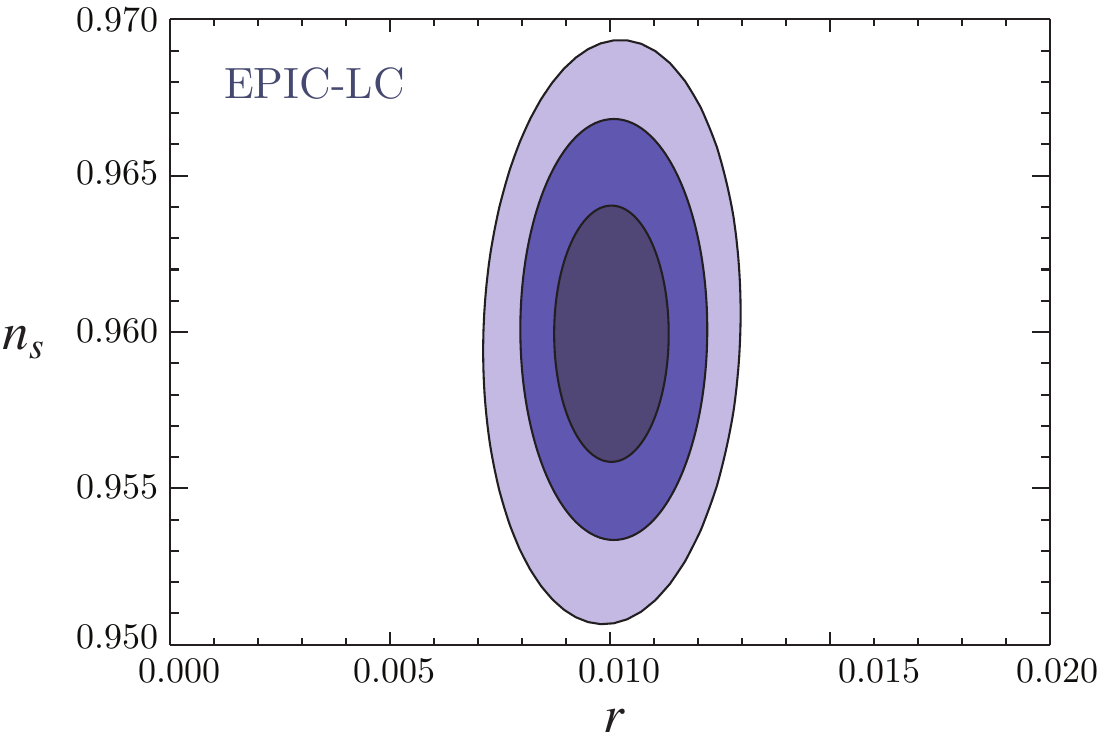}
        \hskip 10pt \includegraphics[width=.48\textwidth]{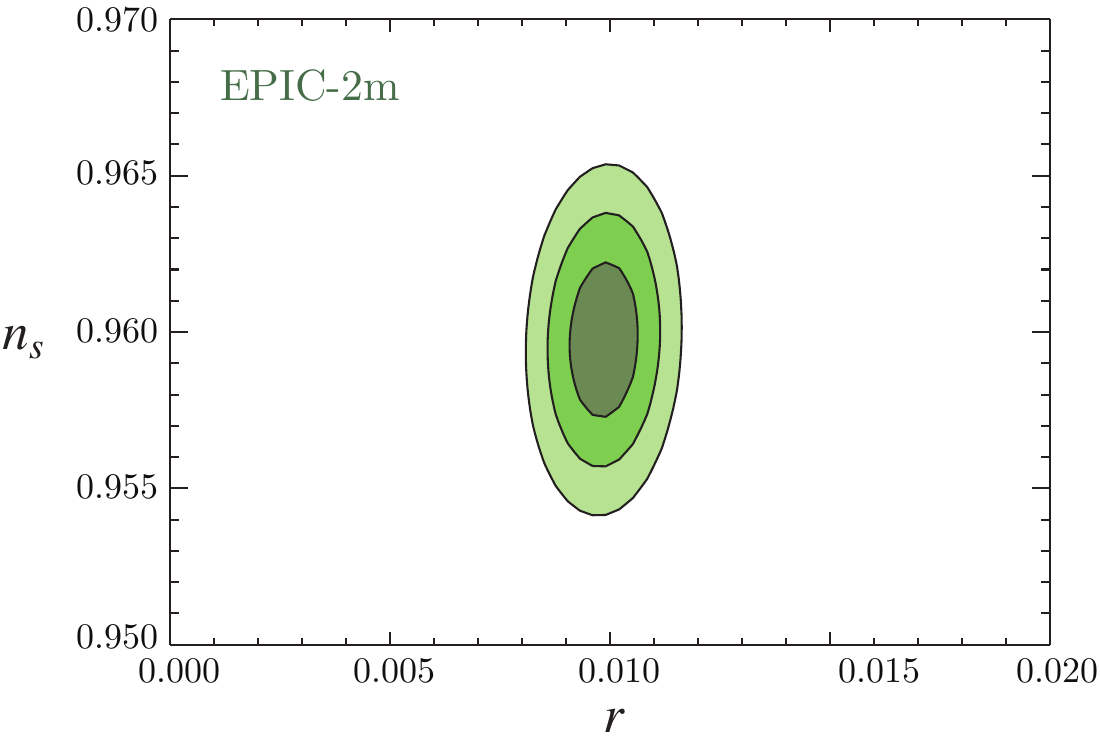}
   \caption{Forecasts of {\sl CMBPol} constraints in the $r$-$n_s$ plane (Copy of Figure \ref{fig:nsrX}). {\it Left}: EPIC-LC with pessimistic foreground option. {\it Right}: EPIC-2m with optimistic foreground option.}
    \label{fig:nsrXX}
\end{figure}

\begin{figure}[h!]
    \centering
        \includegraphics[width=.48\textwidth]{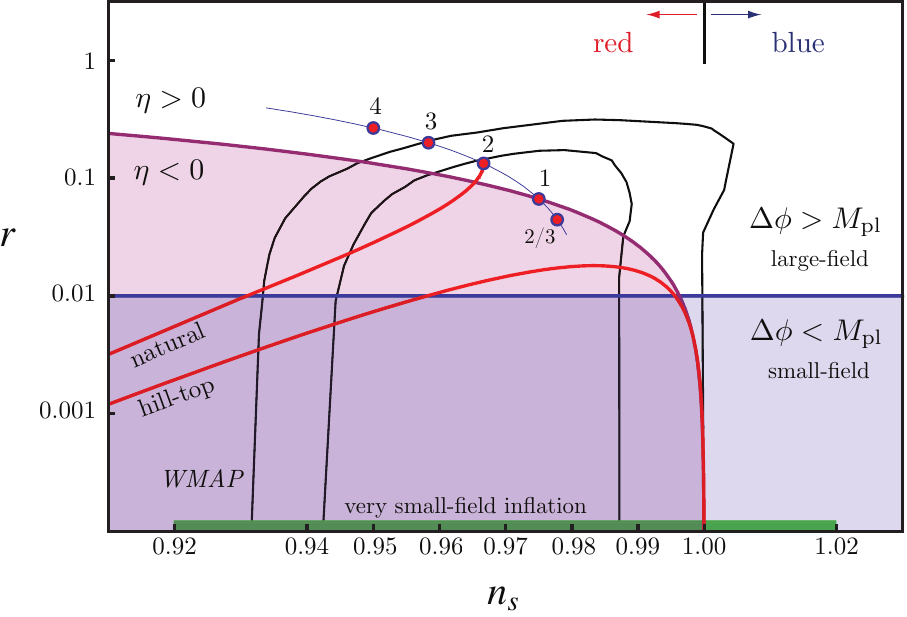}
        \hskip 10pt \includegraphics[width=.48\textwidth]{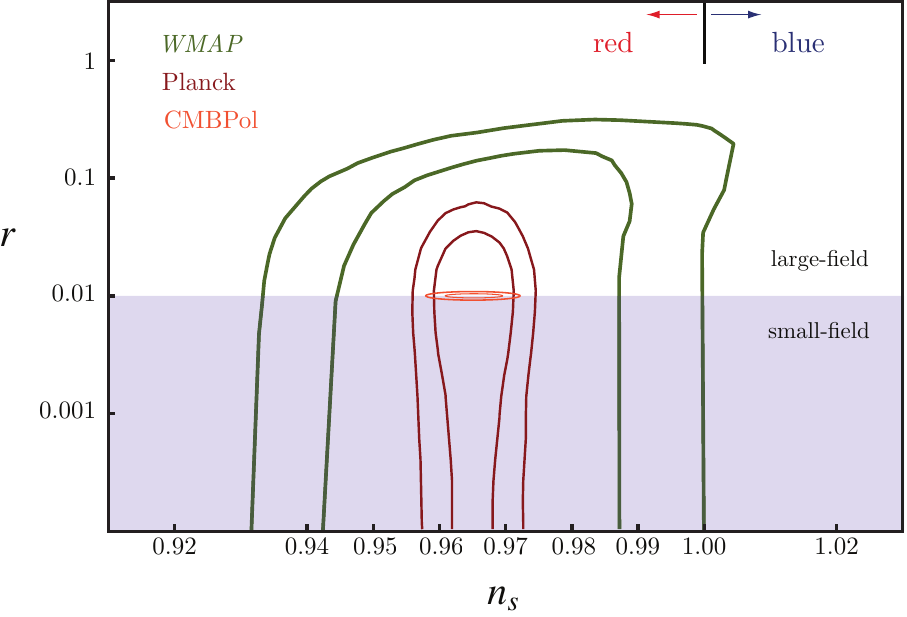}
   \caption{Summary of slow-roll predictions in the $n_s$-$r$ plane (Figure \ref{fig:nsr}) and forecasts of future constraints (Figure \ref{fig:nsr2}).}
    \label{fig:nsrSummary}
\end{figure}

\vskip 6pt
\newpage
\noindent
{\it Final Remarks -- Physics at the Highest Energies and Smallest Distances}

Particle physics is entering a new era. In the next few years, the {\sl Large Hadron Collider} (LHC) at CERN will provide unprecedented information about physics at the TeV scale. This is a tremendous achievement, but a vast range of even higher energies will remain forever unexplored by terrestrial collider experiments. Fundamental questions about the most basic workings of Nature at the highest energy scales -- questions about grand unification, string theory, and the physics of the Planck scale, for example --- must await a more powerful experimental method. Inflation serves as the ultimate particle accelerator, amplifying physical processes from the smallest scales to the very largest. The detection of primordial gravitational waves from inflation would illuminate energies a trillion times higher than those at the LHC and provide a unique window onto the laws of Nature at the highest energy scales.

\section*{Acknowledgements}

The research of the Inflation Working Group was partly funded by NASA Mission Concept Study award NNX08AT71G S01. We also acknowledge the organizational work of the Primordial Polarization Program Definition Team.

\newpage
\appendix

%\section{Taxonomy of Inflationary Models}
\section{Models of Inflation}
\label{sec:taxonomy}

Inflation requires a form of stress-energy which sources a nearly constant Hubble parameter $H= \partial_t \ln
a$. Theoretically, this can arise via a truly diverse set of mechanisms with disparate phenomenology and varied
theoretical motivations.  Recently, a useful model-independent characterization of {\it single-field} models of
inflation and their perturbation spectra has been given \cite{Cheung:2007st,Garriga:1999vw,Chen:2006nt, Weinberg:2008hq}.
Starting from this basic structure, each model of single-field inflation arises as a special limit.  We first review the traditional case of single-field slow-roll inflation (\S\ref{sec:SFSR}).  Next, we present more
general single-field mechanisms for inflation and their density perturbations (\S\ref{sec:GenSF}).  Finally, we give a brief discussion of
multi-field models (\S\ref{sec:MultipleFields}).
For more details on some of the models, we refer the reader to the comprehensive review by Lyth and Riotto~\cite{LythRiotto}.
We discuss inflationary model-building in the context of supergravity and string theory in \S\ref{sec:sugra} and \S\ref{sec:string}, respectively.
In  Appendix \ref{sec:alternatives} we also contrast the predictions of inflation to the potential predictions arising from alternative models of
the early universe.

\subsection{Single-Field Slow-Roll Inflation}
\label{sec:SFSR}

The definition of an inflationary model amounts to a specification of the inflaton action (potential and kinetic terms) and its coupling to gravity.
Single-field models including only first derivative interactions and minimally coupled to gravity are described by the action \cite{Garriga:1999vw,Chen:2006nt} \beq \label{equ:Paction}
S = \frac{1}{2} \int \d^4 x \sqrt{-g} \left[ {\cal R} + 2 P(X,\phi)\right]\, , \qquad \qquad \Mp^{-2} \equiv 8 \pi G =
1\, , \eeq where $X \equiv -\frac{1}{2} g^{\mu \nu} \partial_\mu \phi \partial_\nu\phi$.
Slow-roll inflation then corresponds to the special case of a canonical kinetic term
\beq
P(X, \phi) = X- V(\phi)\, .
\eeq
In this case, the inflationary dynamics is fully specified by the potential $V(\phi)$.
More general single-field models of the type (\ref{equ:Paction}) will be described in the next section.

Single-field slow-roll models of inflation are usefully divided into two classes:
\begin{enumerate}
\item[i)] {\it Large-field inflation}:

Models that
imply a high energy scale for
inflation and involve large field excursions ($\Delta \phi > \Mp$).
\item[ii)] {\it Small-field inflation}:

Models that
imply a low energy scale and small field
excursions ($\Delta \phi < \Mp$).
\end{enumerate}
In the following we present characteristic examples of models of each type.

\newpage
%\addtocontents{toc}{\SkipTocEntry} 
\subsubsection{Large-Field Slow-Roll Inflation}

In Section \ref{sec:Bmodes}, we showed that the amplitude of inflationary gravitational waves (measured by the tensor-to-scalar ratio $r$) relates to the field variation $\Delta \phi = |\phi_{\rm end} - \phi_{\rm cmb}|$ between the end of inflation and the time when CMB scales exited the horizon about 60 $e$-folds before (see Figure \ref{fig:potential}),
\beq
\frac{\Delta \phi}{\Mp} \gtrsim {\cal O}(1) \left(\frac{r}{0.01} \right)^{1/2}\, .
\eeq
An observation of $B$-mode polarization with {\sl CMBPol} ($r>0.01$) would therefore be convincing evidence that i) inflation occurred and ii) Nature realized large-field inflation.
In Section \ref{sec:Bmodes}, we described the fundamental clues that this would provide about the symmetries of the high-energy theory underlying inflation.
In the context of effective field theory the following large-field models have been considered in the inflationary literature:
%Large-field inflation refers to models with super-Planckian field excursions\footnote{Here, $M_{\rm pl}^{-2} = 8 \pi G$ is the reduced Planck mass. $\Delta \phi \equiv \phi_{\rm cmb} - \phi_{\rm end}$ is the field distance between the end of inflation and the time when CMB fluctuations were created about 60 $e$-folds before (see Figure \ref{fig:potential}).} $\Delta \phi > M_{\rm pl}$.
%These models correspond to a tensor amplitude that is {\it observable} with {\sl CMBPol} ($r> 0.01$).\\

\vskip 8pt
\noindent {\it Chaotic Inflation}

The prototype for chaotic inflation \cite{Linde:1983gd} involves a single polynomial term (with $p>0$) \beq
\label{equ:chaotic} V(\phi) = \lambda_p\, \Bigl(\frac{\phi}{\mu} \Bigl)^p\, . \eeq Here, the scale $\mu$ relevant for
higher-dimensional terms in the effective potential corresponds to the mass of heavy states that have been
integrated out in forming the effective potential. By computing the slow-roll parameters corresponding to
(\ref{equ:chaotic}) one easily sees that inflation requires $\phi > \Mp$, and as explained in detail in \S4.2,
we must have $ \mu < \Mp$.  Thus, the absence of ever higher-order terms $(\phi/\mu)^n$ ($n \to \infty$) with
order-one coefficients is tantamount to the presence of a shift symmetry which forbids such terms. Such a shift
symmetry is quite consistent with the radiative stability of the potential (\ref{equ:chaotic}) because the
coefficient $\lambda_p$ must be extremely small to match the {\sl COBE} normalization of the power spectrum; hence the
potential, as well as its coupling to gravity, very weakly breaks the shift symmetry. An example of a
supergravity model where such a symmetry is present and the simplest chaotic inflation potential $\frac{1}{2} m^{2}\phi^{2}$
emerges was proposed in \cite{jap}. We discussed the prospects for UV completing such shift-symmetric models in
\S4.2; a relatively simple mechanism producing chaotic inflation in string theory was recently described in
\cite{Silverstein:2008sg,MSW}. In the relevant range of $\phi$, these models yield a potential of the form
(\ref{equ:chaotic}), but with $p$ in general a fraction of the powers considered in the original chaotic
inflation literature.  As shown in Figure \ref{fig:nsr},
many of these models are observationally distinguishable from each other.

Chaotic inflation models of the form (\ref{equ:chaotic}) make the following predictions
\begin{equation}
r=\frac{8p}{2N_\star}=8 \left({p \over p + 2}\right) \left(1 - n_s\right),
\end{equation}
where $N_\star$ is the number of $e$-folds between the end of inflation and the time
when the observable scale leaves the horizon.

\vskip 8pt
\noindent
{\it Hill-top models with quadratic term}

Typical hill-top models can be expanded as
\begin{equation}
\label{equ:hill:large}
V(\phi) =V_0 \left[ 1 -  \Bigl( \frac{\phi}{\mu} \Bigr)^p \right] + \dots\, , \qquad \phi < \mu\, .
\end{equation}
The potential (\ref{equ:hill:large}) may be considered an approximation to a generic symmetry-breaking potential.
The dots in (\ref{equ:hill:large}) represent higher-order terms that become important near the end of inflation and during reheating.
If $p=2$, the second slow-roll parameter reads,
\begin{equation}
\eta = - 2 \left( \frac{M_{\rm pl}}{\mu}\right)^2 \frac{1}{1 - ( \phi / \mu )^2}~.
\end{equation}
Hence slow-roll requires $\mu > M_{\rm pl}$, and inflation ends when
$\phi \sim \mu > M_{\rm pl}$.  So,
this model can {\it only} be of the large-field type. It predicts the
following relation between $r$, $n_s$ and $N_\star$
\begin{equation}
r = 8 (1 - n_s) \exp\left[- 1 - N_\star \left(1 - n_s \right)\right].
\end{equation}
For $p > 2$, the potential (\ref{equ:hill:large}) can lead to either large-field or
small-field inflation, depending on the value of $\mu$.

\vskip 8pt
\noindent {\it Axion Inflation}

In the context of inflationary model building, pseudo-Nambu Goldstone
bosons (PNGB; axions) have the attractive feature that their potential
is protected by a shift symmetry $\phi \to \phi + \alpha$.  This
symmetry guarantees that to first approximation the PNGB is
massless. However, non-perturbative corrections break the shift
symmetry and generically lead to a potential of the form \beq V(\phi)
= V_0 \left[ 1 - \cos \left(\frac{\phi}{ \mu}\right)\right]\, . \eeq
This potential is a particular case of (\ref{equ:hill:large}).
For $\mu > \Mp$, it gives a successful model of large-field inflation
which is natural in the Wilsonian sense \cite{Freese:1990rb,Adams:1992bn}. A
supergravity version of natural inflation was recently constructed in
\cite{Kallosh:2007ig,Kallosh:2007cc}.

Axions
are generically present in string theory and extra-dimensional theories of gravity.  Nevertheless, early
attempts to derive large-field inflation from such axion fields \cite{ArkaniHamed:2003wu,ArkaniHamed:2003mz} were difficult
to implement in string theory; the resulting effective potentials in many cases have $\mu < M_{\rm pl}$ for
detailed dynamical reasons \cite{Banks:2003sx, Svrcek:2006yi}.

However, further research has produced several promising ideas for making working models of axion inflation. Typical string
models have a large number of axion fields, so there may be a possibility of obtaining the large field excursion
from the combined effect of many axions \cite{Nflation, Easther:2005zr}\ (though the field range is not {\it
parametrically} increased as a function of the number of fields).\footnote{Difficulties with concretely
constructing such `$N$-flation' models are discussed in \cite{Kallosh:2007cc,Grimm:2007hs}; these difficulties may reflect the fact that the problem is naturally complicated by the large number of fields
required by the basic mechanism.} Recently, a reasonably generic string theory mechanism for
large-field inflation has been elucidated. This involves `monodromy' in field space -- a phenomenon arising from
the higher-dimensional branes of string theory which enlarges the periodicity of angular directions (such as
certain D-brane positions and axions) to yield a super-Planckian field range \cite{Silverstein:2008sg, MSW}. The corresponding potential in the case of axion monodromy inflation \cite{MSW} takes the form
\begin{equation}
V(\phi) =\mu^3\phi+\ldots,
\end{equation}
where the leading omitted terms are periodic functions of the angular variables.

\newpage
%\addtocontents{toc}{\SkipTocEntry}
\subsubsection{Small-Field Slow-Roll Inflation}

Small-field inflation refers to models with sub-Planckian field excursions $\Delta \phi < M_{\rm pl}$.
The associated tensor amplitude is most likely {\it unobservable} with {\sl CMBPol} ($r \ll 0.01$).

\vskip 8pt
\noindent
{\it Hill-top models with no quadratic term}

A characteristic small-field potential has the following form
\begin{equation}
\label{equ:hill}
V(\phi) =V_0 \left[ 1 -  \Bigl( \frac{\phi}{\mu} \Bigr)^p \right] + \dots\, , \qquad \phi < \mu \ll   M_{\rm pl} \, , \qquad p > 2 \, .
\end{equation}
This potential is identical to that of (\ref{equ:hill:large}) with
the two restrictions $\mu \ll   M_{\rm pl}$ and $p > 2$ necessary
for small-field inflation; as already mentioned,
this potential may be considered an approximation to a
generic symmetry breaking potential, and the dots in (\ref{equ:hill})
represent higher-order terms that become important near the end of
inflation and during reheating.  The fine-tuning of initial conditions
({\it e.g.}~the initial value of $\dot \phi$) is often more severe for
small-field models than for large-field models (but see {\it e.g.}
\cite{Linde:2004nz}).
For this model, the scalar spectral index is given by
\begin{equation}
n_s-1 = - \frac{2}{N_\star} \frac{p-1}{p-2}~,
\end{equation}
and
there exists an upper bound on the gravitational wave amplitude
\begin{equation}
r < 8 {p \over N_\star  \left(p - 2\right)} \left({8 \pi \over N_\star \, p \left(p - 2\right)}\right)^{p / \left(p - 2\right)}.
\end{equation}

\vskip 8pt
\noindent
{\it Coleman-Weinberg}

Historically, a famous inflationary potential is the Coleman-Weinberg potential \cite{Linde:1981mu, Albrecht:1982wi}
\beq
V(\phi) = V_0 \left[ \left(\frac{\phi}{\mu}\right)^4 \left( \ln \left( \frac{\phi}{\mu}\right) - \frac{1}{4}\right) + \frac{1}{4}\right]\, ,
\eeq
which arises as the potential for radiatively-induced symmetry breaking in electroweak and grand unified theories.
Although the original values of the parameters $V_0$ and $\mu$ based on the $SU(5)$ theory are incompatible with the small amplitude of inflationary fluctuations, the Coleman-Weinberg potential remains a popular phenomenological model.

%\addtocontents{toc}{\SkipTocEntry}
\subsubsection{Hybrid Models}

The hybrid scenario \cite{linde91,linde94,copeland94} frequently
appears in models which incorporate inflation into supersymmetry. In a
typical hybrid inflation model, the effective inflaton potential
receives a constant contribution from a false vacuum energy,
stabilized by interactions of the inflaton field $\phi$ with other fields $\psi$.
When the inflaton passes a critical value, the false vacuum is
destabilized and another field triggers a phase transition to a
lower energy vacuum state, bringing inflation to an end. Topological
defects may be produced in such a phase transition and could provide a
distinctive observational signature of such models. The dynamics which
brings inflation to an end in hybrid models is decoupled from the
inflatonary slow-roll parameters.

During inflation, such models are characterized by potentials
of the form
\begin{equation}
V\left(\phi\right) = V_0 \left[1 + f(\phi/\mu) \right]\, ,
%V\left(\phi\right) = V_0 \left[1 + \left(\frac{\phi}{\mu}\right)^p\right].
\end{equation}
where $f$ is a function which should be compatible with the slow-roll
conditions. A particular case is that of hybrid inflation with a
single polynomial term
\begin{equation}
V\left(\phi\right) = V_0 \left[1 + \left(\frac{\phi}{\mu}\right)^p\right].
\end{equation}
The field value at the end of inflation, $\phi_{\rm end}(\psi)$, is
determined by some other physics, so there is a second free parameter
characterizing the models. Because of this extra freedom, hybrid
models fill a broad region in the $n_s$-$r$ plane. For
$\left({\phi_\star / \mu}\right)\gg 1$ (where $\phi_\star$ is the
value of the inflaton field when there are $N_\star$ {\it e}-foldings
until the end of inflation) one recovers the results of the
large-field models. On the other hand, when $\left({\phi_\star /
\mu}\right)\ll 1$, the dynamics are analogous to small-field models,
except that in some cases -- including $f(\phi)=(\phi/\mu)^p$ -- the field
is evolving toward, rather than away from, a dynamical fixed
point. This distinction is important to the discussion here because
near the fixed point the parameters $r$ and $n_s$ become independent
of the number of {\it e}-folds $N_\star$.

Models of inflation based on global supersymmetry \cite{fterm} or
$D$-term inflation models \cite{dterm1, dterm2} are of the hybrid type
and the potential is of the form
\begin{equation}
V=V_0\left[1+\alpha\log\frac{\phi}{\mu}\right],
\end{equation}
where $\alpha$ is a loop factor. The logarithmic behavior arises from
the fact that the quadratic divergences are canceled thanks to
supersymmetry, leaving only the mild logarithmic dependence. In this
particular example of hybrid inflation, the field is not rolling
towards a dynamical fixed point, and depending on parameter values
the slow-roll conditions can break down before or after the
false vacuum destabilization.

\vskip 8pt
\noindent
{\it Warped D-brane Inflation}

In string theory, a version of hybrid inflation can arise from a
brane-antibrane system in a warped flux compactification of type IIB
string theory \cite{KKLMMT,BDKM}.  The inflaton potential arises from a combination of the Coulomb interaction between the brane and antibrane, and of moduli-stabilizing effects that generate Planck-suppressed operators in the four-dimensional theory (see
\cite{Holographic} for a systematic treatment of these contributions to the
inflaton potential). This class of models does not require -- or allow
\cite{BM} -- a large field range, and no symmetry appears in general
in the direction of the inflaton. Instead, inflation occurs in a small range around a fine-tuned inflection point \cite{BDKM, Pajer, LindeWestphal}.  Ref.~\cite{Holographic} in particular has argued that such potentials arise under rather general circumstances in warped brane-antibrane systems; see \cite{Cline} for a systematic study of the corresponding parameter space. 
%The phenomenology of inflection point inflation was reviewed in \cite{BDKM}, and systematic study of the parameter space
%
These models allow only a very low tensor amplitude,
$r \ll 10^{-4}$ \cite{BM}, but the scalar spectrum can be either red or blue at CMB scales, depending on
%the tuning of
the first derivative of the potential near the inflection point.
The prediction for $r$ is too
small to observe, while $n_s$ depends on the details of the full
string compactification and its effect on the brane-antibrane
potential \cite{BDKM, Pajer, Holographic}; even small-field models are
UV-sensitive in this basic sense.
%It simply requires a one-percent
%cancellation among different contributions to $\eta$.
In this class of
models, the exit from inflation is rather economically accomplished by
the annihilation of the branes, a process which leads to a cosmic
string signature in a subset of models.

\newpage
\subsection{General Single-Field Models}
\label{sec:GenSF}

The classification of models has so far relied on properties of the potential in cases where the inflaton has a standard kinetic term. However, considering all possible interactions that preserve a shift symmetry naturally leads to the inclusion of derivative interactions~\cite{Creminelli:2003iq}, which may {\it a priori} be added to any of the scenarios discussed above. These more general models introduce new parameters, the sound speed $c_s$ and its running, that enter the consistency relation and the scalar spectral index. This was briefly discussed in Section \ref{sec:beyond} and will be presented in more detail below. Perhaps most importantly, non-standard kinetic terms also introduce the possibility of large non-Gaussianity since the derivative interactions may be large without destroying the %flatness of the potential.
slow evolution of the Hubble parameter.

\vskip 8pt
Single-field models including first-derivative interactions are described by the action (\ref{equ:Paction}).
The function $P(X,\phi)$ in (\ref{equ:Paction}) corresponds to the pressure of the scalar fluid
while the energy density is given by \beq \rho= 2 X P_{,X} - P\, . \eeq
Examples of inflation
models where $P(X,\phi)$ takes a non-trivial form are k-inflation \cite{k-Inflation}, DBI inflation \cite{DBI}
and ghost inflation \cite{GhostInflation}. These models are characterized by a
speed of sound \beq c_s^2 \equiv \frac{P_{,X}}{\rho_{,X}} =\frac{P_{,X}}{P_{,X} + 2 X P_{,XX}}\, ,\eeq where $c_s=1$
for a canonical kinetic term and a smaller sound speed indicates a more significant departure from the standard
scenario.

Notice that $X$ has mass dimension four, so that we expect higher powers of $X$ to be suppressed by some scale
$\mu$ as $X^n/\mu^{4n-4}$. The significance of these terms (and the magnitude of non-Gaussianity) depends on the
size of $X$, evaluated on the background classical evolution of the inflaton, compared to the scale $\mu$. For
potential energy dominated inflation, this is no larger than $V(\phi)/\mu^4$. Since $\mu$ is typically the Planck
scale or the string scale, these interactions can often be ignored. Their relevance in some scenarios is another
example of UV sensitivity in inflation ({\it cf.}~Section \ref{sec:Bmodes}).

To calculate observables, it proves convenient to define parameters for the time-variation of the expansion rate
$H(t)$ and the speed of sound $c_s(t)$ \beq \epsilon \equiv - \frac{\dot H}{H^2} = \frac{X P_{,X}}{H^2}\, , \qquad
\tilde \eta \equiv \frac{\dot \epsilon}{\epsilon H}\, , \qquad s \equiv \frac{\dot c_s}{c_s H}\, . \eeq Inflation of
significant duration occurs when $\epsilon$ and $|\tilde\eta|$ are small. Although large $s$ may not necessarily
imply that inflation ends, the analytic expressions given below assume that $|s|\ll1$. Small $s$ is also a
desirable feature in models that match observational bounds on the spectral index.

A non-trivial speed of sound modifies the scalar spectrum \beq \label{equ:PPs} P_s(k) = \left. \frac{1}{8
\pi^2 \Mp^2} \frac{H^2}{c_s \epsilon}\right|_{c_s k = aH}\, . \eeq That is, for a fixed energy scale of
inflation, a small sound speed enhances the scalar perturbations. Scalar fluctuations now freeze out at the
sound horizon, so the r.h.s.~of (\ref{equ:PPs}) is evaluated at $a H = c_s k$. The scale-dependence of the
spectrum is \beq
n_s - 1 = - 2 \epsilon - \tilde \eta - s\, . %\qquad \alpha_s =
\eeq Note that the running of the spectral index, $\alpha_s$, will now involve a new term from $ds/d\ln k$. The
tensor fluctuation spectrum is not affected by the new interactions and so is the same as for slow-roll models
\bea
P_t(k)   &=& \left. \frac{2}{\pi^2} \frac{H^2}{\Mp^2} \label{equ:Ptt} \right|_{k = aH}\, ,\\
n_t &=&  - 2 \epsilon \label{equ:ntt}\, . \eea The r.h.s.~of (\ref{equ:Ptt}) and (\ref{equ:ntt}) is evaluated at
the usual horizon $a H  = k$. We see that for models with $c_s \ne 1$ the consistency relation between $r$ and
$n_t$ is modified to
\beq r = - 8 c_s n_t\, . \eeq

Arguably the most important distinction of small sound speed models is that for $c_s \ll 1$ the scalar
fluctuations are highly non-Gaussian \cite{Chen:2006nt}. For example, the three-point function is largest for
equilateral triangles, with magnitude \beq \label{equ:fNLeq} f_{\rm NL}^{\rm equil} = - \frac{35}{108} \left(\frac{1}{c_s^2} - 1
\right) + \frac{5}{81} \left(\frac{1}{c_s^2}-1 - 2 \Lambda \right)\, , \eeq where \beq \Lambda \equiv \frac{X^2
P_{,XX} + \frac{2}{3} X^3 P_{,XXX}}{X P_{,X} + 2 X^2 P_{,XX}}\, . \eeq

At the time of writing, bounds on the magnitude of non-Gaussianity at CMB scales provide one of the strongest constraints
on these models. Section~\ref{sec:NG} elaborates on those constraints and contrasts the non-Gaussian signal from
single-field models with that from multi-field models (see also Section \ref{sec:MultipleFields}).

A particularly useful example of this type of scenario occurs in brane inflation, where the inflaton is related to the brane position~\cite{Alishahiha_etal04}. The kinetic part of the action, in the limit of small acceleration, is the Dirac-Born-Infeld (DBI) action. In the simplest case\footnote{This case corresponds to a single brane without worldvolume flux, with motion in a single direction along which the background warp factor may vary.} and including an arbitrary potential, this action takes the form of Eqn.~(\ref{equ:Paction}) with
\beq
P(X,\phi)=-h(\phi)\sqrt{1-2Xh^{-1}(\phi)}+h(\phi)\,-V(\phi).
\eeq
The function $h(\phi)$ is the warped brane tension ($\propto\phi^4$ if the background is Anti-de Sitter (AdS) space) so the scale suppressing the kinetic terms is the warped string scale. The square root enforces a speed limit for the brane which allows more $e$-folds of inflation along a steep potential than in standard slow-roll. When the brane is moving near the speed limit, the square root may not be expanded and the non-Gaussianity is significant. The specific form of the action leads to two simplifying relationships: $P_{,X}=c_s$ and $2\Lambda= c_s^{-2}-1$. Then the relationship between the field range and $r$ is the same as in slow-roll (although now $r$ may vary more significantly) and the second term in Eqn.~(\ref{equ:fNLeq}) for $f_{\rm NL}^{\rm equil}$ vanishes.

\subsection{Inflation with Multiple Fields}
\label{sec:MultipleFields}

Invoking two or more scalar fields extends the possibilities for
inflationary models \cite{Silk:1986vc,Kofman:1986wm}, but also
diminishes the predictive power of inflation.

We have already considered models of inflation involving more than one
field: namely, hybrid inflation models. However, in hybrid inflation,
the dynamics of inflation and the generation of primordial
perturbations is still
governed by a single inflaton field: hence, these models can still
be classified as single-field models, with the peculiarity that
the end of inflation is then independent from the breaking of the
slow-roll conditions.

Multi-field models include double inflation
\cite{Kofman:1985aw, Mukhanov:1991rp, Polarski:1992dq,Polarski:1994bk},
thermal inflation \cite{Adams:1997de},
double hybrid inflation \cite{Lesgourgues:1999uc},
curvaton models
\cite{Linde:1996gt,Lyth:2001nq,Moroi01}, inhomogeneous reheating
\cite{Dvali:2003em,Kofman:2003nx} and assisted inflation
\cite{Liddle:1998jc}.  These models relax some of the constraints on
inflation arising from the predictions for cosmological observables.
However, multi-field models can have distinctive observational
signatures such as features in the spectrum of
adiabatic perturbations
\cite{Kofman:1988xg, Salopek:1988qh, Hodges:1989dw,Mukhanov:1991rp, Polarski:1992dq,Polarski:1994bk,Adams:1997de,Lesgourgues:1999uc},
observable isocurvature perturbations
\cite{Linde:1984ti,Kofman:1986wm, Polarski:1994rz,Langlois:1999dw,GarciaBellido:1995qq,Gordon:2000hv},
or large non-Gaussianities.

In models
such as {\em assisted inflation} \cite{Liddle:1998jc} (or
the specific case of assisted quadratic inflation, known as {\em
N-flation} \cite{Nflation}) there may be many fields which evolve
during inflation. In this case one must take into account quantum
fluctuations in all the fields which affect the dynamical evolution
during inflation, or afterwards. In general this leads to additional
sources for primordial density perturbations, while the gravitational
waves still depend only upon the energy scale during inflation. Thus
the consistency equation for the tensor-to-scalar ratio in
single-field inflation becomes an upper bound on the tensor-to-scalar
ratio in multi-field models \cite{Sasaki:1995aw}.

Many multi-field models decouple the creation of density perturbations
from the dynamics during inflation. If the decay of the vacuum energy
at the end of inflation is sensitive to the local values of fields
other than the inflaton then this can generate primordial
perturbations due to {\em inhomogeneous reheating}
\cite{Dvali:2003em,Kofman:2003nx}. In the {\em curvaton} scenario
\cite{Linde:1996gt,Lyth:2001nq,Moroi01}, the inhomogeneous
distribution of a weakly coupled field generates density perturbations
when the field decays into radiation sometime after inflation. The
curvaton scenario can also produce isocurvature density perturbations
in particle species ({\it e.g.}~baryons) whose abundance differs from
the thermal equilibrium abundance at the time when the curvaton decays
\cite{Linde:1996gt, Lyth:2002my}.

Inflation is still required to set up large-scale perturbations from
initial vacuum fluctuations in all these models. But if the primordial
density perturbation is generated by local physics some time after
slow-roll inflation then the local form of non-Gaussianity is no
longer suppressed by slow-roll parameters.  A large value of $f_{\rm
NL}^{\rm local}$ may therefore serve as a useful diagnostic of
inflation models with multiple fields.  Recent work on general
multi-field inflation such as \cite{Langlois:2008qf}\ reveals
interesting features in the power spectrum -- particularly on the
amplitude and shape of the non-Gaussianities -- from the combination
of more generic kinetic terms with multiple fields.

The presence of multiple fields during an inflationary phase is one of the possible sources of deviation from
the consistency relation that holds for single-field models of slow-roll inflation. There exists a
model-independent consistency relation for slow-roll inflation with canonical fields \cite{Wands:2002bn},
\begin{equation}
r
= -8n_t \sin^2 \Delta\, ,
\end{equation}
where for two-field inflaton $\cos\Delta$ is the correlation between the adiabatic and isocurvature
perturbations, which is a directly measurable quantity. More generally $\sin^2\Delta$ parameterizes the ratio
between the adiabatic power spectrum at horizon exit during inflation and that which is observed.
The conversion of non-adiabatic perturbations into curvature perturbations after horizon exit decreases the
tensor-to-scalar ratio for a fixed value of the slow-roll parameter $\epsilon$, which determines the tensor tilt.

This underscores the importance of measuring or constraining the scale-dependence of the tensor power spectrum. Although
it will be hard to measure any scale dependence of the tensors if the single-field consistency relation $n_t=-r/8$ holds, a large tilt would invalidate this consistency relation. A large negative tilt could be consistent
with multiple-field inflation.
%whereas any positive tilt
%would imply violation of the null energy condition during inflation.

\subsection{Inflation and Supersymmetry}
\label{sec:sugra}

Considerable theoretical effort has been devoted to realizing inflation in the context of well-motivated
theories of high-energy physics.  The earliest models of inflation were connected to GUT scenarios, and much
work in the intervening decades has focused on connections between inflation and supersymmetry.

There are three basic motivations for pursuing inflation in a supersymmetric theory.  First, supersymmetry is
the most intensively studied candidate for the physics of the TeV scale -- with several indirect hints from
particle physics pointing in its direction, including quantitative ties to GUT physics -- and it would be
striking if inflation were natural in a supersymmetric extension of the standard model. 
%Unfortunately, this appears not to be the case, barring artificial assumptions about the couplings of the Higgs field to gravity. 
Proposed models of inflation in the MSSM include \cite{Allahverdi:2006iq, Allahverdi:2006we, Murayama:1992ua, Murayama:1993xu}.
We will
know more about the relevance of these models with low-energy supersymmetry after the {\sl Large Hadron Collider}
({LHC}) runs for several years. It could be that in the next decade, we will know that low-energy supersymmetry is
a fact of Nature, or on the other hand that the physics of the TeV scale is not supersymmetric, and hence that
inflationary models with low-energy supersymmetry are irrelevant.

A second motivation, independent of the outcome at the {LHC}, is that supersymmetry might serve as a protective
symmetry that preserves the desired flatness of the inflaton potential.  Indeed, in a non-supersymmetric scalar
field theory without an approximate shift symmetry, loop corrections will be large, driving the physical
inflaton mass up to a value of order of the UV cutoff. (This is avoided, even in the absence of supersymmetry,
in models with a shift symmetry, {\it e.g.}~if the inflaton is an axion.)  Supersymmetry does provide a
considerable degree of radiative stability, but it is fair to say that supersymmetry alone (even in its local
form, supergravity) is not sufficient to ensure adequate flatness.  In particular, an entire class of
supergravity models, those in which the inflationary energy comes from an F-term, visibly suffers from the `eta
problem' described in \S\ref{sec:Bmodes}: dimension-six Planck-suppressed contributions to the potential generically spoil
flatness by rendering $\eta \sim {\cal O}(1)$.  This result is occasionally misinterpreted as indicating that
inflation is unusually difficult to obtain in supergravity.  As explained in \S\ref{sec:Bmodes}, the eta problem is present in
rather general effective quantum field theories coupled to gravity, supersymmetric or not; the problem is simply
harder to ignore in the case of F-term supergravity models. Conversely, although inflation sourced by a D-term
has been advanced as a solution to the eta problem in the context of supergravity, inclusion of generic
Planck-suppressed contributions to the potential is expected to spoil this conclusion, and can be shown to do so
in string theory realizations of D-term inflation. In summary, supergravity does not appear to provide a more
natural source for approximately flat inflaton potentials than non-supersymmetric field theory provides, but
neither are supergravities particularly deficient in this regard.

The third motivation for realizing inflation in supergravity
is that supergravity is the low-energy effective theory descending from supersymmetric compactifications of string theory.
As with the previous motivations, this one is subject to important caveats.  In particular, most limits
of string theory yield a higher scale of supersymmetry breaking.\footnote{In particular, most six-manifolds that admit consistent compactifications of supersymmetric string theories break supersymmetry at the Kaluza-Klein scale.} Nonetheless, much work has been done on the particular class of string compactifications which admit low energy supersymmetry, motivated in part by the exciting
possibility of TeV-scale supersymmetry reviewed above. Within this class of compactifications, it is very
interesting to assess whether the high scales of inflation required to see a tensor signal can coincide with the
low energy scales involved in modeling TeV-scale supersymmetry in string theory.  
In the specific moduli-stabilization scenarios studied to date, it
appears challenging to construct a natural model with low-scale
supersymmetry and detectable primordial tensors \cite{Kallosh:2007wm}.
Low-energy supersymmetry being a leading candidate
for physics beyond the Standard Model of particle physics, it will
be worthwhile to determine whether this result has broader validity
or is instead an artifact of the limited class of configurations
understood at present.  The study of string
compactifications with generic supersymmetry-preserving
ingredients is just beginning, and may lead to progress on
this question.

\subsection{Inflation in String Theory}
\label{sec:string}

As explained in detail in \S\ref{sec:Bmodes}, inflation is sensitive to the ultraviolet completion of gravity.
This strongly motivates formulating inflation within an ultraviolet-complete theory.  String theory, as a
candidate ultraviolet completion of particle physics and gravity, is a natural setting in which to address this
question. The problem is technically challenging in part because of the plethora of gravitationally-coupled
scalar fields, or moduli, descending from the extra dimensions of string theory.  The moduli generically roll
too rapidly for inflation, and must be stabilized as part of the construction of a viable cosmological model;
this difficulty is a specific example of the ultraviolet sensitivity described above.  Much of the progress in
realizing inflation in string theory in recent years has involved the incorporation of methods of moduli
stabilization.

Several ideas for the string theory origin of the inflaton have emerged.  Commonly-studied models rely on
inter-brane separations (brane inflation), geometric moduli, or axions.  Reviews which discuss various subsets
of early models can be found in \cite{HenryTye:2006uv,Kallosh:2007ig,McAllister:2007bg, Burgess:2007pz, Cline:2006hu, Quevedo:2002xw, Linde:2005dd}; other models have emerged
more recently. Some of these models involve {\it mechanisms} for inflation, {\it i.e.}~systematic arguments from
string theory that motivate or protect the near-constancy of the Hubble expansion rate.  Much work remains to systematically map
out the space of robust mechanisms and models.  At this early stage it is already clear that the phenomenology
of string inflation models is very rich: certain classes of current models readily produce tensor modes, others
predict strongly non-Gaussian perturbation spectra, while others yield cosmic superstrings, for example.
Moreover, in certain cases the couplings of the inflaton sector to our low-energy world can be specified,
leading to studies of reheating.

\vskip 8pt
\noindent
{\it Large-field inflation in string theory}

Because observable tensor modes are a powerful probe of Planck-scale physics (see \S\ref{sec:Bmodes}), it is worth
examining what string theory has to say about the large-field models of inflation in which detectable tensors
can arise.  In brief, it is too early to draw a definitive conclusion, but it appears that both large-field and
small-field models of inflation can be reasonably realized in string theory -- via rather different mechanisms
which will be distinguished by upcoming CMB observations.

The earliest mechanisms studied in stabilized string theory vacua were small-field models\footnote{For a review
of early models, see {\it e.g.}~\cite{McAllister:2007bg}.} ({\it e.g.}~generalizations of hybrid inflation using
D3-branes, such as \cite{KKLMMT}), but more recent constructions have revealed explicit mechanisms for
large-field models ({\it e.g.}~generalizations of chaotic inflation, such as \cite{Silverstein:2008sg, MSW}). Some
fine-tuning or significant specialization of the string compactification is inevitably involved in modeling
inflation in string theory; this amounts to making explicit the dependence on Planck-suppressed operators that
will be present in any scenario, string-theoretic or otherwise. However, the existing mechanisms are reasonably
generic in the sense that they each use common features of string compactifications, and models involving axions can be fully ``natural", in the sense of 't Hooft and Wilson. Although both possibilities
for the field range ($\Delta\phi> \Mp$ and $\Delta\phi< \Mp$) -- and different values of $r$ -- have been shown
to arise, the two cases are very different both microscopically and observationally, and detecting or
constraining tensor modes can therefore serve as a {\it powerful selection principle} for inflationary models in
string theory.

For a subset of candidate inflatons in string theory, one can prove that the field range is kinematically
constrained to be sub-Planckian. One example is D3-brane inflation \cite{Dvali:1998pa} in warped throats \cite{KKLMMT,DBI}, where
$\Mp$ and the field range are both constrained by the volume of the compactification \cite{BM}\footnote{Implications of field range limits for eternal inflation appear in \cite{Chen:2006hs}.}. (The case
studied in \cite{Becker:2007ui}\ of wrapped branes on tori and in warped throats is more subtle, with dynamical
backreaction effects becoming important.) Single axions \cite{Banks:2003sx, Svrcek:2006yi}, in the absence of
monodromy (see below), also have sub-Planckian field ranges.  In all such cases, the associated gravitational wave signal is
therefore small, independent of the structure of the potential.

For other candidate inflatons, the field range is kinematically unbounded.  For example, moduli spaces of string
vacua often contain angular directions which are lifted by additional ingredients such as fluxes and wrapped
branes that undergo {\it monodromy} -- not returning to their original potential energy when the system moves
around the angular direction.  This effect has been used to produce string-theoretic
realizations of large-field inflation with detectable tensor signatures, with the first explicit example involving repeated motion of a wrapped D-brane around a circle in a twisted torus \cite{Silverstein:2008sg}, and a further class of models involving repeated motion in the direction of a single axion \cite{MSW}.  An earlier idea was to consider two \cite{Kim:2004rp} or many axions \cite{Nflation}\ to increase the field range in a way
consistent with the sub-Planckian range of each individual axion.  The many axions also renormalize $\Mp$, and
moreover have reduced ranges at weak coupling and large volume, leading to a certain degree of difficulty in
constructing models within a computable regime. 
(The motion of multiple M5-branes \cite{Becker:2005sg} is a related possibility, but it remains necessary to incorporate the effects of moduli stabilization into the dynamics.)
In certain limits of other moduli spaces, kinematically large
field ranges may also occur, as in D3/D7 inflation \cite{DasguptaD3D7} on degenerate tori \cite{Haack:2008yb}, a case which also
provides an arena for concrete small-field inflationary model building, or in `fibre inflation' \cite{fiber}, in which the field range is geometrically limited but may still be large enough to give an ultimately detectable tensor signal.

To date no physical principle has been identified that explains why super-Planckian vevs should be favored or disfavored
in string theory, or which class of models is more generic, or which is more likely from the point of view of
initial conditions.  Future work may shed light on these questions, and it is important to recognize that
systematic exploration of the space of string inflation models has just begun. Moreover, as already emphasized,
genuinely predictive model-building in string theory has become possible due to the UV sensitivity of inflation
combined with the progress in CMB measurements; the data can therefore be used to distinguish the different
mechanisms.

\newpage
\section{Alternatives to Inflation}
\label{sec:alternatives}

As we discussed in \S\ref{sec:inflation}, inflation is a compelling solution of the
homogeneity, flatness, and monopole problems of the standard FRW
cosmology. In addition, quantum fluctuations during inflation provide an elegant mechanism to create the initial seeds for structure formation. One of inflation's most robust predictions is an adiabatic, nearly
scale-invariant spectrum of density perturbations.
This prediction is in
very good agreement with observations, especially considering
the recent evidence for the expected small deviation from
exact scale invariance \cite{WMAP5}.
However, it is disputable whether these observations can be considered a
proof that inflation did occur.
Clearly, a fair evaluation of the status of inflation requires the
consideration of alternatives, in the hope to find experimental distinctions
among different models.

In this Appendix we discuss the theoretical challenges and observational prospects of ekpyrotic/cyclic models
(\S\ref{sec:ekpyrosis}), string gas cosmology (\S\ref{sec:SGC}) and pre-Big Bang models (\S\ref{sec:PBB}). Our
discussion emphasizes the following two aspects:
\begin{enumerate}
\item Each alternative invokes novel and `incompletely understood' physics to solve the problems associated with the standard Big Bang cosmology.  This implies important theoretical challenges that have to be addressed carefully before the models mature into compelling alternatives to inflation.

\item Most or all of the alternatives to inflationary cosmology predict negligible tensors on CMB scales.
This strengthens the case for considering $B$-modes a ``smoking gun" of inflation.  It should be considered an amazing
opportunity to use CMB observations to constrain all known alternatives to inflation.
\end{enumerate}

\subsection{Ekpyrotic/Cyclic Cosmology}
\label{sec:ekpyrosis}

The ekpyrotic model~\cite{Khoury:2001wf, Khoury:2001bz} (see \cite{Lehners:2008vx} for a recent review) was proposed as an alternative to the
inflationary paradigm. Instead of invoking a short burst of
accelerated expansion from an energetic initial state, the
ekpyrotic scenario relies on a cold beginning and a subsequent phase of
slow contraction. This is then followed by a bounce which leads
to the standard expanding, decelerating FRW cosmology. Despite the stark
contrast in dynamics with respect to inflation, %the inflationary case,
the model is claimed to be equally successful at solving the flatness and homogeneity
problems of the standard Big Bang cosmology \cite{Khoury:2001wf, Khoury:2001bz}. In its cyclic extension~\cite{Steinhardt:2001st}, the ekpyrotic phase
occurs an infinite number of times --- our current expansion is to be followed by a contracting ekpyrotic phase, leading to a new hot Big Bang phase, and so on.
A critical evaluation of the ekpyrotic/cyclic scenario can be found in \cite{Kallosh:2001ai, Kallosh:2001du, Lyth:2001pf, Lyth:2001nv, Felder:2002jk, Linde:2002ws, Liu:2002ft, Horowitz:2002mw}.

\vskip 4pt
While the observational predictions of the ekpyrotic model outlined below do not rely on a particular realization
of the bounce, clearly the viability of the scenario hinges on whether a bounce can happen or not.
 Indeed, a bouncing phase requires the
violation of the null energy condition (NEC) and this is usually
associated with catastrophic instabilities. By a deformation of the
ghost condensate theory \cite{ArkaniHamed:2003uy}, an example of a
stable bounce was put forward in \cite{Creminelli:2006xe} and then
used in the new ekpyrotic scenario in
\cite{Buchbinder:2007ad,Creminelli:2007aq}. Although this model is consistent at the level of effective field theory, it is not clear whether it is possible to find a UV completion for it. As we already mentioned, according to \cite{Kallosh:2007ad}, this is a very important issue because the quantization of the new ekpyrotic theory, prior to the introduction of a UV cutoff and a UV completion, leads to a catastrophic vacuum instability.
Despite these theoretical challenges, we will
 highlight the phenomenologically distinct predictions of the ekpyrotic universe.
 %, keeping in mind the theoretical uncertainties that the model faces.

\vskip 4pt
As with inflation, during the contracting phase ekpyrosis relies on a scalar field $\phi$ rolling down a potential $V(\phi)$. Instead of
being flat and positive, however, here $V(\phi)$ must be steep, negative and nearly exponential in form. A
fiducial ekpyrotic potential is
\begin{equation}
V(\phi) = -V_0e^{-\phi/\sqrt{\tilde{\epsilon}} M_{\rm pl}}\,, \label{Vphi}
\end{equation}
where $\tilde{\epsilon} \ll 1$ is the ekpyrotic ``fast-roll" parameter. The Friedmann and scalar field equations
then yield a background scaling solution describing a slowly-contracting universe.

Two drawbacks prevented the original ekpyrotic scenario from becoming
a serious competitor to inflation: the lack of an explicit and
controllable model of a bouncing phase, and the problem of
the generation of a scale-invariant spectrum of perturbations.  The
two issues are clearly related, as the absence of a completely
explicit model prevented full control of the predictions.  The curvature perturbation on
uniform-density hypersurfaces, $\zeta$, has an unacceptably blue
spectrum in the contracting phase. If $\zeta$ remains constant during
the bounce, as can be shown under quite general conditions (see
\cite{Brandenberger:2001bs, Creminelli:2004jg} and references therein), the model is
experimentally ruled out.

The issue of scalar perturbations was addressed in the new ekpyrotic scenario \cite{Lehners:2007ac,Buchbinder:2007ad, Finelli:2002we, Creminelli:2007aq}. Due to an
entropy perturbation generated by a second scalar field, the curvature
perturbation $\zeta$ acquires a scale-invariant spectrum well before
the bounce, which, under the general assumption of
\cite{Creminelli:2004jg}, subsequently goes through the bounce
unscathed and emerges in the hot Big Bang phase with a scale-invariant
spectrum.

An important prediction of this new mechanism for generating
density perturbations in the ekpyrotic model is a substantial level of non-Gaussianity \cite{Creminelli:2007aq,Buchbinder:2007tw, Buchbinder_etal07, Koyama_etal07, Lehners:2007wc}. This
is a consequence of the self-interactions in the steep exponential
potential and of the mechanism of conversion to adiabatic
perturbations. As both these sources of non-Gaussianity act when the
modes are outside of the Hubble radius, the shape of non-Gaussianity is of
the local form. Although the level of
non-Gaussianity is rather model dependent, we can quote $f_{\rm
  NL}^{\rm local} >$ {\it few} as a rough lower bound.

Another generic prediction of ekpyrosis
%, which is in sharp contrast with inflation,
is the absence of a detectable signal of tensor modes
\cite{Khoury:2001wf,Boyle:2003km}.  Inflation predicts scale-invariant primordial gravitational waves, whereas ekpyrosis does not. Intuitively, this traces back to the difference
in dynamics: in the ekpyrotic background the curvature of the universe is slowly growing towards the bounce, and therefore the spectrum is not scale-invariant, but grows towards smaller scales. The tensor spectrum is highly blue ($n_t \approx 3$), resulting in an exponentially small primordial
gravitational wave amplitude for observable wavelengths.
A detection of tensor modes through CMB $B$-mode polarization would therefore rule out the ekpyrotic/cyclic scenarios.
%In its current realization the new ekpyrotic universe is clearly less compelling than inflation as many ingredients must be put together to give a sensible scenario and the central question of the bounce remains open.
Thus, independent of one's opinion about the theoretical status of ekpyrotic cosmology, it is encouraging that observations have the potential to falsify ekpyrosis.
%distinguish between ekpyrosis and inflation.

\subsection{String Gas Cosmology}
\label{sec:SGC}

String gas cosmology (SGC) is a model of early universe cosmology in which the universe initially begins in a
hot, dense state as suggested by Big Bang cosmology (see \cite{Battefeld:2005av} for a review).  All dimensions
are taken to be compact and initially at the string scale, where the theory exhibits a scale-inversion symmetry
$R \rightarrow 1/R$ believed fundamental to string theory \cite{Tseytlin:1991xk}.  In \cite{Brandenberger:1988aj} it was suggested that this was not only a natural initial state for the universe, but
that by taking into consideration the additional winding and momentum modes of the string gas (and their
interactions) one would generically expect three spatial dimensions to `decompactify,' leaving any other
dimensions stabilized at the string scale.  However, a further analysis of the dynamics suggested that
the model would require substantial fine-tuning for the dimensionality argument to work \cite{Easther:2004sd}.
Nonetheless it remains an intriguing avenue to explore other issues of early-time cosmology, including the
generation of primordial tensor mode perturbations.  In fact, it was recently claimed that a spectrum of nearly
scale-invariant cosmological perturbations could be produced from such a string gas phase and would be
observationally distinct from inflationary theory due to a blue-tilted tensor power spectrum
\cite{Nayeri:2005ck,Brandenberger:2006xi,Brandenberger:2006pr}.  However, subsequent work has shown that a
smooth transition between this string gas phase and the standard radiation phase would require either a
violation of the null energy condition (conjectured by some to be impossible in UV complete theories) or stabilization of the
dilaton field (which would destroy the desired scale-inversion symmetry) \cite{Kaloper:2006xw,Kaloper:2007pw}, and there are counter-claims in the literature that the spectrum of scalar perturbations appears to be very blue: instead of the ßat perturbations with $n_s= 1$ one finds a spectrum with $n_s= 5$ \cite{Kaloper:2006xw} (but see \cite{Brandenberger:2008nx}).
Thus, addressing these challenges is an important initial step before SGC can be considered a viable alternative to
inflation for producing primordial tensor perturbations.
%\footnote{Moreover, even if one takes for granted numerous assumptions made in \cite{Nayeri:2005ck,Brandenberger:2006xi,Nayeri:2006uy,Brandenberger:2006vv,Brandenberger:2006pr}, the resulting spectrum of scalar perturbations appears to be very blue: instead of the flat perturbations with $n_{s} = 1$ one finds a spectrum with $n_{s} = 5$ \cite{Kaloper:2006xw}. }

\subsection{Pre-Big Bang Cosmology}
\label{sec:PBB}

Initially motivated by SGC, the pre-Big Bang model (PBB) also attempts to invoke new symmetries and degrees of
freedom expected if our universe is correctly described by string theory \cite{Veneziano:1991ek} (see
\cite{Gasperini:2002bn} for a review).  However, unlike SGC and the conventional hot Big Bang theory, the PBB
model initially begins in a cold, empty state with zero curvature.  Then fluctuations drive the universe into a
period of dilaton-driven super (or pole) inflation, during which the expansion rate is increasing.  This phase
continues until the expansion rate reaches the string scale, at which time the effective theory description
breaks down and corrections from string theory become important.  It is then argued that because string theory
has a natural UV cutoff (set by the string length), new string physics should become important causing the
expansion rate to take a maximum value near the string scale.  From this phase, our radiation dominated universe
is then to emerge, with the PBB supplying adequate initial conditions for the hot Big Bang.

The key challenge for this model is describing the exit from the PBB phase to the radiation dominated universe.
Similar to the challenge facing SGC discussed above, it was shown in
\cite{Brustein:1994kw,Kaloper:1995ey,Kaloper:1995tu} that such an exit requires violation of the null energy
condition.  It has been argued in the literature that this might be reasonable given quantum gravity
corrections, but lack of parametric control and understanding of explicit time dependent solutions in string
theory make this an important open challenge.  Until the issue of the exit from the string phase is better
understood, both the PBB and SGC models lack predictability, making it too early to consider them for alternative
predictions to those of inflation for primordial tensor perturbations.\footnote{Furthermore, according to \cite{Kaloper:1998eg, Buonanno:2001nb}, the PBB scenario does not solve the horizon, flatness and isotropy problems. Until these problems are resolved, it too early to consider the PBB theory a consistent alternative to inflation.}

\newpage
\section{Fisher Methodology}
\label{sec:Fisher}

In Section \ref{sec:CMBPol} we performed a standard Fisher analysis to forecast the errors on inflationary parameters derived from a future satellite experiment. In this Appendix we give details of the Fisher methodology and define the survey parameters of two realistic experimental configurations. Our treatment 
%HVP follows closely 
parallels the approach of Ref.~\cite{LiciaFisher}; we give the relevant equations and definitions for completeness, but direct the reader to Ref.~\cite{LiciaFisher} for further details.

\subsection{Likelihood Function and Parameter Errors}
The Fisher information matrix \cite{Fisher:1935} is defined as
\beq
F_{ij} \equiv \left. \left \langle - \frac{\partial^2 \ln {\cal L}}{\partial \alpha_i \partial \alpha_j} \right \rangle \right|_{{\bf \alpha} = {\bf \bar \alpha}}\, ,
\eeq
where $\ln {\cal L}$ is the likelihood function and $\alpha_i$ denote model parameters.
We consider the following vector of cosmological parameters
$\alpha \equiv \{ r, n_s, n_t, \alpha_s, A_s,
 \tau, \omega_b, \omega_c, h, \Omega_k\}$.
 The Cramer-Rao inequality for the minimum standard deviation of a parameter $\alpha_i$ is
 \beq
 \sigma_{\alpha_i} \ge (F^{-1})_{ii}^{1/2}\, .
 \eeq
 %HVP
 For the fiducial set of parameters we use 
\bea
\bar \alpha &\equiv& \{ r = 0\ {\rm or} \ 0.01\ {\rm or} \ 0.001, n_s=0.963, n_t = - r/8, \alpha_s =0, A_s= 2.41 \times 10^{-9}, \nonumber \\
&& \tau = 0.087 , \omega_b = 0.02273 , \omega_c = 0.1099, h=0.72, \Omega_k =0 \}\, .
\eea
The forecasted errors do not depend strongly on the chosen fiducial model,  
except in the choice of $r$ (because of cosmic variance), since the signal primarily comes from large angular scales.
For this reason we will report results for different fiducial cases where we vary $r$ while keeping the other parameters constant (except adjusting $n_t$ via the consistency relation $n_t=-r/8$).
The pivot scale for $r$, $n_t$, $A_s$, $n_s$ and $\alpha_s$ is $k_\star = 0.05$ Mpc$^{-1}$.

%HVP
For data with partial sky coverage, experimental noise, and foreground subtraction residuals, the likelihood function can be approximated as:
\bea
-2 \ln {\cal L} &=& \sum_\ell (2 \ell+1) \left\{ f_{\rm sky}^{BB } \ln \left( \frac{{\cal C}_\ell^{BB}}{\hat {\cal C}_\ell^{BB}}\right) + \sqrt{f_{\rm sky}^{TT} f_{\rm sky}^{EE}}\,  \ln \left( \frac{{\cal C}_\ell^{TT} {\cal C}_\ell^{EE} -({\cal C}_\ell^{TE})^2}{\hat {\cal C}_\ell^{TT} \hat {\cal C}_\ell^{EE} - (\hat {\cal C}_\ell^{TE})^2}\right) 
\right.   \nonumber \\
&& \hspace{2cm}+\sqrt{f_{\rm sky}^{TT} f_{\rm sky}^{EE}}\
\frac{\hat {\cal C}_\ell^{TT} {\cal C}_\ell^{EE}  + {\cal C}^{TT}_\ell \hat {\cal C}_\ell^{EE} - 2 \hat {\cal C}_\ell^{TE} {\cal C}_\ell^{TE} }{{\cal C}_\ell^{TT} {\cal C}_\ell^{EE} -({\cal C}_\ell^{TE})^2} \nonumber \\
&& \hspace{2cm} \left. +f_{\rm sky}^{BB} \frac{\hat {\cal C}_\ell^{BB}}{{\cal C}_\ell^{BB}} - 2 \sqrt{f_{\rm sky}^{TT} f_{\rm sky}^{EE}} - f_{\rm sky}^{BB}
\right\}\, .
\eea
%HVP
Here, ${\cal C}_\ell^{XY}(\alpha_i)$ are the theoretical angular power spectra, with $X, Y = \{T, E, B\}$. The estimator of the measured angular power spectra, $\hat {\cal C}_\ell^{XY}$, includes a contribution from noise, and the fraction of the sky used for cosmological analysis is $f_{\rm sky}^{XY}$.
%LV
The scaling of the errors with $f_{\rm sky}$ adopted here is valid only if $\sim 70$\% or more of the sky can be used for cosmological analysis. Should the foreground contamination impose more drastic sky-cuts there will be a
%HVP
significant error degradation -- see {\it e.g.}~\cite{Amarie:2005in}. Here, we assume that 80\% of the sky can be used for cosmological analysis. 

The estimated errors also assume that 
there is no effect of leakage of power from $E$ to $B$-modes. By using a large fraction of the sky, the errors on the measured polarization will vary spatially when foreground uncertainty is included, resulting in additional contamination 
of the $B$-mode signal. The analysis of \cite{Amarie:2005in} suggests that this would inflate error bars over those presented here, although initial studies in \cite{DunkleyFGs} indicate that the effect should be small for models with $r=0.01$. For further discussion see Ref.~\cite{DunkleyFGs} .

We treat the weak lensing $B$-mode signal as a Gaussian noise in the Fisher matrix. In all cases we do not assume that lens-cleaning (delensing) can be implemented. Should delensing be possible the constraints will improve.

Residual foregrounds introduce a bias ({\it i.e.}~a systematic error) to constraints on cosmological parameters while noise just introduces a statistical error. We attempt to include both these effects in the reported confidence regions, despite the very different natures of these two terms. To estimate their effects on the final constraints on cosmological parameters, we adopted the ansatz of \cite{LiciaFisher} (this ansatz has been found to reproduce the results of simulations of \cite{DunkleyFGs}). The {\it systematic} uncertainty on the constraints introduced by residual foregrounds can be appreciated by comparing forecasts for the case with no foregrounds (only statistical errors) and the case with foregrounds (with statistical and systematic errors).
The theoretical power spectra ${\cal C_\ell}$ are therefore split into a primordial contribution $C_\ell$,  a contribution from instrumental noise $N_\ell$, and a %HVP
residual foreground term $F_\ell$, which will also be treated as a noise term:
\beq
{\cal C}_\ell = C_\ell + N_\ell + F_\ell\, .
\eeq
The primordial signal $C_\ell$ is computed using the publicly available Code for Anisotropies in the Microwave Background (CAMB)~\cite{CAMB}. We now describe our models for the noise $N_\ell$ and the residual foregrounds $F_\ell$.  %LV

\vskip 8pt
\noindent
{\it Instrumental noise}

%{\bf make sure following discussion corresponds to definitions in the tables.} 
%For our forecasts 
We assume Gaussian beams, where  $\Theta_{\rm FWHM}$ denotes the FWHM of a beam and 
$\sigma_b = 0.425\, \Theta_{\rm FWHM}$.
The noise per multipole is
$n_0 = \sigma_{\rm pix}^2 \Omega_{\rm pix}$,
where $\Omega_{\rm pix}$ and $\sigma_{\rm pix}^2$
are the pixel (beam) solid angle and the variance per pixel, respectively.
In terms of the sky fraction $f_{\rm sky}$, the number of pixels $N_{\rm pix}$, the detector sensitivity $s$, the number of detectors $N_{\rm det}$ and the integration time $t$, we find
\beq
\Omega_{\rm pix} = \Theta_{\rm FWHM} \times \Theta_{\rm FWHM} = 4 \pi \frac{f_{\rm sky}}{N_{\rm pix}}\, , \qquad \sigma_{\rm pix} = \frac{s}{\sqrt{N_{\rm det} t}}\, .
\eeq
With these definitions the noise bias becomes
\beq
N_\ell = \frac{\ell(\ell+1)}{2\pi} \, n_0 \, \exp(\ell^2 \sigma_b^2)\, .
\eeq
For $N_{\rm chan}$ frequency channels the noise bias is reduced by a factor of $1/{N_{\rm chan}}$.
We therefore treat the noise bias as a function of $\{\Theta_{\rm FWHM}, \sigma_{\rm pix}, N_{\rm chan}\}$.

If the different channels have different noise levels we need to generalize the above considerations.
The optimal channel combination then is
\beq
C_\ell = \frac{\sum_{i, j \ge i } w_{ij} C^{ij}_\ell}{\sum_{i,j} w_{ij}}\, , \qquad w_{ij} \equiv \left[N^i_{\rm det} N^j _{\rm det} \frac{1}{2}(1+\delta_{ij})\right]^{-1}\, ,
\eeq
where $i,j$ label the different frequency channels, and $N^i_{\rm det}$ are the number of detectors in frequency channel $\nu_i$.
The resulting noise is given by 
\beq
\left[N_{\rm eff}^{XY}(\ell)\right]^{-2} = \sum_{i \geq j} \Bigl[\left(n_{{\rm fg},i}^{XY}(\ell)+n_i^{XY}(\ell)\right)\left(n_{{\rm fg},j}^{XY}(\ell)+n_j^{XY}(\ell)\right) \frac{1}{2} (1+\delta_{ij})\Bigr]^{-1}\, ,
%\left[ \sum_{i, j \ge i} \frac{1}{N_{{\rm eff},i} N_{{\rm eff}, j} \frac{1}{2} (1+\delta_{ij})}\right]^{-1/2}\, .
\eeq
%[...]
where $i,j$ runs though the channels,  $n_i$ is the instrumental noise bias ({\it i.e.} convolved with the beam) of channel $\nu_i$, and $n_{\rm fg}$ is given by the sum of $n_{\rm dust}+n_{\rm synch}$,
\begin{equation}
n_{{\rm dust,synch},i}^{XY}(\ell)=C_{{\rm residual},i}^{XY}(\ell)+\frac{n_i^{XY}(\ell)}{N_{\rm chan}(N_{\rm chan}-1)/4}\left(\frac{\nu_i}{\nu_{\rm ref}}\right)^{2\alpha}  \, ,
\end{equation}
where $N_{\rm chan} $ is the total number of channels used, and the reference channel $\nu_{\rm ref}$ is the highest and lowest frequency channel  {\it included in the cosmological analysis} for dust and synchrotron respectively. The frequency dependence $\alpha$ for the foreground under consideration is defined in Table \ref{tab:fore}. We define the frequency channels and their associated noise levels for two realistic CMB satellites in \S\ref{sec:specs}.

\vskip 8pt

%\newpage
\noindent
{\it Foreground residuals}

%HVP
Details of the foreground subtraction are discussed in a separate publication \cite{DunkleyFGs}. As described there,
foreground removal is most effectively and optimally carried out in pixel space. Here, we {\it assume} that foreground subtraction can be done correctly down to a given level ({\it i.e.} 1\% in the $C_\ell$ for the optimistic case  and 10\% in the $C_\ell$ for a more pessimistic case). We then use foreground models in harmonic space to propagate the effects of foreground subtraction residuals into the resulting error-bars for the cosmological parameters. Actual cleaning of foregrounds should not be carried out in harmonic space.

%We here describe foreground cleaning in harmonic space. Foreground removal pixel by pixel is described in Dunkley et al.~\cite{DunkleyFGs}.
We focus on the two dominant polarized foregrounds: synchrotron (S) and dust (D).
The residual Galactic contamination is
\beq
F_\ell(\nu) = \sum_{{\rm fg} = S, D} C_\ell^{{\rm fg}, XY}(\nu) \,  \sigma^{{\rm fg}, XY} + N_\ell^{{\rm fg}, XY}(\nu; \nu_{\rm tp})\, .
\eeq
%HVP
Here,  $X,Y$ stand for $\{E,B\}$, $C_\ell^{\rm fg}(\nu)$ is our model for the power spectrum of the synchrotron and dust signals, $\sigma^{\rm fg}$ is the assumed residual (1\% for the optimistic case, 10\% for the pessimistic case), and $N_\ell^{\rm fg}(\nu; \nu_{\rm tp})$ is the noise power spectrum of the foreground template map (created at template frequency $\nu_{\rm tp}$),
%HVP , and $N_\ell^{\rm fg}$ denotes the template noise 
as foreground templates are created by effectively taking map differences and thus are somewhat affected by the instrumental noise. %LV

%[...]

\vskip 6pt
For the scale-dependence of the synchrotron signal we assume
\beq
C_\ell^{S, XY} (\nu) = A_S \left( \frac{\nu}{\nu_0} \right)^{2\alpha_S} \left( \frac{\ell}{\ell_0} \right)^{\beta_S}\, ,
\eeq
%LV
where $\alpha_S = -3$, $\beta_S=-2.6$, $\nu_0 = 30$ GHz, and $\ell_0 = 350$, $A_S=4.7\times 10^{-5}$ $\mu$K$^2$ (corresponding to $0.91 \mu$K$^2$ in $\ell (\ell+1)/(2\pi)C_{\ell}$). This choice matches the synchrotron emission at 23 GHz observed and parameterized by  {\sl WMAP}  \cite{WMAPPage07},  and agrees with the  {\sl DASI}  \cite{DASILeitch} measurements.

%We will make different assumptions for the amplitude $A^S$.

For dust we assume
\beq
C_\ell^{D,XY}  (\nu) = p^2 A_D  \left( \frac{\nu}{\nu_0} \right)^{2\alpha_D} \left( \frac{\ell}{\ell_0} \right)^{\beta_D^{XY}}
\left[ \frac{e^{h \nu_0/kT} -1}{e^{h \nu/kT} -1}\right]^2\, ,
\eeq
where $\alpha_D = 2.2$, %LV
 $\nu_0 = 94$ GHz, 
$\ell_0=10$, $A_D=1.0$ $\mu$K$^2$, $\beta_D^{XY} =-2.5$.
 %$\ell_0 = 900$, $A_D=1.2\times 10^{-5}$ $\beta_D^{XY} =-2.6$.
The intensity of the dust, given by $A_D$, is estimated to be $1.0$ $\mu$K$^2$ at $\ell_0=10$ from the {\sl IRAS} dust map extrapolated to 94 GHz by Ref.~\cite{FDS}. The dust polarization fraction, $p$, is estimated to be 5\%, motivated by the fact that even a very weak Galactic magnetic field of $\sim$ 3 $\mu$G already gives a 1\% polarization \cite{Padoan:2001} and that  {\sl Archeops} \cite{Benoit_Archeops} finds an upper limit for the diffuse dust component of a 5\% dust polarization fraction at $\ell = 900$. This is also consistent with {\sl WMAP} observations \cite{WMAPPage07, Gold:2008kp}, and with the {\sl Planck} sky model that has been derived from these observations \cite{DunkleyFGs}. However, including possible depolarization effects due to the Galactic magnetic field, there is around an order of magnitude uncertainty in the observed dust polarization fraction, which could reasonably lie in the approximate range $\sim$1\% to $\sim$10\%. For more discussion see \cite{DunkleyFGs} and \cite{FraisseFGs}. Recent studies by \cite{Draine:2008hu} suggest an upper limit of $p \sim$15\%. The normalization used here yields $\ell (\ell+1)/(2\pi)C_{\ell} \sim 0.04$ $\mu$K$^2$ at $\ell_0=10$ for $p =5$\%. 

% {\bf add some words of explanation or a reference where this is explained in the fg paper}.
%LV
The $\ell$-dependence, however,  is quite uncertain. The slope for polarization may not be the same as that for the temperature, $\beta_D^{TT} =-2.5$. The work of Refs.~\cite{Lazarian_Prunet:2002, PrunetLazarian:1999} seems to indicate that  any modulation of the  density field by  the magnetic field orientation would  always flatten the spectrum. Measurements of  starlight polarization \cite{Fosalba_starlight} indicate
$\beta_D^{EE} = -1.3$, $\beta_D^{BB} =-1.4$, $\beta_D^{TE} =-1.95$.
We will  thus also examine in some cases how constraints improve  for   a more optimistic case with the flatter spectrum of $\beta_D^{EE} = -1.3$, $\beta_D^{BB} =-1.4$, $\beta_D^{TE} =-1.95$ (foreground option B). In this case $A_D=1.2\times 10^{-4}$~$\mu$K$^2$ at $\ell_0 = 900$, $\nu_0 = 94$ GHz.

\vskip 6pt
%The  noise power spectrum of the template map is computed as
%\beq
%N_\ell^{{\rm fg}, XY}(\nu; \nu_{\rm tp}) = N^{{\rm fg}, XY}_\ell \left(\frac{\nu }{\nu_{\rm tp}} \right)^{2 \langle \alpha_{\rm fg} \rangle}\, .
%\eeq
%HVP
%For $N^{{\rm fg}, XY}_\ell$ we assume a white noise spectrum equal to the noise spectrum of one of the channels reduced by $\frac{1}{2} N_{\rm chan}(N_{\rm chan}-1)/2$. In particular, for dust we use the highest frequency channel {\it included in the cosmological analysis} and for synchrotron the lowest.

We summarize the foreground parameterization\footnote{The Greek letter $\beta$ is used in this text as in \cite{LiciaFisher}, to quantify the angular dependence of the foreground power spectra. We note that in many foreground analyses this letter is used to quantify the frequency dependence of the foregrounds. The companion {\sl CMBPol} document on Foreground Removal \cite{DunkleyFGs} uses $m$ in place of $\beta$ for the angular dependence of the foregrounds, and $\beta$ in place of $\alpha$ for the frequency dependence.} in Table \ref{tab:fore}, again emphasizing that the simple foreground models above are only used for the purpose of propagating the effects of foreground residuals into the estimated uncertainties on the cosmological parameters. 

\begin{table}[h!]
\begin{center}
\caption{Assumptions about foreground subtraction.}
\vspace{0.2cm}
\begin{tabular}{ || l | p{3.cm}| p{3.cm}| p{3.cm}  || }
\hline
\hline
 parameter & {\bf synchrotron} & {\bf dust}  & {\bf dust B} \\
 \hline
 \hline
 $A_{S,D}$ &  $4.7\times 10^{-5}$ $\mu$K$^2$ & $1.0$ %1.2\times 10^{-5} $
 $\mu$K$^2$ & $1.2\times 10^{-4} $ $\mu$K$^2 $\\
$p$ (dust only)& -- & 5\%& 5\%\\
$\nu_0$ &  30 GHz  & 94 GHz & 94 GHz\\
$\ell_0$ &  350 & 10%900
& 900\\
$\alpha$ & $-3$ & 2.2 & 2.2 \\
$\beta^{EE}$&$-2.6$ & $-2.5$ & $-1.3$\\
$\beta^{BB}$&$-2.6$ & $-2.5$ & $-1.4$\\
$\beta^{TE}$&$-2.6$ & $-2.5$ &$-1.95$\\
\hline
subtraction & & & \\
{\bf Opt}imistic & 1\% &1\%& 1\%\\
{\bf Pess}imistic & 10\% &10\%& 10\%\\
  \hline
  \hline
\end{tabular}
\label{tab:fore}
\end{center}
\end{table}

%LV

\newpage
\subsection{Ideal Experiment}

For comparison with the (semi-)realistic satellite experiments described below, we here quote for reference the parameter constraints derived from an {\it ideal} experiment.
The reference experiment covers the full sky ($f_{\rm sky}=1$), with {\it no} instrumental noise ($N_\ell =0$) and {\it no} foregrounds ($F_\ell =0$) up to $\ell_{\rm max} = 1500$.
Results are shown in Table \ref{tab:ideal} and are taken directly from Ref.~\cite{LiciaFisher}.

\begin{table}[h]
\begin{center}
\begin{tabular}{|| c | c | c | c | c | c || }
\hline
\hline
\hspace{2cm} & {\small $r$} & {\small $\Delta r$} & {\small $\Delta n_s$} & {\small $\Delta n_t$} & {\small $\Delta \alpha_s$} \\
\hline
\hspace{2cm} & {\small 0.01} & {\small 0.001} & {\small 0.0017} & {\small 0.056} & {\small 0.003} \\
{\small L} & {\small 0.03} & {\small 0.0027} & {\small 0.0017} & {\small 0.047} & {\small 0.0036} \\
  & {\small 0.1} & {\small 0.006} & {\small 0.002} & {\small 0.035} & {\small 0.0035} \\
  \hline
\hspace{2cm} & {\small 0.01} & {\small 0.000021} & {\small 0.0021} & {\small 0.0019} & {\small 0.0038} \\
{\small NL} & {\small 0.03} & {\small 0.000063} & {\small 0.0021} & {\small 0.0019} & {\small 0.0038} \\  
  \hline
  \hline
\end{tabular}
\caption{1$\sigma$ errors for an ideal experiment, including lensing (L), and with no lensing (NL) \cite{LiciaFisher}.}
\label{tab:ideal}
\end{center}
\end{table}

%\newpage
%\addtocontents{toc}{\SkipTocEntry}
\subsection{Realistic Satellite Experiments}
\label{sec:specs}

We forecast the expected observational constraints on inflationary parameters from different types of space-based experiments.
For each experiment we specify the spectral range and resolution, the spatial resolution, the collecting area, the field of view, as well as assumptions about foreground subtraction and instrumental noise (see Tables \ref{tab:fore}, \ref{tab:LCspecs} and \ref{tab:2mspecs}). When computing forecasts in the presence of foregrounds we use only the five central frequencies of each experimental setup. This is motivated by the fact that, effectively, the statistical power of  the highest and lowest frequencies is entirely used to characterize the foregrounds themselves. We present our results in Tables   \ref{Tab:forecasts_cosm} and \ref{Tab:forecasts_r}.

\begin{table}[h!]
\begin{center}
\begin{tabular}{|| l | l | l    ||}
\hline
\hline
{\small Freq (GHz)} & {\small beam FWHM (arcmin)} & {\small $\delta T$ ($\mu$K arcmin)} \\
\hline
\hline
{\small 30} & {\small 155} & {\small 44.12} \\
\hline
{\small 40} & {\small 116} & {\small 15.27} \\
{\small 60} & {\small 77} & {\small 8.23} \\
{\small 90} & {\small 52} & {\small 3.56} \\
{\small 135} & {\small 34} & {\small 3.31} \\
{\small 200} & {\small 23} & {\small 3.48} \\
\hline
{\small 300} & {\small 16} & {\small 5.94} \\
  \hline
  \hline
\end{tabular}
\caption{Experimental specifications for the low-cost (EPIC-LC) {\sl CMBPol} mission. The highest and lowest frequencies are excluded from the analysis when we consider the realistic case with  foregrounds, but included in the idealized case of no foregrounds. $\delta T$ is for the Stokes $I$ parameter; the corresponding sensitivities for the Stokes $Q$ and $U$ parameters are related to this by a factor of $\sqrt{2}$.
}
\label{tab:LCspecs}
\end{center}
\end{table}

%\addtocontents{toc}{\SkipTocEntry}
%\subsubsection{Epic-2m: Experimental Specifications}
\begin{table}[h!]
\begin{center}

\begin{tabular}{|| l | l | l    ||}
\hline
\hline
{\small Freq (GHz)} & {\small beam FWHM (arcmin)} & {\small $\delta T$ ($\mu$K arcmin)} \\
\hline
\hline
{\small 30} & {\small 26} & {\small 13.58} \\
\hline
{\small 45} & {\small 17} & {\small 5.85} \\
{\small 70} & {\small 11} & {\small 2.96} \\
{\small 100} & {\small 8} & {\small 2.29} \\
{\small 150} & {\small 5} & {\small 2.21} \\
{\small 220} & {\small 3.5} & {\small 3.39} \\
\hline
{\small 340} & {\small 2.3} & {\small 15.27} \\
  \hline
  \hline
\end{tabular}
\caption{Experimental specifications for the mid-cost (EPIC-2m) {\sl CMBPol} mission. The highest and lowest frequencies are excluded from the analysis when we consider the realistic case with  foregrounds, but included in the idealized case of no foregrounds. $\delta T$ is for the Stokes $I$ parameter; the corresponding sensitivities for the Stokes $Q$ and $U$ parameters are related to this by a factor of $\sqrt{2}$.}
\label{tab:2mspecs}
\end{center}
\end{table}

%\newpage
\subsection{Forecasts}
Here we report complete forecast tables.   In Table \ref{Tab:forecasts_cosm} we give the constraints on other cosmological parameters including the
 %HVP
scalar spectral index $n_s$ and its running $\alpha_s$ for the mid-cost set up (EPIC-2m).  
Table  \ref{Tab:forecasts_r} shows the forecasts for the parameters $r$ and
 $n_t$, 
 %HVP
 the constraints on which completely rely
 on the $B$-mode polarization measurements.
In the absence of foregrounds {\sl CMBPol} can reach constraints similar to those of an ideal experiment for $r \gtrsim 0.001$.  These results illustrate the importance of accurate foreground subtraction: the key conclusion to be drawn from     Table  \ref{Tab:forecasts_r} is that our ability to detect a primordial tensor background with $r\lesssim 0.01$   depends critically on the detailed properties of the polarized foregrounds that exist in the universe, and 
our ability to subtract them at the $1\%$ level or better in the power ({\it i.e.} $10\%$ level in the amplitude).
            
The forecasts in Table \ref{Tab:forecasts_r} are averages of the results of two independent implementations of the Fisher algorithm (\cite{Adshead:2008vn, LiciaFisher}).  In low signal-to-noise regimes the Fisher approach is unlikely to provide reliable forecasts -- for these situations we do not give quantitative forecasts.   In the absence of foregrounds, {\sl CMBPol} will provide constraints comparable to those of an ideal experiment for $r \gtrsim 0.001$.  Recall that if we impose the inflationary consistency condition, the tensor spectrum  is specified by just one parameter, $r$, and this single parameter can be tightly constrained.  If we do not impose this prior, fits to $r$ and $n_t$ permit only weak null tests of the consistency condition.  In particular, if $r \lesssim 0.01$, $|n_t|$ is very much smaller than the forecast constraint -- even for EPIC-2m and perfect foreground subtraction.

\newpage

\begin{table}[h!]
\begin{center}
 \begin{tabular}{||c|c|c|c||}
 \hline
 \hline
              & no FG                                  & Opt FG            &    Pess FG \\
              \hline
$\Delta w_b$       & $5.8\times 10^{-5}$  &$5.9\times 10^{-5}$    &  $5.9\times 10^{-5}$  \\
$\Delta w_c$       &  $0.00020 $                & $0.00022$                  & $0.00030$ \\
$\Delta \exp(-2\tau) $&$0.0028$             &  $0.0031$                    &$0.0046$ \\
$\Delta h$            & $0.0010$                   &  $0.0011$                    & $0.0014$\\
$\Delta (A_s/2.95\times10^{-9})$       &  $0.0029$                  &  $0.0031$                     &$0.0041$\\
$\Delta n_s$        & $0.0016$                   & $0.0016$                     &$0.0017$\\
$\Delta \alpha_s$&$0.0036$                   & $0.0036$                     &$0.0036$\\
\hline
\hline
\end{tabular}
\caption{Forecasted constraints on cosmological parameters, applying the consistency relation. 
We only report forecasts for the EPIC-2m set up. Error bars for EPIC-LC are comparable with those from {\sl Planck}. }
\label{Tab:forecasts_cosm}
\end{center}
\end{table}

\begin{table}[h!]
\begin{center}
\begin{tabular}{|| c | l | c  c | c || c  c | c ||}
\cline{3-8}
\cline{3-8}
\cline{3-8}
          \multicolumn{2}{c}{} &  \multicolumn{3}{|c||}        { EPIC-LC} &  \multicolumn{3}{c||}  {EPIC-2m} \\
%            &                     &            LC                       &     &                            &2m                               &                   &                          \\ 
            \hline
            \hline
            &                     & $ \Delta r$              & $\Delta n_t $& $\Delta r$         & $\Delta r $                 &$\Delta n_t$    &$\Delta r$    \\
\hline
no FG   & $r=0$         &           \text{--}        &        \text{--}      &          \text{--}        & $5.0\times 10^{-5} $&$0.20$ & $3.3 \times 10^{-5} $ \\
            & $r=0.001$ &$6.9\times 10^{-4} $&$ 0.18 $&$ 2.3\times 10^{-4} $&$ 5.7\times 10^{-4} $&$ 0.17 $&$ 2.1\times 10^{-4}$ \\
                        & $r=0.01$&$    0.0017$ & $0.076$ & $5.4\times 10^{-4}$ & $0.0015$ &$ 0.072 $& $4.8\times 10^{-4}$ \\
\hline            
%Opt fg   & $r=0$         &                                    &              &                                     & $$                                 &$$          & $ $ \\
 Opt FG  & $r=0.001$ & $ 0.0022 $&$ 1.1 $&$ 5.2\times 10^{-4} $&$ 0.0018 $&$ 0.93 $&$ 4.1\times 10^{-4}$   \\
             & $r=0.01$   &$ 0.0029 $&$ 0.15$&$ 6.6\times 10^{-4} $&$0.0025 $&$ 0.13 $&$ 5.4\times 10^{-4}$ \\
\hline
\       
%Pess fg & $r=0$         &                                    &              &                                     & $$                             &$$             & $$ \\
Pess FG & $r=0.001$ & $\text{--} $&$ \text{--} $&$ 8.0\times 10^{-4} $&$ \text{--} $&$ \text{--} $&$ 6.3\times 10^{-4} $ \\
             & $r=0.01$    &  $\text{--} $&$ \text{--} $&$  9.2\times 10^{-4} $&$ 0.0049$ &$ 0.28$ & $7.4\times 10^{-4}$ \\
\hline
 Opt FG B& $r=0.001$ &$8.6\times 10^{-4} $&$ 0.26 $&$ 3.5\times 10^{-4} $&$ 6.7\times 10^{-4} $&$ 0.22 $&$ 3.0\times 10^{-4} $ \\
             & $r=0.01$    &$0.0018 $&$ 0.085 $&$ 6.0\times 10^{-4} $&$ 0.0016 $&$ 0.078 $&$ 5.0\times 10^{-4} $ \\
\hline 
Pess FG B &$r=0.001$& $\text{--} $&$ \text{--} $&$ 6.4\times 10^{-4} $&$0.0016 $&$ 0.81 $&$ 5.2\times 10^{-4}$ \\
             & $r=0.01$    &$ 0.0029 $&$ 0.15 $&$ 7.8\times 10^{-4} $&$ 0.0025 $&$ 0.14 $& $6.5\times 10^{-4} $\\
\hline
\hline
\end{tabular}
\caption{Forecasted constraints on tensor modes. Results are presented for EPIC-LC and EPIC-2m with optimistic and pessimistic foreground assumptions and two different models for the scale-dependence of the dust polarization. Cases where there was no predicted detection and the Fisher approach is unreliable are denoted by dashes, and a quantitative forecast is not presented for these cases.}
\label{Tab:forecasts_r}
\end{center}
\end{table}

\newpage
\section{List of Acronyms}
\label{sec:acronyms}

\begin{table}[h!]
\begin{center}
\begin{tabular}{|| c | c ||}
\hline \hline
 {\small \bf Acronym} & {\small \bf Definition and Comments}   \\
\hline
{\small CMB} & {\small Cosmic Microwave Background} \\
\hline
{\small CDM} & {\small Cold Dark Matter}  \\
\hline
{\small $\Lambda$CDM} & {\small Concordance Cosmology}\\
\hline
 {\small ISW} & {\small Integrated Sachs-Wolfe Effect}   \\
\hline
{\small SZ} & {\small Sunyaev-Zel'dovich Effect} \\
\hline
{\small BAO} & {\small Baryon Acoustic Oscillations} \\
\hline
{\small LSS} & {\small Large-Scale Structure}  \\
\hline
{\small SN} & {\small Supernovae}  \\
\hline
 {\small GR} & {\small General Relativity} \\
\hline
{\small FRW} & {\small Friedmann-Robertson-Walker}  \\
 \hline
 {\small SVT} & {\small Scalar-Vector-Tensor}  \\
 \hline
{\small QM} & {\small Quantum Mechanics}  \\
 \hline
 {\small SM} & {\small Standard Model}  \\
 \hline
 {\small EFT} & {\small Effective Field Theory}  \\
 \hline
  {\small QFT} & {\small Quantum Field Theory}  \\
 \hline
{\small UV} & {\small Ultraviolet}  \\
 \hline
 {\small TeV} & {\small $10^{12}$ eV; energy scale probed by LHC}  \\
 \hline
 {\small GUT} & {\small Grand Unified Theory}  \\
 \hline
{\small PNGB} & {\small Pseudo-Nambu-Goldstone-Boson}  \\
 \hline
 {\small $TT$} & {\small Temperature Autocorrelation}  \\
 \hline
{\small $TE$} & {\small Temperature-Polarization Crosscorrelation}  \\
 \hline
{\small $EE$} & {\small $E$-mode Autocorrelation}  \\
 \hline
{\small $BB$} & {\small $B$-mode Autocorrelation}  \\
 \hline
 {\small C.L.} & {\small Confidence Limit}  \\
  \hline
 {\small FWHM} & {\small Full Width at Half Maximum}  \\
\hline
 {\small FG} & {\small Foreground}  \\
 \hline
 {\small Pess FG} & {\small Pessimistic Foreground Level}  \\
 \hline
 {\small Opt FG} & {\small Optimistic Foreground Level}  \\
 \hline
 \hline
\end{tabular}
\caption{ \label{table:acc} Common acronyms in physics and cosmology.}
\end{center}
\end{table}

\begin{table}[h!]
\begin{center}
\begin{tabular}{|| c | c ||}
\hline
\hline
\multicolumn{2}{|| l ||}{\bf \small Space-based} \\
\hline
% {\small \bf Accronym} & {\small \bf Definition and Comments}   \\
%\hline
{\small COBE} & {\small Cosmic Background Explorer} \\
\hline
{\small RELIKT-1} & {\small --} \\
\hline
{\small WMAP} & {\small Wilkinson Microwave Anisotropy Probe}  \\
\hline
{\small Planck} & {\small Planck Satellite}\\
\hline
 {\small SDSS} & {\small Sloan Digital Sky Survey}   \\
\hline
{\small 2dFGRS} & {\small Two Degree Galaxy Redshift Survey} \\
\hline
{\small CMBPol} & {\small Future CMB Polarization Satellite} \\
\hline
{\small EPIC} & {\small Experimental Probe of Inflationary Cosmology}  \\
\hline
{\small EPIC-LC} & {\small EPIC-low cost}  \\
\hline
 {\small EPIC-2m} & {\small EPIC-mid cost} \\
\hline
 {\small SPOrt} & {\small Sky Polarization Observatory} \\
 \hline
  {\small BBO} & {\small Big Bang Observer}  \\
  \hline
\hline
\multicolumn{2}{|| l ||}{\bf \small Balloon} \\
\hline
 {\small BOOMERanG} & {\small --}\\
\hline
 {\small Archeops} & {\small --}\\
\hline
 {\small MAXIMA} & {\small Millimeter Anisotropy eXperiment IMaging Array}\\
\hline
{\small SPIDER} & {\small --}  \\
 \hline
 {\small EBEX} & {\small $E$ and $B$ Experiment}  \\
\hline
\hline
\multicolumn{2}{|| l ||}{\bf \small Ground-based} \\
\hline
{\small ACT} & {\small Atacama Cosmology Telescope}  \\
 \hline
 {\small SPT} & {\small Southpole Telescope }  \\
 \hline
{\small AMI} & {\small Arcminute Imager}  \\
 \hline
  {\small SZA} & {\small Sunyaev-Zel'dovich Array}  \\
 \hline
 {\small ACBAR} & {\small Arcminute Cosmology Bolometer Array Receiver}  \\
 \hline
 {\small DASI} & {\small Degree Angular Scale Interferometer}\\
\hline
 {\small CBI} & {\small Cosmic Background Imager}\\
\hline
 {\small PolarBEAR} & {\small Polarization of Background Radiation}  \\
 \hline
%{\small SPUD} & {\small xxx}  \\
 %\hline
 {\small Clover} & {\small $C_\ell$-over}  \\
 \hline
{\small BICEP} & {\small Background Imaging of Cosmological Extragalatic Polarization}  \\
 \hline
  {\small QUIET} & {\small $Q$/$U$ Imaging ExperimenT}  \\
 \hline
   {\small QUaD} & {\small QUEST at DASI}  \\
 \hline
   {\small CAPMAP} & {\small --}  \\
 \hline
 % {\small Pappa} & {\small Primordial Anisotropy Polarization Pathfinder Array}  \\
 %\hline
   {\small VSA} & {\small Very Small Array}  \\
    \hline
   {\small LHC} & {\small Large Hadron Collider}  \\
 \hline
\hline
\end{tabular}
\caption{ \label{table:acc} Common acronyms for cosmological experiments; mostly limited to CMB experiments.}
\end{center}
\end{table}

\newpage
%--------------------------------------------------------------------------------------------------------------
% BIBLIOGRAPHY

\bibliographystyle{h-physrev3.bst}
\bibliography{InflationWhite}

\end{document}